\documentclass[twocolumn,prd,nofootinbib,preprintnumbers]{revtex4}

\pdfoutput=1
\usepackage{graphicx}
\usepackage{amsmath,bm}
\usepackage[T1]{fontenc}
\usepackage{rotating}
\usepackage{float}
\usepackage{aas_macros}
\usepackage{booktabs}
\floatstyle{plaintop}

\usepackage{color}
\definecolor{naviBlue}{RGB}{0,0,128}
\usepackage[colorlinks  = true,
            linkcolor   = naviBlue,
            urlcolor    = naviBlue,
            citecolor   = naviBlue,
            anchorcolor = naviBlue]{hyperref}
\newcommand{\secref}[1]{\hyperref[sec::#1]{SECTION~\ref*{sec::#1}}}
\newcommand{\subsecref}[1]{\hyperref[subsec::#1]{SECTION.~\ref*{subsec::#1}}}
\newcommand{\figref}[1]{\hyperref[fig::#1]{FIG.$\,$\ref*{fig::#1}}}
\newcommand{\tabref}[1]{\hyperref[tab::#1]{TABLE$\,$\ref*{tab::#1}}}
\newcommand{\eqnref}[1]{\hyperref[eqn::#1]{Eq.$\,$(\ref*{eqn::#1})}}

\newcommand{\diff}{\mathrm{d}}

\newcommand{\av}[1]{\big\langle #1 \big\rangle}	
\newcommand{\eq}{\mathrm{eq}}

\def\beq{\begin{equation}}
\def\eeq{\end{equation}}
\definecolor{darkgreen}{RGB}{0,170,0}
\definecolor{darkgray}{RGB}{110,110,108}
\newcommand{\bea}{\begin{eqnarray}}
\newcommand{\eea}{\end{eqnarray}}
\newcommand{\be}{\begin{equation}}
\newcommand{\ee}{\end{equation}}
\newcommand{\nn}{\nonumber}

\usepackage{array}
\newcolumntype{C}{>{$}c<{$}} 	

\definecolor{purple}{RGB}{160,0,160}
\definecolor{plotpink}{RGB}{205,0,180}
\definecolor{plotcyan}{RGB}{0,215,215}
\definecolor{darkcyan}{RGB}{17,155,155}

\definecolor{plotblue}{RGB}{0,0,235}
\definecolor{plotorange}{RGB}{245,140,0}
\definecolor{darkorange}{RGB}{210,100,0}
\definecolor{plotgreen}{RGB}{30,130,0}
\definecolor{plotred}{RGB}{240,0,0}
\definecolor{darkred}{RGB}{180,0,0}
\definecolor{darkergreen}{RGB}{0,130,3}
\definecolor{darkgreen}{RGB}{0,160,0}
\definecolor{lblue}{RGB}{100,130,205}
\definecolor{ourbrown}{RGB}{151,105,56}
\definecolor{darkblue}{RGB}{10,10,145}
\definecolor{gray}{RGB}{90,90,90}
\definecolor{graycyan}{RGB}{90,113,113}
\definecolor{dgraycyan}{RGB}{90,113,113}
\definecolor{darkgraycyan}{RGB}{66,84,84}
\definecolor{darkH}{RGB}{11,65,188}

\newcommand{\MG}{\textsc{\mbox{MadGraph5\_aMC@NLO}}}

\newcommand{\avb}[1]{\big\langle #1 \big\rangle}

\newcommand{\BSF}{\text{BSF}}
\newcommand{\vrel}{v_\text{rel}}
\def\beq{\begin{equation}}
\def\eeq{\end{equation}}

\newcommand{\E}{\mathrm{e}}

\newcommand{\ie}{\emph{i.e.}}
\newcommand{\eg}{\emph{e.g.}}
\newcommand{\cf}{\emph{cf.}}
\renewcommand{\vec}[1]{\boldsymbol{#1}}

\begin{document}

\title{
Bound-state effects on dark matter coannihilation: \\Pushing the boundaries of conversion-driven freeze-out
}

\author{Mathias Garny}
\affiliation{Physik Department T31, Technische Universit\"at M\"unchen,
James-Franck-Stra\ss e 1,
D-85748 Garching, Germany}
\author{Jan Heisig}
\affiliation{Institute for Theoretical Particle Physics and Cosmology, RWTH Aachen University, Sommerfeldstra\ss e 16, D-52056 Aachen, Germany}
\affiliation{Centre for Cosmology, Particle Physics and Phenomenology (CP3),  Universit\'e catholique de Louvain, Chemin du Cyclotron 2, B-1348 Louvain-la-Neuve, Belgium}
\preprint{TUM-HEP 1379/21}
\preprint{TTK-21-52}

\begin{abstract}
Bound-state formation can have a large impact on the dynamics of dark matter freeze-out in the early Universe, in particular for colored coannihilators. 
We present a general formalism to include an arbitrary number of excited bound states in terms of an effective annihilation cross section, taking bound-state formation, decay and transitions into account, and derive analytic approximations in the limiting cases of no or efficient transitions. Furthermore, we provide explicit expressions for radiative bound-state formation rates for states with arbitrary principal and angular quantum numbers $n,\ell$ for a mediator in the fundamental representation of $SU(3)_c$, as well as electromagnetic transition rates among them in the Coulomb approximation. 
We then assess the impact of bound states within a model with Majorana dark matter and a colored scalar $t$-channel mediator. We consider the regime of coannihilation as well as conversion-driven freeze-out (or coscattering), where the relic abundance is set by the freeze-out of conversion processes. We find that the region in parameter space where the latter occurs is considerably enhanced into the multi-TeV regime. For conversion-driven freeze-out, dark matter is very weakly coupled, evading direct and indirect detection constraints but leading to prominent signatures of long-lived particles that provide great prospects to be probed by dedicated searches at the upcoming LHC runs.
\end{abstract}

\maketitle

\section{Introduction}\label{sec:intro}

Thermal freeze-out of dark matter has proved to be a successful framework for explaining the
measured dark matter abundance in the Universe. However, the sizeable couplings of dark matter to the 
Standard Model (SM) particles required in the simplest realizations of this mechanism have been 
put under pressure by experimental null-results at colliders~\cite{Kahlhoefer:2017dnp}, 
direct~\cite{Undagoitia:2015gya} and indirect~\cite{Gaskins:2016cha} detection
experiments. Hence, fulfilling the relic density constraint often requires the exploration of 
`exceptional'~\cite{Griest:1990kh} regions, \eg~the region where coannihilation 
effects increase the effective annihilation rate~\cite{Edsjo:1997bg}. 

Such effects commonly occur in models 
with a so-called $t$-channel mediator, where the mediator is taken to be odd under the
$Z_2$-parity that stabilizes dark matter and for relatively small mass splittings between the mediator
and the dark matter particle. Prominent and well-studied examples are the sfermion coannihilation regions 
in the minimal supersymmetric standard model (MSSM), see \eg~\cite{Ellis:1999mm,Boehm:1999bj,Ellis:2001nx}. They have, in turn, motivated a wide range of phenomenological studies of $t$-channel mediators in the simplified model framework 
exploring different spin assignments and a wide range of coupling strengths~\cite{Garny:2015wea,Ibarra:2015nca,Delgado:2016umt,Garny:2018icg,Arina:2020udz,Arina:2020tuw}.

While the presence of coannihilating mediators can increase the effective dark matter annihilation rate, 
toward small mass splittings, mediator pair-annihilation alone can become so efficient that
dark matter is rendered under-abundant (seemingly) independent of the dark matter coupling. However, this conclusion is only valid if dark matter remains in chemical 
equilibrium with the mediator during freeze-out through efficient conversions. Dropping this assumption
opens up a cosmologically viable part of the parameter space where the relic density is set by conversion-driven 
freeze-out~\cite{Garny:2017rxs} (or coscattering~\cite{DAgnolo:2017dbv}).\footnote{We use the term \emph{conversion-driven freeze-out} here as the mechanism is not restricted to scattering processes. In general, conversions can proceed via (inverse) decays and scatterings~\cite{Garny:2017rxs}.} In this scenario, thermal decoupling is initiated by the breakdown of efficient conversions between 
dark matter and the coannihilating partner(s). The required coupling is several orders of magnitude smaller 
than the one required to initiate the breakdown of dark matter pair-annihilations. This is due to the significantly 
larger number density for light standard-model initial states in
conversion processes with respect to the dark matter number density.

The boundary between the coannihilation and conversion-driven freeze-out region marks a significant change in the phenomenology
within the parameter space of a given model. While the former is characterized by sizeable couplings that give rise to observable 
signals in conventional dark matter searches, the latter is largely immune to constraints from (in)direct detection  
but predicts long-lived particles with typical lifetimes of the order of millimeters to meters to be searched for at the LHC\@.
The conversion-driven freeze-out region was unexplored terrain for a long time, and often 
flagged \emph{under-abundant} when displaying the viable parameter space in terms of masses, see~\eg~\cite{Garny:2015wea}. Recently, it has been studied in various contexts~\cite{Junius:2019dci,Brummer:2019inq,Maity:2019hre,Blekman:2020hwr,Belanger:2021smw,Herms:2021fql} and often constitutes one of a few regions still allowed within a given model~\cite{Arina:2020tuw}.

\smallskip
For electrically and color-charged coannihilators -- interacting via massless force carriers -- non-perturbative effects such as Sommerfeld enhancement and, in particular, bound-state formation can play an important role in dark matter freeze-out. Radiative bound-state formation has been studied for a variety of dark matter models and for general unbroken Abelian and non-Abelian gauge theories~\cite{Petraki:2015hla,Asadi:2016ybp,Mitridate:2017izz,Harz:2018csl,Binder:2020efn}. The latter is related to earlier results for quarkonium formation inside the quark-gluon plasma obtained in potential nonrelativistic quantum chromodynamics (pNRQCD), see \emph{e.g.}~\cite{Brambilla:2011sg,Yao:2018sgn}.
Recently, next-to-leading-order (NLO) finite temperature corrections of the general singlet-adjoint dipole interactions have been computed~\cite{Binder:2021otw}. 

While it has been shown that bound-state formation effects provide sizeable corrections to the effective annihilation cross section for a coannihilation scenario with a colored mediator~\cite{Liew:2016hqo,Biondini:2018pwp,Harz:2018csl}, it has widely been overlooked
that their effects become considerably more relevant for scenarios with small dark matter couplings such as conversion-driven freeze-out.
As a consequence of the small coupling, freeze-out is a prolonged process and the mediator annihilation down to significantly smaller temperatures (\emph{i.e.}~later times) becomes important. This increases the relevance of bound-state effects further prolonging the freeze-out process.
Furthermore, studies have focussed on the effect of the ground state, while excited bound states become (increasingly) relevant toward smaller temperatures.

\smallskip
In this work, we extend the study of bound-state effects in several aspects:
\begin{itemize}
\item First, we revisit the formulation of the underlying Boltzmann equations in the presence of excited bound states and derive a general framework for incorporating their effects in terms of an effective annihilation cross section of the coannihilator. This general form requires not only the knowledge of bound-state formation and decay rates but also the transition between the various excited states. However, we formulate two meaningful limiting cases considering fully efficient or non-efficient transitions, the latter of which is considered as a (conservative) benchmark scenario. This part is model independent and applies to any set of bound states in general.
\item Secondly, focussing on the case of a colored mediator in the fundamental representation of $SU(3)_c$, we derive general expressions for the bound-state formation rates of arbitrary $n, \ell$ (the principal and angular momentum quantum numbers of the bound state, respectively) and estimates for the transition in some cases. Furthermore, we discuss the impact of higher-order corrections to the bound-state formation and decay rates.
\item Finally, we assess the impact of bound states for a colored $t$-channel mediator model and study the phenomenological consequences of bound-state effects in the coannihilation and conversion-driven freeze-out region. In particular, we observe a drastic shift in the boundary between the two regimes, greatly enlarging the latter region. 
These considerations allow us to assess the importance of the various corrections in the prescription of bound-state effects studied here. 
\end{itemize}

\smallskip

The remainder of the paper is structured as follows. In Sec.~\ref{sec:model} we introduce the considered benchmark model and review the Boltzmann equations that describe both the coannihilation and conversion-driven freeze-out scenario. In Sec.~\ref{sec:bound} we develop our general formalism to include bound states and discuss various limiting cases analytically. In Sec.\,\ref{sec:rates} we compute the involved rates for a colored mediator in the fundamental representation of $SU(3)_c$. Section~\ref{sec:sol} is dedicated to the phenomenological application before concluding in Sec.~\ref{sec:conclusion}. Appendices~\ref{sec:bsfapp} and \ref{sec:NLOcorr} contain further details of the computation of bound-state formation cross sections and discuss NLO QCD effects, respectively.

\smallskip

\section{Model and conversion-driven freeze-out}\label{sec:model}

We consider a simplified $t$-channel model with a singlet Majorana fermion $\chi$ providing the dark matter candidate, and a colored scalar mediator $\tilde q$ that exhibits a Yukawa coupling involving $\chi$ and a right-handed SM quark $q$,
\be
  {\cal L} = \lambda_\chi \tilde q \bar q_R \chi + \text{h.c.}
\ee
The scalar mediator $\tilde q$ transforms as a triplet under $SU(3)_c$, as a singlet under $SU(2)_L$, and has hypercharge that is identical to the one of right-handed SM quarks.
It gives rise to a $t$-channel annihilation diagram for a pair of $\chi$ particles, and the corresponding process $\chi\chi\to \bar q q$ leads to a relic abundance of $\chi$ via thermal freeze-out.

If the masses of $\chi$ and $\tilde q$ are of comparable size, coannihilation processes need to be taken into account as well, in particular mediator pair annihilation,
which dominantly proceeds via the process $\tilde q \tilde q^\dag\to gg$. (Annihilation into a pair of quarks is $p$-wave suppressed.) Being a pure QCD process, its cross section is entirely determined by the strong coupling $\alpha_s$.
Indeed, this contribution can be so large that the $\chi$ relic density falls below the measured dark matter abundance, independently of the value of $\lambda_\chi$~\cite{Garny:2015wea}.

However, this conclusion hinges on the assumption that $\chi$ and $\tilde q$ are in chemical equilibrium during the freeze-out process, \ie~that the corresponding conversion rates are large compared to the Hubble expansion rate $H$ during the freeze-out process. Since the rates of all conversion processes necessarily involve some power of the coupling $\lambda_\chi$, the assumption of chemical equilibrium can be violated if the coupling strength is small enough. In that case, the conversions have to be included along with (co-) annihilation processes in the Boltzmann equations.
This scenario is known as conversion-driven freeze-out~\cite{Garny:2017rxs}, or coscattering~\cite{DAgnolo:2017dbv}.

In general, for the minimal $t$-channel model considered here, the coupled set of Boltzmann equations reads~\cite{Garny:2017rxs}
\begin{widetext}
\begin{eqnarray}
  \frac{\diff Y_{\chi }}{\diff x} &= &\frac{1}{ 3 H}\frac{\mbox{d} s}{\mbox{d} x}
  \left[\,\avb{\sigma_{\chi\chi}v}\left(Y_{\chi}^2-Y_{\chi}^{\eq\,2}\right)+\avb{\sigma_{\chi\tilde q} v}\left(Y_{\chi}Y_{\tilde q}-Y_{\chi}^{\eq}Y_{\tilde q}^{\eq}\right)
  \phantom{\frac{Y_{\tilde q}^{\eq}}{Y_{\chi}^{\eq}}}
  \right. \nonumber \\
  &&\left. -\frac{\Gamma_{\text{conv}} }{s}\left(Y_{\tilde q}-Y_{\chi}\frac{Y_{\tilde q}^{\eq}}{Y_{\chi}^{\eq}}\right)
 -\frac12\avb{\sigma_{\tilde q\tilde q^\dagger\rightarrow \chi\chi}v}\left(Y_{\tilde q}^2-Y_{\chi}^2\frac{Y_{\tilde q}^{\eq\,2}}{Y_{\chi}^{\eq\,2}}\right) \right] \label{eq:BMEchi}\,,\\
  \frac{ \diff Y_{\tilde q } }{\diff x} &= & \frac{1}{ 3 H}\frac{\mbox{d} s}{\mbox{d} x}
  \left[ \,\frac{1}{2}\avb{\sigma_{\tilde q\tilde q^\dagger}v}\left(Y_{\tilde q}^2-Y_{\tilde q}^{\eq\,2}\right)+\avb{\sigma_{\chi \tilde q}v}\left(Y_{\chi}Y_{\tilde q}-Y_{\chi}^{\eq}Y_{\tilde q}^{\eq}\right) 
  \phantom{\frac{Y_{\tilde q}^{\eq}}{Y_{\chi}^{\eq}}}
   \right. \nonumber \\
  &&\left. +\frac{\Gamma_{\text{conv}}}{s}\left(Y_{\tilde q}-Y_{\chi}\frac{Y_{\tilde q}^{\eq}}{Y_{\chi}^{\eq}}\right)
  +\frac12\avb{\sigma_{\tilde q\tilde q^\dagger\rightarrow \chi\chi}v}\left(Y_{\tilde q}^2-Y_{\chi}^2\frac{Y_{\tilde q}^{\eq\,2}}{Y_{\chi}^{\eq\,2}}\right) \right]\label{eq:BMEsqu}\,,
\end{eqnarray}
\end{widetext}
where $x=m_\chi/T$ and $Y_i= n_i/s$, with number density $n_i$ and entropy density $s$, with
 \beq
 \frac{1}{ 3 H}\frac{\mbox{d} s}{\mbox{d} x} = -\sqrt{\frac{8}{45 }} \pi M_\text{pl} \frac{m_\chi}{x^2} \sqrt{g_\star}\,,
 \eeq
where $M_\text{pl}\simeq 2.4 \times 10^{18}\,$GeV is the reduced Planck mass.
$Y_{\tilde q}$ represents the summed contribution of the mediator and its anti-particle, 
\be
  Y_{\tilde q} \equiv (g_{\tilde q}+g_{\tilde q^\dag})\frac{1}{s}\int \frac{\diff^3p}{(2\pi)^3} f_{\tilde q}(p)\,,
\ee
leading to the various factors $1/2$. Here, $g_{\tilde q}=g_{\tilde q^\dag}=N_c=3$, and $f_{\tilde q}$ is the distribution function that is assumed to
be identical for particles and antiparticles as well as all colors.
The processes in the first line of each equation denote the usual (co-)annihilation processes into SM particles.

The conversion terms in the second line of each equation can be split into processes of the form
$\tilde q\to \chi$ and $\tilde q\tilde q^\dagger\rightarrow \chi\chi$.
The former case requires accompanying SM particles, and can be further decomposed
into $1\to 2$ and $2\to 2$ processes,
\be\label{eq:GamConv}
  \Gamma_\text{conv} = \Gamma_{\tilde q \to \chi q} + \Gamma_{\tilde q X\rightarrow \chi Y}\,,
\ee
with
\beq
   \Gamma_{\tilde q \to \chi q} \equiv \Gamma \av{\frac{1}{\gamma}} = \Gamma \frac{K_1\left( m_{\tilde q} /T\right) }{ K_2 \left( m_{\tilde q} /T \right) }\,,
\eeq
where $\Gamma$ is the decay rate at rest, and
\bea
 \Gamma_{\tilde q X\rightarrow \chi Y} &=&  \sum_{k,l}\big \langle\sigma_{\tilde q k  \rightarrow \chi l}v \big\rangle \,n_k^{\eq}= \sum_{k,l} \frac{g_k}{4 \pi^2 m_{\tilde q}^2 K_2(m_{\tilde q}/T)}\nn\\
 && \times \int \diff s \,\sqrt{s} \,p_\text{in}^2(s) \,\sigma_{\tilde q k  \rightarrow \chi l}(s) K_1(\sqrt{s}/T)\,,
\eea
where $n_i^\eq=T/(2\pi^2)\,g_im_i^2K_2(m_i/T)$ and $K_i$ denote modified Bessel functions of the second kind.
Depending on kinematic constraints further $1\to 3$, $1\to 4$ or $2\to 3$ process can be relevant, especially for a coupling to top quarks, $\tilde q=\tilde t$~\cite{Garny:2018icg}.
In the following, we focus on a coupling to bottom quarks, $\tilde q=\tilde b$, and include the processes stated in eq.~\eqref{eq:GamConv}.

The set of Boltzmann equations can describe both coannihilations in and out of chemical equilibrium, with well-known simplifications being possible in the former case
by summing both equations~\cite{Edsjo:1997bg}. Out of chemical equilibrium, the coupled set of equations needs to be solved.
However, since the coupling $\lambda_\chi$ is small in this case, all terms except for the ones involving $\avb{\sigma_{\tilde q\tilde q^\dagger}v}$ and $\Gamma_\text{conv}$
can be neglected for conversion-driven freeze-out. The former process is considerably Sommerfeld enhanced for small relative velocities, due to the attractive potential generated by
gluon exchange in the color singlet configuration of the $\tilde q\tilde q^\dagger$ pair~\cite{Ibarra:2015nca}. In addition, the same potential leads to the formation of bound states~\cite{Mitridate:2017izz,Biondini:2018pwp,Harz:2018csl,Biondini:2018ovz}. 
In this work, we improve the computations of the relic density in the coannihilation and conversion-driven freeze-out scenario by considering bound-state effects, including an exploration of the role of excited states. 

\section{Including bound states}\label{sec:bound}

Within the $t$-channel model, bound states of $\tilde q\tilde q^\dag$ pairs in the color singlet configuration exist
and can contribute to the freeze-out process.
We consider an extension of the Boltzmann equation by including a set of bound states ${\cal B}_i$, enumerated by an abstract index $i$, and with $g_{{\cal B}_i}$ internal degrees of freedom.
Within the model considered here, the bound states are characterized by their $n$ and $\ell$ quantum numbers, $i\equiv (n,\ell)$ and $g_{{\cal B}_{n\ell}}=2 \ell+1$, but the discussion
in this section applies to any set of bound states in general.

We add a Boltzmann equation for the yield $Y_{{\cal B}_i}=n_{{\cal B}_i}/s$ for each bound state, taking into account ionization (or equivalently breaking) into an unbound $\tilde q\tilde q^\dag$ pair via gluon or photon absorption, direct decay of the bound state into SM particles, and transitions between two bound states. In addition, the collision term in the Boltzmann equation of the mediator $\tilde q$ picks up an extra term  due to ionization and its inverse process, recombination [or equivalently bound-state formation (BSF)]. The changes in the Boltzmann equations compared to eqs.~\eqref{eq:BMEchi} and \eqref{eq:BMEsqu} are given by
\begin{widetext}
\begin{eqnarray}
  \frac{ \diff Y_{{\cal B}_i}}{\diff x}& = &  \frac{1}{ 3 H s}\frac{\mbox{d} s}{\mbox{d} x}
  \Bigg[
  \Gamma_\text{ion}^i \left(Y_{{\cal B}_i}-Y_{{\cal B}_i}^{\eq}\, \frac{Y_{\tilde q}^2}{Y_{\tilde q}^{\eq\,2}}\right) 
  + \Gamma_\text{dec}^i \left(Y_{{\cal B}_i}-Y_{{\cal B}_i}^{\eq} \right) 
  -\sum_{j\neq i}\, \Gamma_\text{trans}^{j\to i} \left(Y_{{\cal B}_j}-Y_{{\cal B}_i}\, \frac{Y_{{\cal B}_j}^\eq}{Y_{{\cal B}_i}^\eq}\right) 
  \Bigg]\label{eq:BMEboundBnneq1}\,, \\
  \frac{ \diff Y_{\tilde q } }{\diff x} &=&   \left( \frac{ \diff Y_{\tilde q } }{\diff x}\right)_{\text{Eq.\,\eqref{eq:BMEsqu}}} 
  +\frac{1}{ 3 H s}\frac{\mbox{d} s}{\mbox{d} x}
  \sum_i\frac{1}{2}\avb{\sigma_{\BSF,i} v}\left(Y_{\tilde q}^2-Y_{\tilde q}^{\eq\,2}\frac{Y_{{\cal B}_i}}{Y_{{\cal B}_i}^{\eq}}\right)\,.\label{eq:BMEboundtqnneq1}
\end{eqnarray}
\end{widetext}
The ionization rate $\Gamma_\text{ion}^i$ is related to the thermally averaged recombination cross section $\avb{\sigma_{\BSF,i} v}$ via the Milne relation
\be\label{eq:Milne}
  \Gamma_\text{ion}^i = \frac{s}{4} \,\frac{Y_{\tilde q}^{\eq\,2}}{Y_{{\cal B}_i}^{\eq}} \avb{\sigma_{\BSF,i} v} \,,
\ee
originating from the detailed balance condition in thermal equilibrium. Indeed, the Milne relation ensures that the ionization and recombination terms
drop out in the sum $\diff(Y_{\tilde q}+2\sum_i Y_{{\cal B},i})/\diff x$, consistent with the conservation of the total number of $\tilde q$ and $\tilde q^\dag$ in the absence of decays.
Note that in the non-relativistic limit
\be
  \frac{s}{4} \,\frac{Y_{\tilde q}^{\eq\,2}}{Y_{{\cal B}_i}^{\eq}} \simeq \frac{g_{\tilde q}^2}{g_{{\cal B}_i}}\left(\frac{T m_{\tilde q}^2}{2\pi m_{{\cal B}_i}}\right)^{3/2}\,\E^{-E_{{\cal B}_i}/T}\,,
\ee
where $E_{{\cal B}_i}=2m_{\tilde q}-m_{{\cal B}_i}>0$ is the binding energy, and we used that $Y_{\tilde q}$ denotes the yield of the sum of $\tilde q$ and $\tilde q^\dag$.
In addition, detailed balance requires
\be\label{eq:transMilne}
  \Gamma_\text{trans}^{i\to j} = \Gamma_\text{trans}^{j\to i}\frac{Y_{{\cal B}_j}^\eq}{Y_{{\cal B}_i}^\eq}\,.
\ee
Also, here, we can see that transition terms drop out when summing the Boltzmann equations for all bound states, as required.

Before discussing explicit expressions for corresponding rates in Sec.\,\ref{sec:rates}, we investigate generic features of the coupled set of equations.

\subsection{Single bound state}\label{sec:single}

We first recall the case of a single bound state ${\cal B}$. In a typical cosmological setting, the ionization and decay rates (mediated by the strong interaction)
are much larger than $H$. In this case, the density of bound states almost instantaneously adjusts to a quasi-stationary number (from the point of view of cosmological versus strong interaction timescales)
that can be obtained by setting the left-hand side of the Boltzmann equation for ${\cal B}$ to zero, turning it into an algebraic equation~\cite{Ellis:2015vaa}.
For the case of a single bound state (dropping the index $i$ and transition terms), one obtains
\begin{equation}\label{eq:ybapp}
  \frac{Y_{\cal B}}{Y_{\cal B}^{\eq}}= \frac{ \Gamma_\text{ion}\, Y_{\tilde q}^2/Y_{\tilde q}^{\eq\,2}  + \Gamma_\text{dec} }{\Gamma_\text{ion} + \Gamma_\text{dec}}\,.
\end{equation}
Inserting this relation in eq.~\eqref{eq:BMEboundtqnneq1} yields the same form as eq.~\eqref{eq:BMEsqu} but with the substitution
\begin{equation}\label{eq:sigmaveff}
  \avb{\sigma_{\tilde q\tilde q^\dagger}v}\to \avb{\sigma_{\tilde q\tilde q^\dagger}v}_\text{eff} = 
  \avb{\sigma_{\tilde q\tilde q^\dagger}v}+\avb{\sigma_\BSF v}\frac{\Gamma_\text{dec}}{\Gamma_\text{ion} + \Gamma_\text{dec}}\,.
\end{equation}
This means it is sufficient to solve the Boltzmann equations for $\tilde q$ and $\chi$, while the impact of the bound state is captured by replacing the $\tilde q\tilde q^\dag$ annihilation cross section
by the effective cross section.

In the limit $H\ll \Gamma_\text{dec}\ll \Gamma_\text{ion}$ the ionization and recombination processes establish equilibrium between the bound state and unbound $\tilde q$ (ionization equilibrium). The corresponding rates
therefore drop out of the effective cross section, which only depends on the decay rate $\Gamma_\text{dec}$, as can be seen using the Milne relation, eq.~\eqref{eq:Milne},
\be\label{eq:sigeffsingleioneq}
  \avb{\sigma_{\tilde q\tilde q^\dagger}v}_\text{eff} \to
  \avb{\sigma_{\tilde q\tilde q^\dagger}v}
  + \frac{g_{{\cal B}}}{g_{\tilde q}^2}\left(\frac{2\pi m_{{\cal B}}}{T m_{\tilde q}^2}\right)^{3/2}\,\E^{E_{{\cal B}}/T}\,\Gamma_\text{dec}\,.
\ee
The effective cross section increases exponentially with falling temperature, due to the energetic preference for bound states in equilibrium.
This increase stops once the ionization rate, which itself becomes exponentially suppressed at low temperatures, falls below the decay rate, and ionization equilibrium breaks down.
Therefore, at low enough temperatures, the regime $H\ll \Gamma_\text{ion}\ll \Gamma_\text{dec}$ becomes relevant, for which
\be
  \avb{\sigma_{\tilde q\tilde q^\dagger}v}_\text{eff} \to
  \avb{\sigma_{\tilde q\tilde q^\dagger}v}
  + \avb{\sigma_\BSF v}\,.
\ee
In that limit, any bound state that forms decays almost immediately, and therefore the effective cross section is only sensitive to the recombination cross section $\avb{\sigma_\BSF v}$.

\subsection{Multiple bound states}\label{sec:multi}

Let us now generalize the previous findings to a set of bound states. When assuming as before that all relevant ionization, decay and transition rates are much larger than $H$,
we obtain a set of coupled algebraic equations for the yields $Y_{{\cal B}_i}$ from setting the left-hand sides of the Boltzmann equations~\eqref{eq:BMEboundBnneq1} to zero. It can be written as
\begin{equation}\label{eq:ybappn}
  \frac{Y_{{\cal B},i}}{Y_{{\cal B},i}^\eq}=\frac{ \Gamma_\text{ion}^i}{\Gamma^i}\frac{ Y_{\tilde q}^2}{Y_{\tilde q}^{\eq\,2}}  + \frac{\Gamma_\text{dec}^i}{\Gamma^i}
  +\sum_{j\neq i}\frac{\Gamma_\text{trans}^{i\to j}}{\Gamma^i} \frac{Y_{{\cal B},j}}{Y_{{\cal B},j}^\eq}\,,
\end{equation}
where we used eq.~\eqref{eq:transMilne} and introduced the total width of ${\cal B}_i$,
\be
  \Gamma^i = \Gamma_\text{ion}^i + \Gamma_\text{dec}^i + \sum_{j\neq i}\Gamma_\text{trans}^{i\to j}\,.
\ee
From the structure of the Boltzmann equation, it is \emph{a priori} not clear whether the impact of bound states can be captured by an effective cross section
when inserting the solution to eq.~\eqref{eq:ybappn} into the Boltzmann equation~\eqref{eq:BMEboundtqnneq1} for $\tilde q$. However, this turns out to
be the case in general. To see it, we rewrite eq.~\eqref{eq:ybappn} in the form
\be\label{eq:yi}
   y_i-1-\sum_{j\not=i} \frac{\Gamma_\text{trans}^{i\to j}}{\Gamma^i}(y_j-1)= \frac{\Gamma_\text{ion}^i}{\Gamma^i}(y^2-1)\,,
\ee
where we defined $y_i\equiv Y_{{\cal B},i}/Y_{{\cal B},i}^\eq$ and $y\equiv  Y_{\tilde q}/Y_{\tilde q}^{\eq}$.
Introducing the matrix
\be
  M_{ij}\equiv\delta_{ij}-\frac{\Gamma_\text{trans}^{i\to j}}{\Gamma^i}\,,
\ee
the solution for the bound-state abundances reads
\be
  y_i = 1 +\sum_j(M^{-1})_{ij}\frac{\Gamma_\text{ion}^j}{\Gamma^j}(y^2-1)\,.
\ee
Inserting it in the Boltzmann equation~\eqref{eq:BMEboundtqnneq1} for $\tilde q$ indeed yields a contribution that has the form of the
annihilation term, involving in particular a factor $y^2-1$. Therefore, provided the rates are large compared to the expansion rate,
the impact of a set of bound states can in general be captured by an effective cross section, given by
\begin{equation}\label{eq:effgeneral}
  \avb{\sigma_{\tilde q\tilde q^\dagger}v}_\text{eff} = 
  \avb{\sigma_{\tilde q\tilde q^\dagger}v}+\sum_{i} \avb{\sigma_{\BSF,i} v}R_i\,,
\end{equation}
with
\be\label{eq:Ri}
  R_i \equiv 1-\sum_j(M^{-1})_{ij}\frac{\Gamma_\text{ion}^j}{\Gamma^j}
\ee
The effective cross section, eq.~\eqref{eq:effgeneral}, describes the impact of an arbitrary number of bound states on the $\tilde q$ abundance, which can
all individually be populated by recombination processes, decay into SM particles, and undergo a network of transitions among them, with the corresponding rates entering in the determination of $R_i$.
For a given setup, the $R_i$ can be determined numerically.
Nevertheless, it is instructive to study two limiting cases analytically.

\subsubsection{No transition limit}\label{sec:notrans}

In the limit $\Gamma_\text{trans}^{i\to j}\ll \Gamma_\text{dec}^i , \Gamma_\text{ion}^i$ we can neglect the
transition terms, such that $M_{ij}\to\delta_{ij}$ becomes the unity matrix, and the total width depends
only on ionization and decay rates. The effective cross section becomes
\begin{equation}\label{eq:effnfirst}
  \avb{\sigma_{\tilde q\tilde q^\dagger}v}_\text{eff} = 
  \avb{\sigma_{\tilde q\tilde q^\dagger}v}+\sum_{i} \avb{\sigma_{\BSF,i} v}\frac{\Gamma^{i}_\text{dec}}{\Gamma^{i}_\text{ion} + \Gamma^{i}_\text{dec}}\,.
\end{equation}
In the absence of transitions, each bound state therefore gives a contribution to the effective cross section that is analogous to the case for a single bound state, see eq.~\eqref{eq:sigmaveff}.
In particular, each summand exhibits the limiting cases of ionization equilibrium ($\Gamma^{i}_\text{ion}\gg \Gamma^{i}_\text{dec}$) or instantaneous decay ($\Gamma^{i}_\text{ion}\ll \Gamma^{i}_\text{dec}$)
in close analogy to the case of a single bound state.

\subsubsection{Efficient transition limit}\label{sec:efftrans}

In the limit $\Gamma_\text{trans}^{i\to j}\gg \Gamma_\text{dec}^i , \Gamma_\text{ion}^i$, we expect that the transitions establish chemical equilibrium among the bound states,
\begin{equation}\label{eq:BSchemeq}
  \frac{Y_{{\cal B},j}}{Y_{{\cal B},i}}\to \frac{Y_{{\cal B},j}^\eq}{Y_{{\cal B},i}^\eq}\simeq \E^{(E_{{\cal B}_j}-E_{{\cal B}_i})/T}\,,
\end{equation}
which is indeed a solution to eq.~\eqref{eq:ybappn} in that limit. The most straightforward way to derive the effective cross section in that limit
is to proceed similarly to the case of coannihilations~\cite{Edsjo:1997bg}, introducing
\begin{equation}
  Y_{\cal B} = \sum_i Y_{{\cal B},i}\,,
\end{equation}
and summing up all Boltzmann equations~\eqref{eq:BMEboundBnneq1} for the ${\cal B}_i$, such that the transition terms drop out.
Using \eqref{eq:BSchemeq} to write
\be\label{eq:YBichemeq}
 Y_{{\cal B},i}=Y_{{\cal B}} \frac{ Y_{{\cal B},i}^\eq }{ Y_{{\cal B}}^\eq }\,,
 \ee
one obtains
\begin{equation}\label{eq:YBeff}
  \frac{ \diff Y_{\cal B}}{\diff x}=  \frac{1}{ 3 H s}\frac{\mbox{d} s}{\mbox{d} x}
  \left[ 
  \Gamma_\text{ion}^\text{eff} \left(Y_{\cal B}-Y_{\cal B}^{\eq} y^2 \right) 
  + \Gamma_\text{dec}^\text{eff} \left(Y_{\cal B}-Y_{\cal B}^{\eq} \right) 
  \right]\,,
\end{equation}
with effective ionization and decay rates
\begin{equation}
\Gamma_\text{ion/dec}^\text{eff} =  \frac{ \sum_i \Gamma_\text{ion/dec}^i\,Y_{{\cal B},i}^\eq}{Y_{\cal B}^\eq}
\end{equation}
Setting again the left-hand side of the Boltzmann equation~\eqref{eq:YBeff} to zero, and inserting the resulting algebraic expression together with eq.~\eqref{eq:YBichemeq}
into eq.~\eqref{eq:BMEboundtqnneq1} yields
\begin{equation}\label{eq:sigmaveffeff}
  \avb{\sigma_{\tilde q\tilde q^\dagger}v}_{\text{eff}} = 
  \avb{\sigma_{\tilde q\tilde q^\dagger}v}+\avb{\sigma_\BSF v}_{\!\text{sum}}\;\frac{\Gamma_\text{dec}^\text{eff}}{\Gamma_\text{ion}^\text{eff} + \Gamma_\text{dec}^\text{eff}}\,,
\end{equation}
where $\avb{\sigma_\BSF v}_{\!\text{sum}}=\sum_i \avb{\sigma_{\BSF,i} v}$.
The result is similar in form to the case of a single bound state, eq.~\eqref{eq:sigmaveff}, but with the ionization and decay rates replaced by a
thermal average over all bound states and the recombination cross section replaced by the sum.

It turns out that obtaining this result directly from the general expression eq.~\eqref{eq:effgeneral} is tedious.
The reason is that naively neglecting the ionization and decay rates in the total width would lead to a singular matrix $M_{ij}$.
However, by carefully expanding the abundances around the chemical equilibrium solution $y_i=$const., and treating
$\Gamma_\text{ion}^i/\Gamma^i$ and $\Gamma_\text{dec}^i/\Gamma^i$ as small, one ultimately arrives at the same expression, eq.~\eqref{eq:sigmaveffeff}.

We also note that using the Milne relation, eq.~\eqref{eq:Milne}, for each bound state, one finds
\begin{equation}\label{eq:Milneeff}
  \Gamma_\text{ion}^\text{eff} = \frac{s}{4} \,\frac{Y_{\tilde q}^{\eq\,2}}{Y_{{\cal B}}^{\eq}} \avb{\sigma_{\BSF} v}_{\!\text{sum}}\,,
\end{equation}
\ie~the summed recombination cross section and the effective ionization rate satisfy a generalized Milne relation.
This implies that, in analogy to the case of a single bound state, within the regime of ionization equilibrium ($\Gamma_\text{ion}^\text{eff} \gg \Gamma_\text{dec}^\text{eff}$),
the effective cross section becomes independent of the recombination cross section, and only depends on the effective decay rate.
In the opposite limit $\Gamma_\text{ion}^\text{eff} \ll \Gamma_\text{dec}^\text{eff}$ of almost instantaneous decay, the decay rate drops out, and the effective
cross section depends only on $\avb{\sigma_\BSF v}_{\!\text{sum}}$.

\subsubsection{Ionization equilibrium}\label{sec:ioneq}

The limit of ionization equilibrium is somewhat orthogonal to the two limiting cases considered above.
When ionization and recombination processes are assumed to be efficient enough to establish ionization equilibrium, the
effective cross section approaches the universal form
\be\label{eq:sigeffioneq}
  \avb{\sigma_{\tilde q\tilde q^\dagger}v}_\text{eff} \to
  \avb{\sigma_{\tilde q\tilde q^\dagger}v}
  + \sum_i \frac{g_{{\cal B}_i}}{g_{\tilde q}^2}\left(\frac{2\pi m_{{\cal B}_i}}{T m_{\tilde q}^2}\right)^{3/2}\,\E^{E_{{\cal B}_i}/T}\,\Gamma_\text{dec}^i\,,
\ee
which is a straightforward generalization of eq.~\eqref{eq:sigeffsingleioneq} and independent of ionization rates $\Gamma_\text{ion}^i$ as well as transition rates
$\Gamma_\text{trans}^{i\to j}$. The reason is that efficient ionization and recombination processes establish chemical equilibrium with the unbound $\tilde q$ particles in that case for each bound state. This means, in turn, that they are in chemical equilibrium among each other, such that the transition processes play no role for their relative abundances in that limit. This result agrees with the finding in~\cite{binderPhD}, in which a set of bound states in ionization equilibrium is considered.

Indeed, it is easy to see that eq.~\eqref{eq:sigeffioneq} follows from both the effective cross section in either the limiting case of no transitions or the case of efficient transitions when assuming in addition that $\Gamma_\text{ion}^i \gg \Gamma_\text{dec}^i$. Moreover, the fact that eq.~\eqref{eq:sigeffioneq} is even valid independently of the size of transition rates can be seen by noticing that the derivation presented in Sec.\,\ref{sec:efftrans} relies only on the assumption of chemical equilibrium among the bound states, which is satisfied in ionization equilibrium. 

Therefore, as long as ionization equilibrium holds, the effective cross section is only sensitive to the bound-state decay rates, independently of the size of transition and ionization rates.

In a realistic setup, the limiting assumptions made above may be too restrictive and at best hold only for a subset of bound states and a subset of the corresponding ionization, decay or transition processes. In this case, the effective cross section can be computed using the general result, eq.~\eqref{eq:effgeneral}.

\section{Rates}\label{sec:rates}

While the discussion in the previous section was generic, we focus on the set of bound states and ionization, decay and transition rates that are relevant for the scalar mediator $\tilde q$ that carries hypercharge and transforms under the fundamental representation of $SU(N_c)$ with $N_c=3$ in the following.

A heavy ($m_{\tilde q}\gg \Lambda_\text{QCD}$), non-relativistic $\tilde q \tilde q^\dag$ pair can be described by two wave functions $\psi^{[\bm{R}]}$, one for the color octet ($[\bm{8}]$) and one
for the color singlet ($[\bm{1}]$) configuration. They obey a Schr{\"o}dinger equation with kinetic energy $\vec{p}_\text{rel}^2/(2\mu)$, where
\be
  \mu=m_{\tilde q}/2\,,
\ee
is the reduced mass, and potential in Coulomb approximation~\cite{Harz:2018csl}
\be
  V_{[\bm{R}]}(r) = - \frac{\alpha^\text{eff}_{[\bm{R}]}}{r}\,,
\ee
with effective coupling strength
\be
  \alpha^\text{eff}_{[\bm{R}]} = \alpha_s\frac{ C_2^{[\bm{3}]} + C_2^{[\overline{\bm{3}}]} - C_2^{[\bm{R}]} }{2}\,.
\ee
Here $C_2^{[\bm{R}]}$ denotes the quadratic Casimir of $SU(N_c)$ with
$C_F=C_2^{[\bm{3}]}=(N_c^2-1)/(2N_c)=4/3$ and $C_A=C_2^{[\bm{8}]}=N_c=3$, and $\alpha_s=g_s^2/(4\pi)$ is related to the strong coupling constant.
Thus,
\bea
  \alpha^\text{eff}_{[\bm{1}]} &=&  C_F\alpha_s = \frac43 \alpha_s\,, \nn\\
  \alpha^\text{eff}_{[\bm{8}]} &=&  (C_F-C_A/2)\alpha_s  =  -\frac16\alpha_s\,.
\eea
The singlet configuration feels an attractive potential, while it is repulsive for the octet.
Therefore, bound states 
\be
  {\cal B}_{n\ell} \equiv {\cal B}_{n\ell}^{[\bm{1}]}
\ee
exist for the singlet only. Note that we treat the $m$ quantum number as an internal degree of freedom of the bound state in the Boltzmann equation,
and therefore label the bound states by $n$ and $\ell$ only. 

In the Coulomb approximation, the bound states are described by
hydrogen-like wave functions $\psi^{[\bm{1}]}_{n\ell m}$, with the fine-structure constant replaced by $\alpha^\text{eff}_{[\bm{1}]}$
and the electron mass by the reduced mass $\mu$. On the other hand, unbound scattering states $\psi^{[\bm{R}]}_{\vec p_\text{rel}}$ exist
for both the octet and singlet, with wave functions containing the respective effective coupling strength (see App.\,\ref{sec:bsfapp}). 

\subsection{Ionization and recombination}\label{sec:bsf}

The leading-order QCD process for bound-state formation is
\be
  (\tilde q \tilde q^\dag)^{[\bm{8}]}\to {\cal B}_{n\ell}^{[\bm{1}]}+g\,,
\ee
where the initial state corresponds to a scattering state in the octet configuration due to color conservation.
The matrix element can be computed within pNRQCD analogously to hydrogen recombination~\cite{Bethe:1957ncq,Yao:2018sgn},
with a dipole interaction Hamiltonian of the form $g_s \omega \vec{r}\cdot\vec{E}$
where $\vec{E}=t^a\vec{E}^a$ is the color-electric field, $\vec{r}$ is the relative coordinate, and 
\be
  \omega=E_{{\cal B}_{n\ell}}+\frac{p_\text{rel}^2}{2\mu}=E_{{\cal B}_{n\ell}} + \frac12\mu v_\text{rel}^2\,,
\ee
is the energy difference of initial and final state, which corresponds to the energy of the emitted gluon in the non-relativistic limit.

The thermally averaged ionization (or breaking) rate and recombination (or bound-state formation) cross section are given by~\cite{Harz:2018csl}
\bea
  \Gamma_\text{ion}^{n\ell} &=& 
  \frac{g_{\tilde q}^2 \mu^3}{(2\pi)^3 g_{{\cal B}_i}}\int \diff^3\vrel  \,f_g(\omega)\, \sigma_{\BSF,{n\ell}} \vrel\,,
  \label{eq:GammaBreak}\nn\\
  \langle\sigma_{\BSF,{n\ell}}v\rangle &=&
  \left(\frac{\mu}{2\pi T}\right)^{3/2} \int \diff^3 \vrel \, \exp \left(-\frac{\mu \vrel^2}{2T}\right) \times \nn\\
  && [1+f_g(\omega)]\, \sigma_{\BSF,{n\ell}} \vrel \,,
  \label{eq:sigmaBSFav}
\eea
which can be checked to satisfy the Milne relation, eq.~\eqref{eq:Milne}, with $f_g(\omega)=1/(\E^{\omega/T}-1)$.
The recombination cross section 
can be expressed as~\cite{Yao:2018sgn} 
\be
  \sigma_{\text{BSF},n\ell} v_\text{rel} =\frac{\omega}{2\pi N_c^2}(2\ell+1)|{\cal M}|^2\,,
\ee
with the matrix element for the QCD process given by
\be
    |{\cal M}|^2_{(\tilde q \tilde q^\dag)^{[\bm{8}]}\to {\cal B}_{n\ell}^{[\bm{1}]}+g} = \frac23 g_s^2C_F \omega^2 |\langle\psi_{n\ell}^{[\bm{1}]}|\vec r|\psi_{\vec p_\text{rel}}^{[\bm{8}]}\rangle|^2\,,
\ee
and
\be
 |\langle\psi_{n\ell}^{[\bm{1}]}|\vec r|\psi_{\vec p_\text{rel}}^{[\bm{8}]}\rangle|^2
 =\frac{1}{2\ell+1}\sum_m |\langle\psi_{n\ell m}^{[\bm{1}]}|\vec r|\psi_{\vec p_\text{rel}}^{[\bm{8}]}\rangle|^2\,.
\ee
One can also consider the analogous electromagnetic process,
\be
  (\tilde q \tilde q^\dag)^{[\bm{1}]}\to {\cal B}_{n\ell}^{[\bm{1}]}+\gamma\,,
\ee
that proceed from a color singlet scattering to bound state
and matrix element obtained from the electromagnetic dipole interaction,
\be
    |{\cal M}|^2_{(\tilde q \tilde q^\dag)^{[\bm{1}]}\to {\cal B}_{n\ell}^{[\bm{1}]}+\gamma} = \frac23 e^2Q_{\tilde q}^2 \omega^2 |\langle\psi_{n\ell}^{[\bm{1}]}|\vec r|\psi_{\vec p_\text{rel}}^{[\bm{1}]}\rangle|^2\,,
\ee
where $Q_{\tilde q}=1/3$ is the electric charge of $\tilde q$.

\begin{table*}
  \centering
    \begin{tabular}{l@\quad l@\quad l@\qquad l}
    \toprule
      $n$&$\ell$&&$s^\text{BSF}_{n\ell}(\zeta_s,\zeta_b)$\\ 
    \midrule
      1&0 & 1s & $(-2 \zeta_b + \zeta_s)^2 (1 + \zeta_s^2)$ \\
      2&0 & 2s & $8 (1 + \zeta_s^2) \Big(4 \zeta_b^3 + 4 \zeta_s - 9 \zeta_b^2 \zeta_s + 4 \zeta_b (-2 + \zeta_s^2)\Big)^2$ \\
      2&1 & 2p & $2\Big(3 \zeta_b^6 - 36 \zeta_b^5 \zeta_s + 16 \zeta_b^3 \zeta_s (11 - 19 \zeta_s^2)  
                 + 12 \zeta_b^4 (-2 + 13 \zeta_s^2) + 64 \zeta_s^2 (4 + 2 \zeta_s^2 + \zeta_s^4) $\\
         &&    & $- 64 \zeta_b \zeta_s (10 + 5 \zeta_s^2 + 4 \zeta_s^4) + 48 \zeta_b^2 (9 + 2 \zeta_s^2 + 8 \zeta_s^4)\Big)$ \\
      3&0 & 3s & $3 (1 + \zeta_s^2) \Big(243 \zeta_s + \zeta_b (-486 + 324 \zeta_b^2 - 22 \zeta_b^4 + 123 \zeta_b (-6 + \zeta_b^2) \zeta_s 
                 + 36 (9 - 5 \zeta_b^2) \zeta_s^2 + 72 \zeta_b \zeta_s^3)\Big)^2$\\
      3&1 & 3p & $24 \Big(\zeta_b^{10} - 28 \zeta_b^9 \zeta_s + 12 \zeta_b^7 \zeta_s (97 - 121 \zeta_s^2) + 
                  4 \zeta_b^8 (-15 + 73 \zeta_s^2) + 8748 \zeta_s^2 (4 + 2 \zeta_s^2 + \zeta_s^4) $\\
              &&& $- 36 \zeta_b^5 \zeta_s (579 - 379 \zeta_s^2 + 218 \zeta_s^4) + 
                 18 \zeta_b^6 (159 - 309 \zeta_s^2 + 239 \zeta_s^4) $\\
              &&& $+ 2916 \zeta_b \zeta_s (-30 - 7 \zeta_s^2 - 8 \zeta_s^4 + 2 \zeta_s^6) - 
                 324 \zeta_b^3 \zeta_s (-285 + 69 \zeta_s^2 - 106 \zeta_s^4 + 14 \zeta_s^6) $\\
              &&& $+    108 \zeta_b^4 (-225 + 364 \zeta_s^2 - 237 \zeta_s^4 + 77 \zeta_s^6) + 
                 243 \zeta_b^2 (243 - 230 \zeta_s^2 + 78 \zeta_s^4 - 96 \zeta_s^6 + 4 \zeta_s^8)\Big) $\\
      3&2 & 3d & $48 (1 + \zeta_s^2) \Big(20 \zeta_b^6 - 180 \zeta_b^5 \zeta_s + 36 \zeta_b^3 \zeta_s (34 - 29 \zeta_s^2) + 
                 27 \zeta_b^4 (-8 + 23 \zeta_s^2) - 324 \zeta_b \zeta_s (33 + 5 \zeta_s^2 + 2 \zeta_s^4) $\\
              &&& $+    54 \zeta_b^2 (126 - \zeta_s^2 + 20 \zeta_s^4) + 81 \zeta_s^2 (53 + 2 \zeta_s^2 (5 + \zeta_s^2))\Big)$\\
      \bottomrule
    \end{tabular}
  \caption{Polynomials $s^\text{BSF}_{n\ell}(\zeta_s,\zeta_b)$ entering the bound-state formation cross section.}
  \label{tab:snl}
\end{table*}

We evaluate the strong couplings entering in the scattering ($s$) and bound ($b$) state wave function
at renormalization scale of the typical momentum transfer related to bound and scattering states, respectively, using the notation
\bea\label{eq:alphaeff}
  \alpha_b^\text{eff} &=& \alpha_{[\bm{1}]}^\text{eff}(\mu_{\overline{\text{MS}}}=\mu\alpha_b^\text{eff}/n )\,,\nn\\
  \alpha_s^\text{eff} &=& \left\{\begin{array}{lll}
    \alpha_{[\bm{8}]}^\text{eff}(\mu_{\overline{\text{MS}}}=\mu v_\text{rel}) && (\tilde q \tilde q^\dag)^{[\bm{8}]}\to {\cal B}_{n\ell}^{[\bm{1}]}+g\,, \\
    \alpha_{[\bm{1}]}^\text{eff}(\mu_{\overline{\text{MS}}}=\mu v_\text{rel}) && (\tilde q \tilde q^\dag)^{[\bm{1}]}\to {\cal B}_{n\ell}^{[\bm{1}]}+\gamma \,.
  \end{array}\right.\nn\\
\eea
For the strong coupling that enters via the interaction Hamiltonian we choose $\alpha_s^\text{BSF}=\alpha_s(\mu_{\overline{\text{MS}}}=\omega)$, evaluated at the gluon momentum scale.
In contrast to~\cite{Harz:2018csl}, we use an identical scale choice for couplings entering either via Abelian or non-Abelian vertices. Within pNRQCD, the latter manifest themselves
exclusively by the $C_A$ contribution to $\alpha_s^\text{eff}$ for the gluonic recombination process. For the binding energy we use
\be
  E_{{\cal B}_{n\ell}} = \frac12\mu \frac{(\alpha^\text{eff}_b)^2}{n^2}\,.
\ee
Using a partial wave decomposition as well as an integral representation for the hypergeometric function entering the scattering-state wave function, and the generating function of the Laguerre polynomials contained in the bound-state wave function, we arrive at the following expressions for the bound-state formation cross sections via the strong and electromagnetic processes (see App.~\ref{sec:bsfapp} for details)
\bea\label{eq:sigmaBSFnl}
  \sigma_{\text{BSF},n\ell}^{(\tilde q \tilde q^\dag)^{[\bm{8}]}\to {\cal B}_{n\ell}^{[\bm{1}]}+g} v_\text{rel} 
   &=&\frac{\pi\alpha_s^\text{BSF}\alpha_b^\text{eff}}{\mu^2}\frac{2^7C_F}{3N_c^2}S^\text{BSF}_{n\ell}(\zeta_s,\zeta_b)\,,\nn\\
  \sigma_{\text{BSF},n\ell}^{(\tilde q \tilde q^\dag)^{[\bm{1}]}\to {\cal B}_{n\ell}^{[\bm{1}]}+\gamma} v_\text{rel} 
  &=&\frac{\pi\alpha_\text{em}\alpha_b^\text{eff}}{\mu^2}\frac{2^7Q_{\tilde q}^2}{3N_c^2}S^\text{BSF}_{n\ell}(\zeta_s,\zeta_b)\,,\nn\\
\eea
where we defined
\be
  \zeta_b=\frac{\alpha_b^\text{eff}}{v_\text{rel}},\quad
  \zeta_s=\frac{\alpha_s^\text{eff}}{v_\text{rel}}\,,
\ee
and
\bea
  S^\text{BSF}_{n\ell}(\zeta_s,\zeta_b) &=& \frac{1}{2^6\zeta_b}(1+\zeta_b^2/n^2)^3\nn\\
  && {} \times \left[(\ell+1)|I_R|^2_{\ell'=\ell+1}+\ell|I_R|^2_{\ell'=\ell-1}\right]\,.\nn\\
\eea
Here, $\ell'$ corresponds to the partial wave of the scattering state, which is constrained by the usual selection rule, and
the radial part of the wave function yields the overlap integral 
\bea\label{eq:IR}
  I_R &=& \sqrt{\frac{2\pi\zeta_s}{1-\E^{-2\pi\zeta_s}}}\frac{(-1)^{\ell+\ell'}n^{\ell'+1}2^{\ell-\ell'}}{\sqrt{(n-\ell-1)!(n+\ell)!}}\nn\\
  && \frac{\zeta_b^{\ell+3/2}}{\sqrt{\zeta_s^2\times(1+\zeta_s^2)\times\cdots\times({\ell'}^2+\zeta_s^2)}}\nn\\
  && \left(\frac{d}{dt}\right)^{n-\ell-1}\left(\frac{d}{d\zeta_b}\right)^{\ell+\ell'+3}\nn\\
  && \frac{\left(1+\left(\frac{\zeta_b}{n}\frac{1+t}{1-t}\right)^2\right)^{\ell'}}{(1+t)^{\ell+\ell'+3}(1-t)^{\ell-\ell'-1}}\E^{-2\zeta_s \text{arccot}\left(\frac{\zeta_b}{n}\frac{1+t}{1-t}\right)}\Bigg|_{t=0}\,.\nn\\
\eea
This expression can be easily evaluated numerically.
The result has the structure
\bea
   S^\text{BSF}_{n\ell}(\zeta_s,\zeta_b) &=& s^\text{BSF}_{n\ell}(\zeta_s,\zeta_b) \frac{2\pi\zeta_s}{1-\E^{-2\pi\zeta_s}}\nn\\
   && {} \times \frac{\zeta_b^{2\ell+2}}{(\zeta_b^2+n^2)^{2n+1}}\E^{-4\zeta_s \text{arccot}\left(\frac{\zeta_b}{n}\right) }\,,\nn\\
\eea
where $s^\text{BSF}_{n\ell}$ is a polynomial, with explicit expressions for $n\leq 3$ given in Tab.\,\ref{tab:snl}.
For the $1s$ ground state our result agrees with~\cite{Harz:2018csl}, and for the $2s$ state it agrees with~\cite{Yao:2018sgn,Binder:2021otw}.
The result for $2p$ differs from the one given in~\cite{Yao:2018sgn,Binder:2021otw} (by a factor $3$ for the $s$-wave contribution with $\ell'=0$,
and a factor $3/2$ for the $d$-wave contribution with $\ell'=2$) but matches the result for hydrogen when translated to the
electromagnetic case~\cite{Bethe:1957ncq}.

  \begin{figure*}
  \centering
  \includegraphics[width=0.49\textwidth]{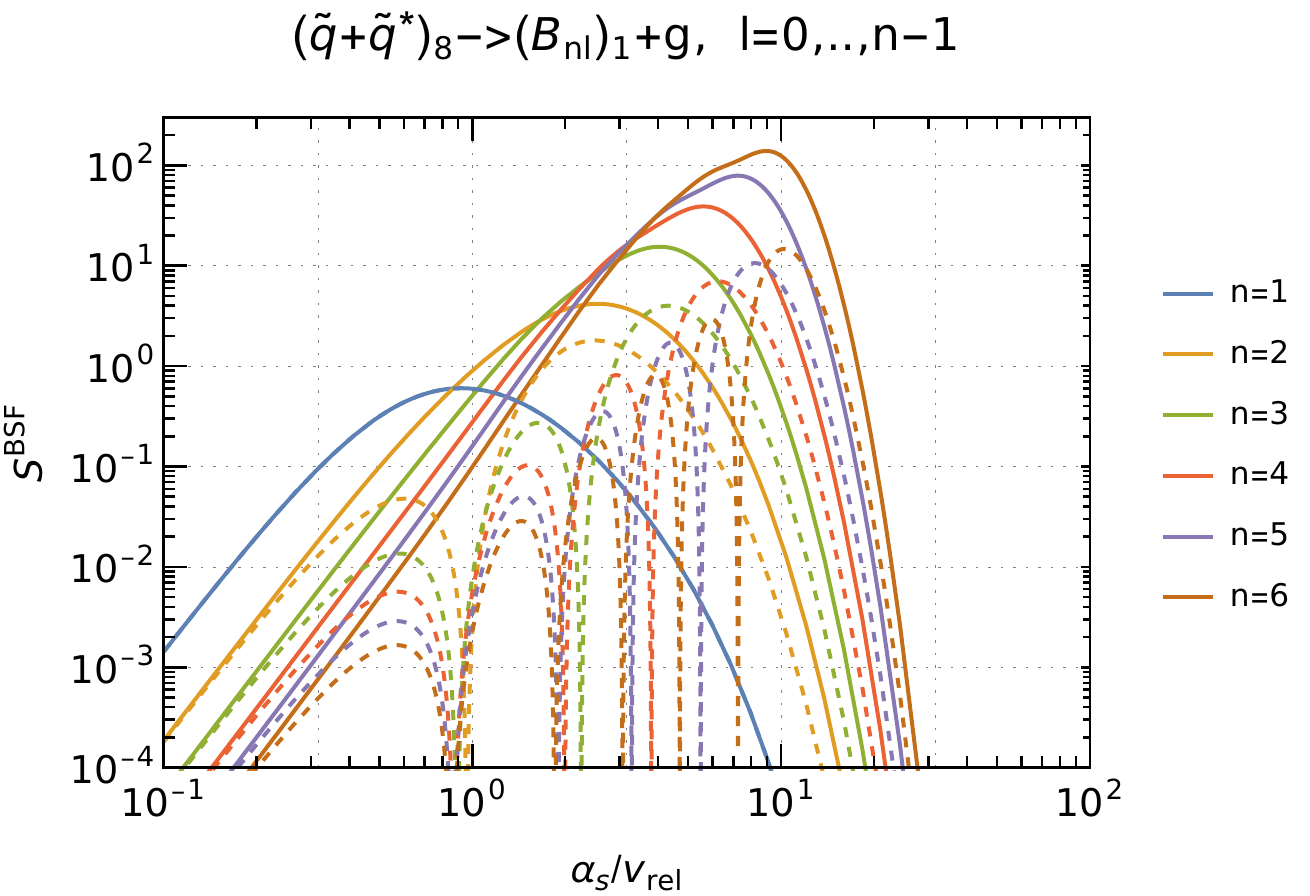}
  \includegraphics[width=0.49\textwidth]{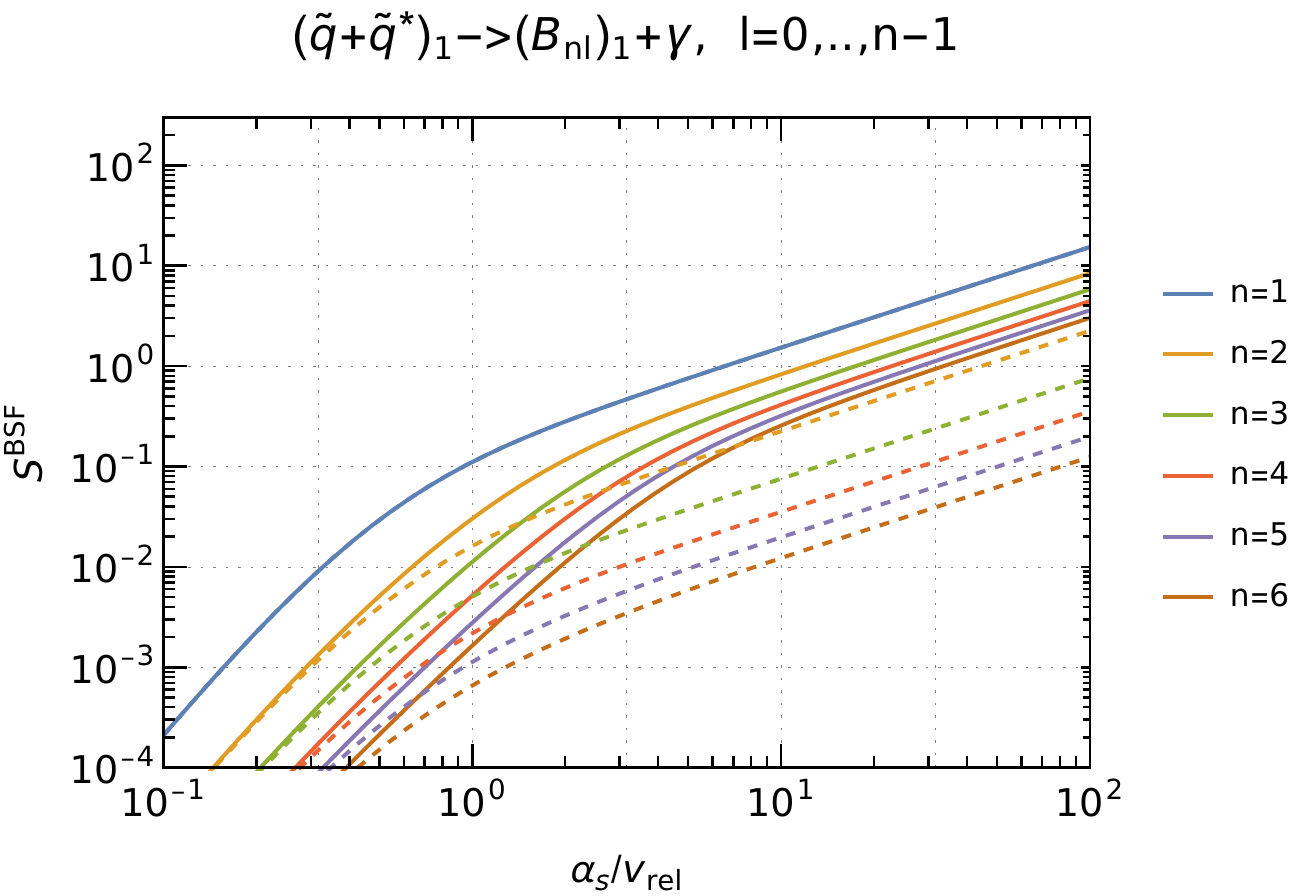}
  \caption{Bound-state formation cross section, eq.~\eqref{eq:sigmaBSFnl}, for the strong (left) and elecromagnetic process (right). We show the sum $\sum_\ell S^\text{BSF}_{n\ell}(\zeta_s,\zeta_b)$ (solid lines) as well as the $\ell=0$ contribution only (dashed lines). The various colors correspond to the principal quantum numbers $n=1,\dots,6$, as given in the legend. 
  For large $\alpha_s/v_\text{rel}$ the cross section of the strong process is Sommerfeld suppressed due to the repulsive interaction of the $\tilde q\tilde q^\dag$ pair in the octet representation, while it is Sommerfeld enhanced for the electromagnetic process, involving a scattering wave function in the color singlet configuration.
}
  \label{fig:SBSF}
\end{figure*}

We show the functions $S^\text{BSF}_{n\ell}(\zeta_s,\zeta_b)$, which are proportional to the bound-state formation cross section, in Fig.\,\ref{fig:SBSF}.
For the figure we assume that the strong coupling entering in $\zeta_s$ and $\zeta_b$ is evaluated at a common renormalization scale, such that $S^\text{BSF}_{n\ell}$ depends only on the ratio $\alpha_s/v_\text{rel}$. Furthermore, we show the sum over all $\ell=0,\dots,n-1$ for a given $n$ (solid lines), as well as the results for the $s$-orbital with $\ell=0$ (dashed lines).
For $\alpha_s/v_\text{rel}\ll 1$ the bound-state formation cross section scales as $(\alpha_s/v_\text{rel})^{4+2\ell}$ for all $n$. The limit is given by
\bea
  S^\text{BSF}_{n\ell}(\zeta_s,\zeta_b)\nn&&\!\!\! \to\\
  &&\frac{2^{2\ell}\zeta_b^{4+2\ell}}{((2\ell+1)!!)^2}\Bigg[\frac{1-\delta_{\ell,0}}{4\ell}\left((3\ell+1)\frac{\zeta_s}{\zeta_b}-3\ell\right)^2\nn\\
  && +\,(\ell+1)\left((\ell+1)\frac{\zeta_s}{\zeta_b}-(\ell+2)\right)^2\Bigg]\nn\\
  && \times \,\frac{\prod_{j=0}^\ell(n^2-j^2)}{n^{5+2\ell}}\,,\qquad \zeta_{s,b}\to 0\,,
\eea
where the second line arises from the $\ell'=\ell+1$ contribution, and the first from $\ell'=\ell-1$ exists only for $\ell>0$.
The contribution from $\ell=0$ orbitals therefore dominates for $\alpha_s/v_\text{rel}\ll 1$, as can also be seen by the convergence of solid and dashed lines for each $n\geq2$ in Fig.~\ref{fig:SBSF} in that limit.

In the opposite limit $\alpha_s/v_\text{rel}\gg 1$,
\bea\label{eq:SBSFlarge}
   S^\text{BSF}_{n\ell}(\zeta_s,\zeta_b) &\to& \frac{2\pi\zeta_s}{1-\E^{-2\pi\zeta_s}} f^\text{BSF}_{n\ell}\left(\frac{\zeta_s}{\zeta_b}\right),\qquad |\zeta_{s,b}|\to\infty \,,\nn\\
\eea
where
\be
  f^\text{BSF}_{n\ell}\left(\frac{\zeta_s}{\zeta_b}\right) = \E^{-\frac{4n\zeta_s}{ \zeta_b} } \, \frac{ s^\text{BSF}_{n\ell}|_{4n-2\ell} }{\zeta_b^{4n-2\ell}}\,.
\ee
Here $s^\text{BSF}_{n\ell}|_{4n-2\ell}$ corresponds to the polynomial obtained when keeping only the terms with maximal combined power in $\zeta_s$ and $\zeta_b$ in $s^\text{BSF}_{n\ell}(\zeta_s,\zeta_b)$, being $4n-2\ell$, such that
$f^\text{BSF}_{n\ell}$ depends only on the ratio $\zeta_s/\zeta_b=\alpha_s^\text{eff}/\alpha_b^\text{eff}$.
Up to the different renormalization scale at which the effective couplings are evaluated, $f^\text{BSF}_{n\ell}$ approaches a constant for $\alpha_s/v_\text{rel}\gg 1$.

The behavior at small relative velocities is therefore governed dominantly by the first factor in eq.~\eqref{eq:SBSFlarge}.
It exhibits a qualitatively different behavior depending on the sign of $\zeta_s$. For $(\tilde q \tilde q^\dag)^{[\bm{8}]}\to {\cal B}_{n\ell}^{[\bm{1}]}+g$, the repulsive potential relevant for the initial state implies $\zeta_s<0$, leading to an exponential suppression for small relative velocities, $S^\text{BSF}_{n\ell} \to 2\pi|\zeta_s| \E^{-2\pi|\zeta_s|} f^\text{BSF}_{n\ell}$.
For the electromagnetic process $(\tilde q \tilde q^\dag)^{[\bm{1}]}\to {\cal B}_{n\ell}^{[\bm{1}]}+\gamma$, both the initial- and final-state wave function are sensitive to the attractive color singlet potential, such that in particular $\zeta_s>0$, and $S^\text{BSF}_{n\ell} \to 2\pi \zeta_s f^\text{BSF}_{n\ell}$ grows with $\zeta_s\propto \alpha_s/v_\text{rel}$. 

The different shape of $S^\text{BSF}_{n\ell}$ for the two processes can clearly be seen in Fig.~\ref{fig:SBSF}. For the electromagnetic process, the combined contribution from all angular momentum states $\sum_\ell S^\text{BSF}_{n\ell}$ decreases with increasing values of $n$, for all velocities $v_\text{rel}$. 
On the other hand, for the strong process the exponential suppression at large $\zeta_s$ leads to a maximum of $S^\text{BSF}_{n\ell}$. Its position shifts to higher values of 
$\alpha_s/v_\text{rel}$ for excited states with increasing $n$. In addition, the value at the maximum increases with $n$. This indicates that excited levels become more and more relevant the smaller the relative velocity, i.e. the lower the temperature that is relevant for determining the relic density.

\subsection{Decay}\label{sec:decay}

The leading decay process is due to annihilation of the constituents of the bound state into a pair of gluons, ${\cal B}_{n\ell}\to gg$.
Here, we briefly review the derivation of the decay rate following~\cite{Petraki:2015hla}, provide an expression for general $n$ (for $\ell=0$) and discuss the role of higher-order corrections.

For a generic $1\to N$ decay process, ${\cal B}_{n\ell}\to X_1X_2\dots X_N$
the matrix element ${\cal M}_{n\ell}$ can be related to the usual Feynman matrix element for the process $\tilde q(k_1,i)+\tilde q^\dag(k_2,j)\to X_1(p_1)+\dots+X_N(p_N)$,
with color indices in the initial state contracted with $P_{ij}^s=\delta_{ij}/\sqrt{N_c}$, which we denote by ${\cal M}^s(k_1,k_2,\{p_j\})$, via
\be\label{eq:decayMnlm}
  {\cal M}_{n\ell m} = \int\frac{\diff^3q}{(2\pi)^3}\frac{\psi_{n\ell m}(q)}{\sqrt{2N_q}}{\cal M}^s( K/2+q, K/2-q,\{p_j\})\,,
\ee
with $N_q\to \mu$ in the nonrelativistic limit, and bound-state wave function $\psi_{n\ell m}\equiv \psi_{n\ell m}^{[\bm{1}]}$ in momentum space, normalized such that $\int \diff^3x |\psi_{n\ell m}(x)|^2=1$ in position space.
Here, $K$ is the four-momentum of the bound state.
The bound-state decay rate is given by
\be
  \Gamma_\text{dec}^{n\ell} =  \frac{1}{2 m_{{\cal B}_{n\ell}}} \frac{1}{S!} \int \diff\text{LIPS}(K;\{p_j\}) \,\overline{|{\cal M}_{n\ell }|^2} \,,
\ee
where $m_{{\cal B}_{n\ell}}=2m_{\tilde q}-E_{{\cal B}_{n\ell}}\simeq 2m_{\tilde q}$, and $\overline{|{\cal M}_{n\ell }|^2}=\frac{1}{2\ell+1}\sum_m \sum_{g_{X_j}} |{\cal M}_{n\ell m}|^2$ is averaged over the $2\ell+1$ states with different $m$, and summed over final-state degrees of freedom. Furthermore, the usual factor $1/S!$ is included if $S$ particles in the final state are of identical type. For two-body decays at rest, the integration over the Lorentz-invariant phase space (LIPS) reduces to a factor $1/(8\pi)$.

At leading order in the small relative momentum $q$ and in the non-relativistic expansion, 
\be
  {\cal M}_{n\ell m} = Z_{n\ell m}{\cal M}^s( K/2, K/2,\{p_j\})\,,
\ee
with on-shell four-momentum $K^2=m_{{\cal B}_{n\ell}}^2$, and
\be
  Z_{n\ell m}=\sqrt{\frac{|\psi_{n\ell m}(x=0)|^2}{2\mu}}\,.
\ee
The wave-function at the origin is non-zero for orbitals with $\ell=0$ only, while the decay of bound states with orbital angular momentum would require keeping further terms in the expansion of ${\cal M}^s$ in $q$, leading to a suppression of the decay rate of order $q^2/K^2\sim E_{n\ell}/m_{\tilde q}$. We therefore focus on the decay of $\ell=0$ states in the following.

For the leading process ${\cal B}_{n\ell}\to g(p_1,a,\mu)g(p_2,b,\nu)$, 
\be
  {\cal M}^s_{n00} = Z_{n00}\frac{ig_s^2}{\sqrt{N_c}}\epsilon^{*\mu}\epsilon^{*\nu}\delta_{ab}\left(\frac{g^{\mu\nu}p_1\cdot p_2-p_2^\mu p_1^\nu}{p_1\cdot p_2}\right)\,,
\ee
and including a factor $1/2$ due to the identical particles in the final state yields
\bea
  \Gamma_\text{dec}^{{\cal B}_{n,\ell=0}\to gg} &=& C_F\frac{\pi(\alpha_s^\text{ann})^2}{m_{\tilde q}^2}|\psi_{n00}(0)|^2 \nn\\
  &=&  \frac{1}{8n^3} C_F(\alpha_s^\text{ann})^2(\alpha_b^\text{eff})^3 m_{\tilde q}\,,
\eea
where we used $|\psi_{n00}(0)|^2$=$\mu^3(\alpha_b^\text{eff})^3/\pi/n^3$, and
\be
  \alpha_s^\text{ann} = \alpha_s(\mu_{\overline{\text{MS}}}=m_{\tilde q})\,.
  \label{eq:scaledec}
\ee
The result agrees with~\cite{Harz:2018csl} for $n=1$.

The decay rate can also be obtained from an effective operator that describes the interaction of an $\ell=0$ bound state with a pair of gluons,
\be
  {\cal L}_\text{eff} = -\frac{g_s^2}{4\sqrt{N_c}}F(Q) G^a_{\mu\nu}G_a^{\mu\nu} \Phi_n\,,
  \label{eq:BSeffVertex}
\ee
with a scalar field $\Phi_n$ that describes the ${\cal B}_{n,\ell=0}$ bound state, and a
form factor $F(Q)\equiv Z_{n00}/Q^2$, where $Q^2\equiv p_1\cdot p_2$.
The coefficient can be obtained by matching the matrix element for the two-body decay in the full and effective description.

Note that the matching does \emph{not} require the gluons to be on-shell. Accordingly, the effective operator, eq.~\eqref{eq:BSeffVertex}, can also be used to compute the $2\to 2$ scattering processes of the form ${\cal B} q \to g q$. Implementing the effective operator in \MG~\cite{Alwall:2014hca}, we checked that $2\to 2$ processes can only compete with the bound-state decay for very early times, $x\lesssim5\!-\!10$, for which the mediator is still in thermal equilibrium with the SM plasma. Hence, these processes are negligible for the dynamics of dark matter freeze-out considered here.

In contrast, NLO corrections to the two-body decay rate are potentially relevant since in ionization equilibrium, the impact of bound states on the effective cross section is determined predominantly by their decay rate, see the discussion in Sec.~\ref{sec:bound}.
Following earlier results in the context of quarkonium~\cite{Barbieri:1979be,Hagiwara:1980nv,Petrelli:1997ge}, the virtual and real corrections to the ${\cal B}_{10}\to gg$ decay rate at ${\cal O}(\alpha_s)$ have been computed for the comparable case of stoponium~\cite{Martin:2009dj}, resulting in:
\bea
\frac{\Gamma_\mathrm{dec}^\mathrm{NLO}}{\Gamma_\mathrm{dec}^\mathrm{LO}} &= &1+\frac{ \alpha_s^\text{ann}}{\pi} \bigg[ \frac{b_0}{2} 
\ln \biggl( \frac{\mu_{\overline{\text{MS}}}^2}{4m_{\tilde{q}}^2} \biggr) + 
\left( \frac{199}{18} - \frac{13 \pi^2}{24} \right) C_A 
\nonumber \\
 & & + \left( -\frac{7}{2} - \frac{\pi^2}{8} 
 + \left( \frac{1}{2} - \frac{\pi^2}{8} \right) \delta_{4\tilde q} 
 \right) C_F 
\nonumber \\
 & & + \left( -\frac{16}{9} n_f
- \frac{1}{3} \ln(2) \right) T_F \bigg] \,,\label{eq:BdecNLO}
\eea
where $n_f$ is the number of light quarks, $T_F=1/2$, and $b_0=11/3\, C_A - (1/3 + 4/3\, n_f) T_F$. The parameter $\delta_{4\tilde q} $ is either 1 or 0 depending on whether or not the four-point interaction of $\tilde q$ is introduced. In the simplified model considered here, $\delta_{4\tilde q} =0$, while in the MSSM, $\delta_{4\tilde q} =1$.

For the scale choice \eqref{eq:scaledec}, we find that the correction \eqref{eq:BdecNLO} is reduced to a few percent rendering the leading-order (LO) and NLO predictions fully compatible with each other within scale uncertainties, see App.\,\ref{sec:NLOcorr} for a detailed discussion. For definiteness, we will consider the LO decay rate in the main results in the following.

The decay of bound states with $\ell>0$ is suppressed compared to those with $\ell=0$. Nevertheless, due to the large number of such states, it would be interesting to include them, which is, however, beyond the scope of this work. We note that the two-body decay into a pair of gluons is forbidden for $\ell=1$ states due to the Landau-Yang theorem. A decay channel that is possible for these states is into a pair of electrically charged particles, via an intermediate photon or $Z$-boson. Note that an analogous process with an intermediate gluon is forbidden by color conservation.
Furthermore, the decay rate into $g\bar q q$ via an intermediate off-shell gluon also vanishes for $\ell=1$, as can be checked by expanding the matrix element in eq.~\eqref{eq:decayMnlm} to first order in the relative momentum $q$. However, a decay into three gluons could be mediated by the strong interactions.

\subsection{Transitions}\label{sec:transitions}

Since bound states exist only in the color singlet configuration, transitions between energy levels cannot proceed via single gluon emission or absorption.
In this work, we do not consider transitions involving two gluons, which can be mediated by the strong interaction. Instead, we provide a lower bound
on the size of transition rates by considering the electromagnetic process
\be
  {\cal B}_{n\ell} \to {\cal B}_{n'\ell'} + \gamma\,,
\ee
which is allowed by color and charge conservation. The transition matrix element squared obtained from the electric dipole interaction is given by~\cite{Bethe:1957ncq}
\be
  |{\cal M}|^2 = \frac23 e^2Q_{\tilde q}^2 \omega^2 |\langle\psi_{n\ell}|\vec r|\psi_{n'\ell'}\rangle|^2\,,
\ee
where $\omega=|E_{n\ell}-E_{n'\ell'}|$ is the photon energy.
The matrix element is averaged over $m$ and $m'$,
\be
 |\langle\psi_{n\ell}|\vec r|\psi_{n'\ell'}\rangle|^2= \sum_{m,m'}  \frac{|\langle\psi_{n\ell m}|\vec r|\psi_{n'\ell' m'}\rangle|^2}{(2\ell+1)(2\ell'+1)}\,.
\ee
The transition rate from higher to lower energy levels is given by Fermi's golden rule
\bea
  \Gamma_\text{trans}^{n\ell\to n'\ell'} &=& \frac{\omega}{2\pi}(2\ell'+1)|{\cal M}|^2 \nn\\
  &=& \frac43 \alpha_\text{em}Q_{\tilde q}^2(2\ell'+1)\omega^3|\langle\psi_{n\ell}|\vec r|\psi_{n'\ell'}\rangle|^2\,.
\eea
The rate of the inverse process of photoabsorption can be obtained from the detailed balance condition eq.~\eqref{eq:transMilne}.
Using the hydrogen-like wave functions and the generating function of the Laguerre polynomials (see App.~\ref{sec:bsfapp}) we find
\be
  |\langle\psi_{n\ell}|\vec r|\psi_{n'\ell'}\rangle|^2 = \frac{\ell'\delta_{\ell',\ell+1}+\ell\delta_{\ell,\ell'+1}}{(2\ell+1)(2\ell'+1)}|I_R^\text{trans}|^2\,,
\ee
where
\bea
  I_R^\text{trans}&=&\frac{N_{n\ell}(\kappa)N_{n'\ell'}(\kappa')(3+\ell+\ell')!}{(n-\ell-1)!(n'-\ell'-1)!} 
  \left(\frac{d}{dt}\right)^{n-\ell-1}\nn\\
  && \left(\frac{d}{dt'}\right)^{n'-\ell'-1}
  \frac{(1-t)^{-2\ell-2}(1-t')^{-2\ell'-2}}{\left(\frac{\kappa}{n}\frac{1+t}{1-t}+\frac{\kappa'}{n'}\frac{1+t'}{1-t'}\right)^{4+\ell+\ell'}}\Bigg|_{t=t'=0}\,,\nn\\
\eea
with
\be
N_{n\ell}(\kappa)=\kappa^{3/2}\sqrt{\frac{4(n-\ell-1)!}{n^4(n+\ell)!}}\left(\frac{2\kappa}{n}\right)^\ell\,.
\ee
Here 
\be
  \kappa\equiv \mu\alpha_b^{\rm eff}\big|_{n\ell}\,,\qquad \kappa'\equiv \mu\alpha_b^{\rm eff}\big|_{n'\ell'}\,,
\ee
with effective strong coupling defined as in eq.~\eqref{eq:alphaeff} and evaluated for the respective energy level as indicated by the subscript. They differ only in the scale choice of the strong coupling constant, related to the typical Bohr momentum of the two energy levels.
We checked agreement with various explicit expressions given for specific $n'$ and $\ell,\ell'$  and all $n$ as well as for $n=n'$ in~\cite{Bethe:1957ncq}, when translating the result to
the analogous hydrogen transition rates.

\begin{figure*}
  \centering
  \includegraphics[width=0.45\textwidth, trim= {0cm 0cm 0cm 0cm}, clip]{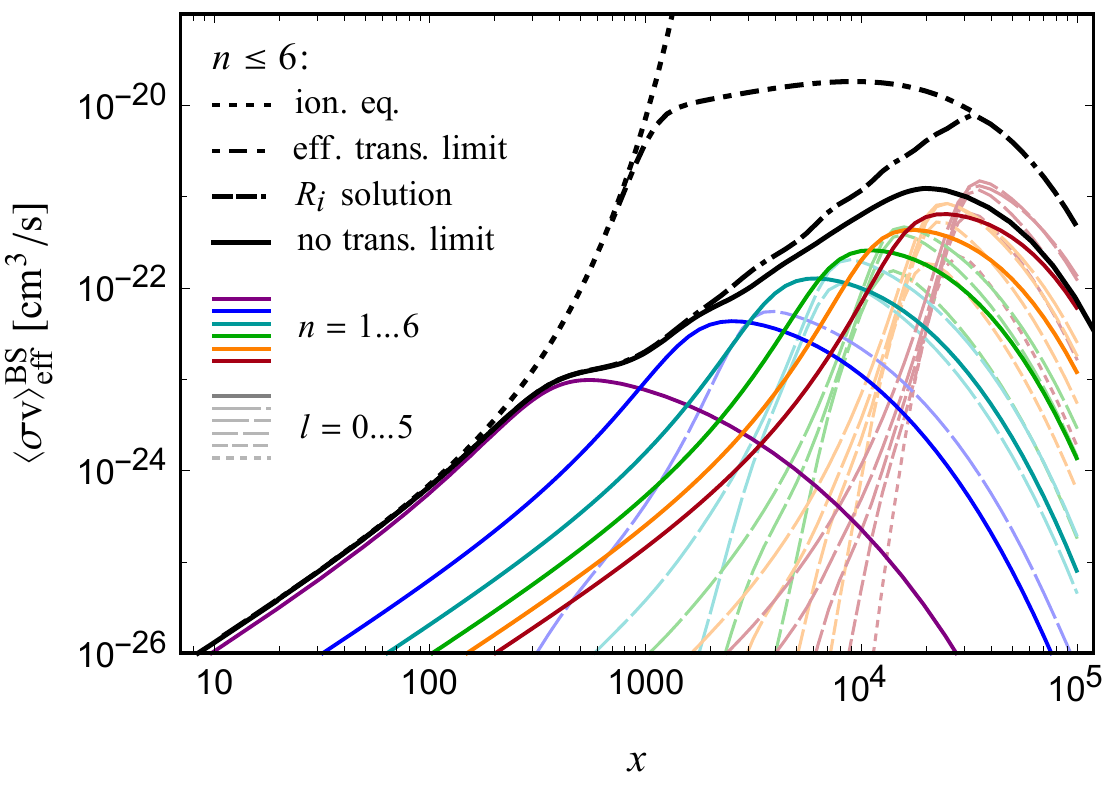}
    \hspace*{5mm}
  \includegraphics[width=0.45\textwidth, trim= {0cm 0cm 0cm 0cm}, clip]{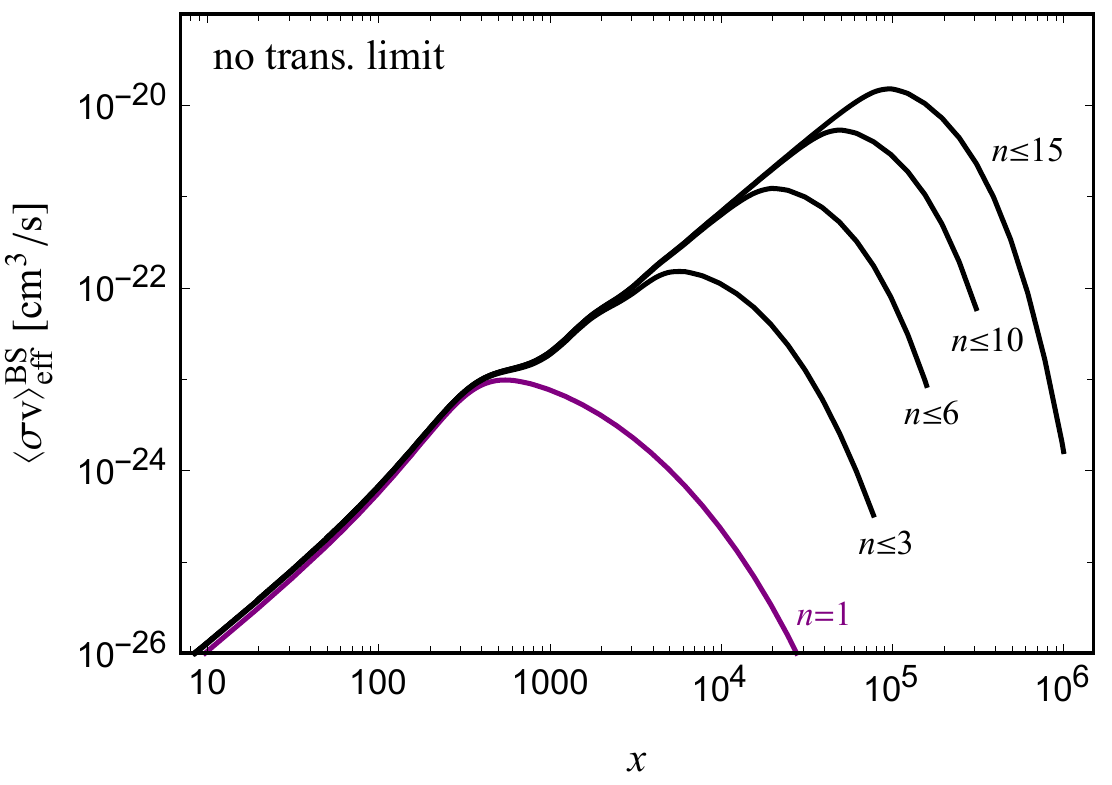}
  \caption{
 Contribution to the effective, thermally averaged mediator annihilation cross section from bound states, $\avb{\sigma_{\tilde q\tilde q^\dagger} v}_\text{eff}^\text{BS}$, for $m_\chi=1000\,$GeV and $m_{\tilde q}=1020$\,GeV, as a function of $x=m_\chi/T$. The left panel shows the result, eq.~\eqref{eq:effgeneral}, when including all recombination, decay and transition rates discussed in Sec.\,\ref{sec:rates} for $n\leq 6$ and $\ell\leq n-1$ (``$R_i$-solution''). In addition, the limits of efficient transitions, eq.~\eqref{eq:sigmaveffeff}, no transitions, eq.~\eqref{eq:effnfirst}, and ionization equilibrium, eq.~\eqref{eq:sigeffioneq}, are shown, in addition to the individual contributions from all $n,\ell$ levels. The right panel shows the no-transition limit, including bound states up to $n\leq 1,3,6,10$ and $15$, respectively. }
  \label{fig:sigeff}
\end{figure*}

\subsection{Effective cross section}\label{sec:sigeff}

Using the ionization, decay and transition rates discussed above we can compute the effective cross section, eq.~\eqref{eq:effgeneral}, that encapsulates the impact of bound states on the freeze-out dynamics. The contribution to the effective cross section, eq.~\eqref{eq:effgeneral}, due to bound states, 
\be
  \avb{\sigma_{\tilde q\tilde q^\dagger} v}_\text{eff}^\text{BS} \equiv \avb{\sigma_{\tilde q\tilde q^\dagger}v}_\text{eff}-\avb{\sigma_{\tilde q\tilde q^\dagger}v}\,,
\ee
is shown in Fig.\,\ref{fig:sigeff} for various approximations as a function of $x=m_\chi/T$. In the left panel, we include bound states up to $n=6$ and for all $\ell\leq n-1$.

The contributions from individual $n,\ell$ states to the effective cross section are indicated by the colored and gray lines in the left panel of Fig.\,\ref{fig:sigeff}.
For small $x$, the $\ell=0$ states dominate. The reason is that in this limit, ionization equilibrium holds, and the effective cross section is determined by the decay rate, see eq.~\eqref{eq:sigeffioneq}. For large $x$, the contribution from each $n,\ell$ level becomes suppressed due to a combination of two effects: 
$(i)$ the suppression due to the repulsive interaction in the scattering state discussed in Sec.\,\ref{sec:bsf}, and
$(ii)$ Boltzmann suppression for $T\ll E_{{\cal B}_{n\ell}}$. Consequently, each individual contribution features a maximum. Its position shifts to the right for higher $n$.
This implies that excited states dominate the effective cross section for large $x$. The larger $x$, the higher $n$ have to be taken into account to obtain a converged result for the total effective cross section.

The line labeled ``$R_i$-solution'' shows the total result obtained when including all rates as given above using the general expression eq.~\eqref{eq:effgeneral} for the effective cross section. For comparison, we show the limit of efficient transitions, eq.~\eqref{eq:sigmaveffeff}, as well as the limit of no transitions, eq.~\eqref{eq:effnfirst}. For small $x$, that is, large enough temperature, all results agree and approach the ionization equilibrium result, eq.~\eqref{eq:sigeffioneq}, that is also shown. The effective cross section can in this limit be written as
\be\label{eq:sigeffioneq2}
  \avb{\sigma_{\tilde q\tilde q^\dagger} v}_\text{eff}^\text{BS} =
  \avb{\sigma_{\tilde q\tilde q^\dagger} v}_\text{eff}^{\text{BS},n\ell=10}\,
  \times \sum_{n\geq 1}  \frac{\E^{(E_{{\cal B}_{n0}}-E_{{\cal B}_{10}})/T}}{n^3}\,,
\ee
with the sum approaching $\zeta(3)\simeq 1.202$ for small $x$.\footnote{Note that this limit is slightly exceeded in our numerical results since $\alpha_b^\text{eff}$ also depends on $n$ due to the running of the strong coupling.}
That is, in ionization equilibrium, excited states lead to a $20\%$ correction to the effective cross section.
The factor in front of the sum is the ground-state contribution to eq.~\eqref{eq:sigeffioneq}.

The impact of excited states is much larger for large $x$, where they give the dominant contribution. The precise value depends in this regime on the recombination as well as transition rates.
The efficient transition limit provides an upper bound on the effective cross section (since all $\ell$ orbitals contribute), while the limit of no transitions provides a lower limit (only the bound states with a sizeable decay rate into SM particles contribute, being $\ell=0$ orbitals in our approximation). The actual effective cross section is therefore expected to lie in between these two limits. The ``$R_i$-solution'' result taking into account the electromagnetic transition rates considered in this work can only be considered as illustrative since additional processes mediating further transitions are expected to play an important role. We therefore conservatively adopt the no-transition limit in our numerical analysis in the following.

The effective cross section in the no-transition approximation eq.~\eqref{eq:effnfirst} is shown in the right panel of Fig.\,\ref{fig:sigeff}. We show the result summed up to some maximum $n$, for $n=1,3,6,10,15$, respectively. While each individual contribution becomes suppressed at large $x$, the summed result continues to grow with increasing $x$. The decline at very large $x$ is due to the restriction to $n\leq 15$. For $x\ll 10^5$, we consider the effective cross section with $n\leq 15$ as converged. We leave an exploration of the full result including transitions to future work, and use the no-transition limit with $n\leq15$ as the default choice in the following. For a discussion of the impact of a certain class of higher-order corrections (related to collisional ionization and recombination processes and the associated virtual contributions) computed in~\cite{Binder:2021otw,Binder:2020efn} as well as to bound-state decay we refer to App.\,\ref{sec:NLOcorr}.

\section{Viable parameter space}\label{sec:sol}

\begin{figure}
  \centering
  \includegraphics[width=0.45\textwidth]{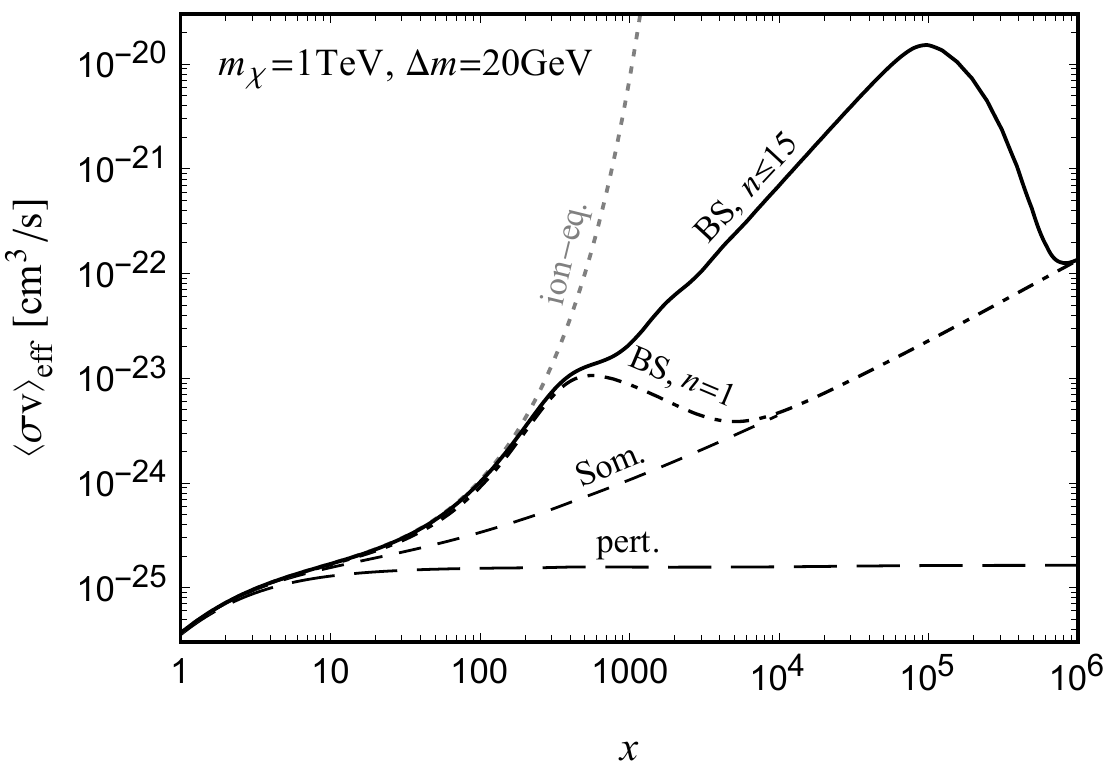}
  \caption{
Effective mediator annihilation cross section, eq.~\eqref{eq:effgeneral}, including the contribution from direct annihilation without (`pert.') and with Sommerfeld enhancement (`Som.'), as well as with the additional contribution from bound states separately considering the ground state only (`BS, $n=1$') and including excited states up to $n= 15$ (`BS, $n\leq15$'). The gray dotted curve show the case of ionization equilibrium (`ion-eq'). The parameters are $m_\chi=1000\,$GeV and $m_{\tilde q}=1020$\,GeV.
}
  \label{fig:sigmav_eff}
\end{figure}

\begin{figure*}
  \centering
\includegraphics[width=0.444\textwidth]{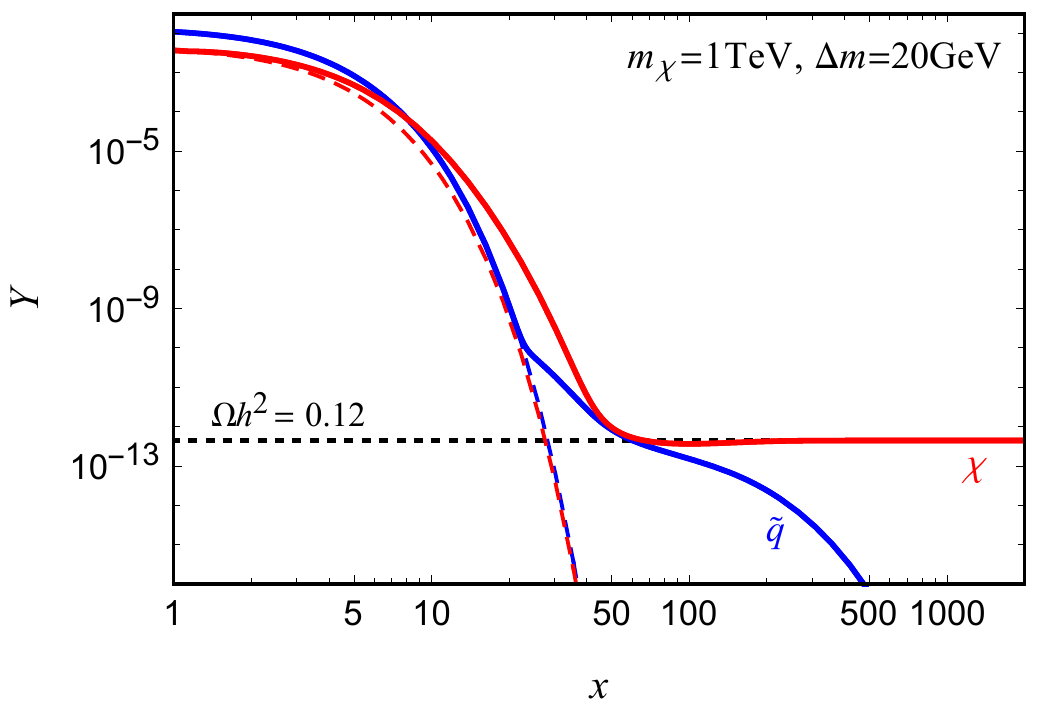}
\hspace{3ex}
  \includegraphics[width=0.45\textwidth]{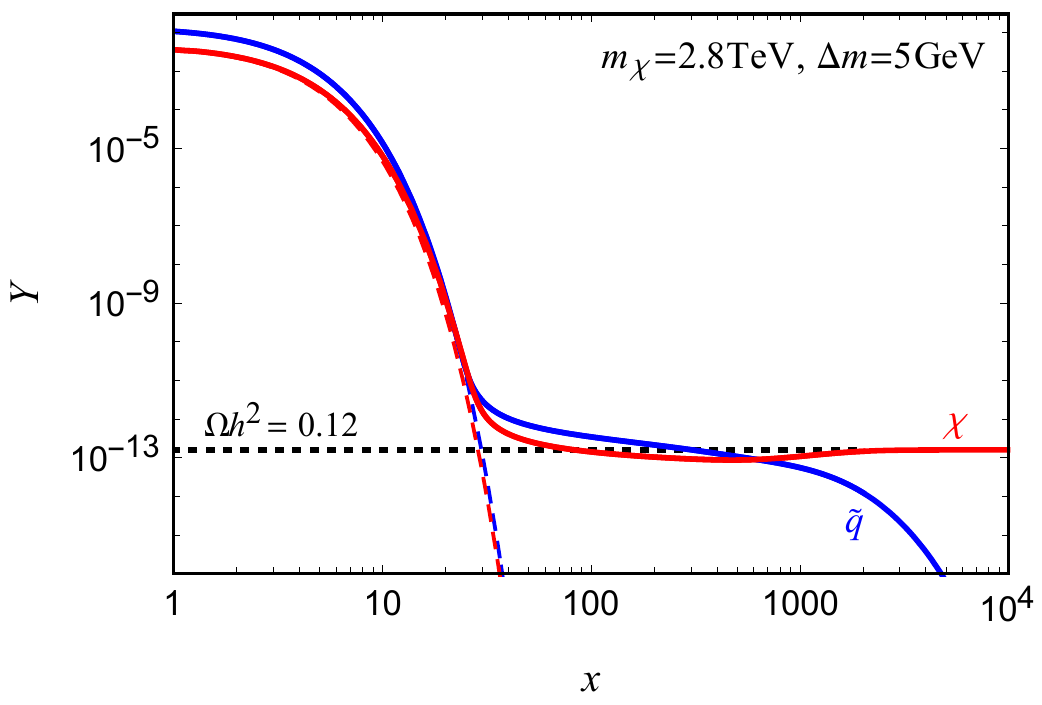}
  \caption{
 Evolution of the dark matter yield $Y_\chi$ and the mediator abundance $Y_{\tilde q}$ with $x=m_\chi/T$ for two benchmark points within the regime of conversion-driven freeze-out, and when including Sommerfeld enhancement as well as bound states up to $n=15$. Dashed lines show the equilibrium abundances, and solid lines show the solution of the coupled Boltzmann equations. 
 }
  \label{fig:benchmark}
\end{figure*}

To determine the relic abundance, we solve the coupled set of Boltzmann equations~\eqref{eq:BMEchi} for $Y_\chi$ and~\eqref{eq:BMEsqu} for  $Y_{\tilde q}$. We compute the involved conversion and annihilation cross sections, $\sigma_{\tilde q k  \rightarrow \chi l}(s)$ and
$\sigma_{\chi\chi}(s),\sigma_{\chi\tilde q}(s), \sigma_{\tilde q\tilde q^\dagger}(s)$, respectively, with \MG~\cite{Alwall:2014hca}. We take into account the leading conversions in $\alpha_\mathrm{s}$ and regularize the soft divergence occurring in the process ${\tilde q} g  \rightarrow \chi b$ (see the discussion in~\cite{Garny:2017rxs}) by introducing a thermal mass for the gluon~\cite{le_bellac_1996}.
To include the impact of bound states, we
replace the annihilation cross section of $\tilde q\tilde q^\dag$ pairs by the effective cross section, eq.~\eqref{eq:effgeneral}. 
In addition, we include Sommerfeld enhancement in the contribution from direct mediator annihilation as described in~\cite{Garny:2017rxs}.
Figure~\ref{fig:sigmav_eff} exemplifies the effective cross section. The long-dashed curve (`pert.') shows the perturbative direct annihilation cross section while the short-dashed (`Som'.), dot-dashed (`BS, $n=1$') and solid (`BS, $n\leq 15$') curves display the effective cross section after successively including Sommerfeld enhancement, bound-state formation effects of the ground state and excited bound states up to $n=15$ (in the \emph{no transition} limit), respectively. In the following, we choose the latter for our main results. We also show the effective cross section under the assumption of ionization equilibrium in the limit of large $n$ as the gray dotted curve (`ion-eq').

For two benchmark points in the conversion-driven freeze-out scenario, the evolution of the abundances is shown in Fig.\,\ref{fig:benchmark}. Because of the small coupling $\lambda_\chi$, the $\chi$ particle cannot annihilate efficiently by itself, and its abundance is reduced only due to conversions into $\tilde q$. While the colored mediator $\tilde q$ starts to depart from thermal equilibrium at $x\gtrsim 25$, the $\chi$ abundance already significantly exceeds the equilibrium value at this time. Subsequently, for $x\gtrsim 25$, conversion processes -- which are on the edge of being efficient -- gradually transform $\chi$ into $\tilde q$ particles.
This leads to a prolonged duration of the freeze-out dynamics, which can last until $x\sim {\cal}{O}(10^2-10^3)$. The mediator $\tilde q$ continues to annihilate and is in addition depleted due to bound-state formation.

\begin{figure*}
  \centering
\includegraphics[width=0.49\textwidth,trim= {0.52cm 0.85cm 0.55cm 1.3cm},clip]{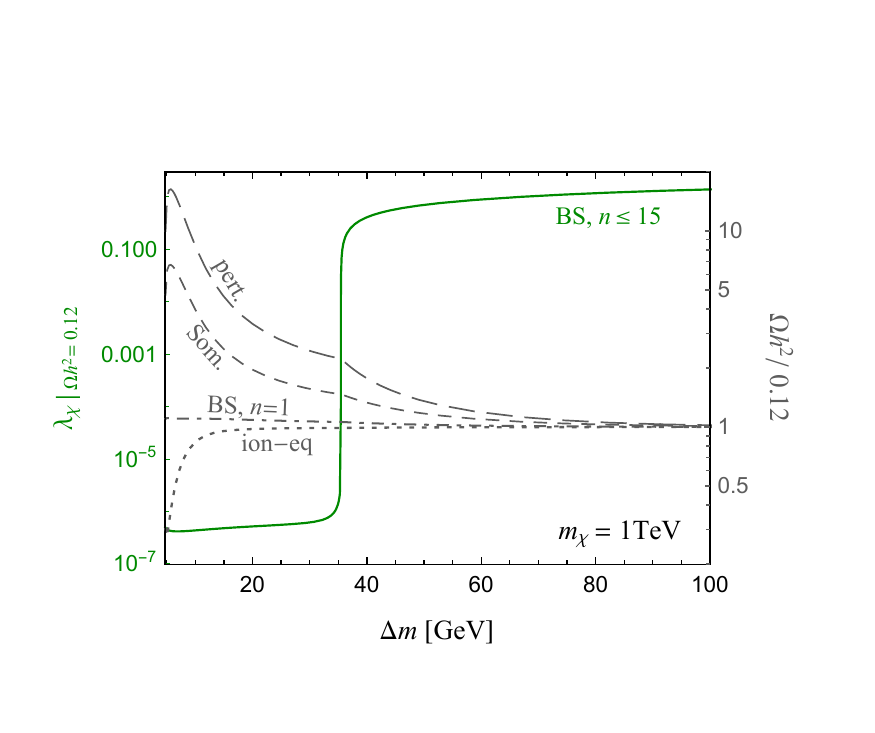}
\includegraphics[width=0.49\textwidth,trim= {0.52cm 0.85cm 0.55cm 1.3cm},clip]{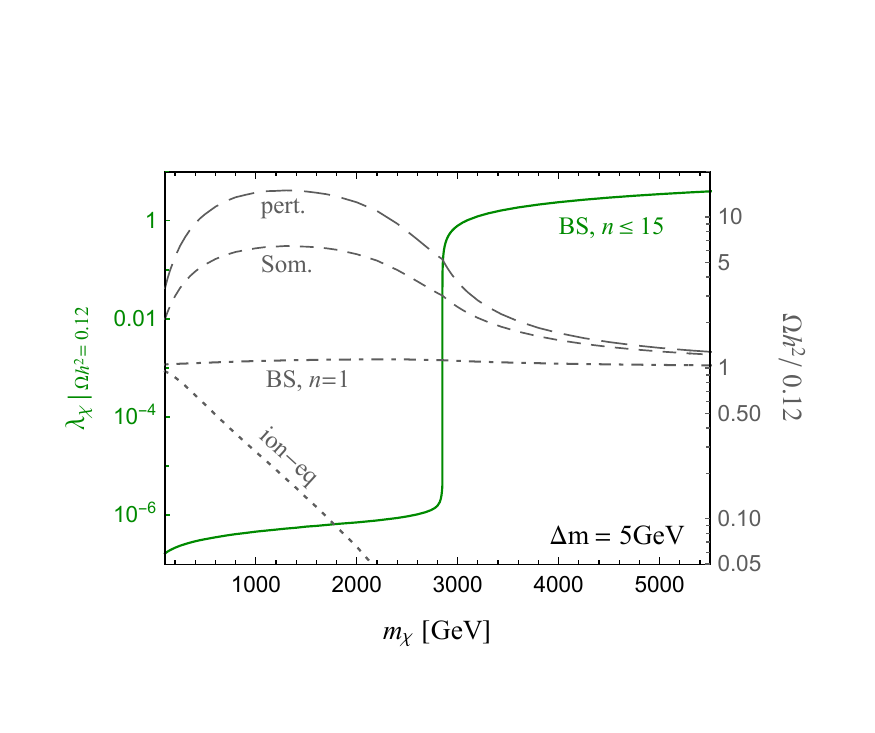}
  \caption{
The two figures show one-dimensional slices of the parameter space, varying the mass splitting $\Delta m=m_{\tilde q}-m_\chi$ for fixed $m_\chi=1$\,TeV (left panel) as well as varying the mass $m_\chi$ for fixed $\Delta m=5$\,GeV (right panel).
  Each figure shows two quantities: The value of the coupling $\lambda_\chi$ that provides the correct relic abundance when taking into account Sommerfeld enhancement and bound states up to $n=15$ is shown as the green solid curve, with values as given on the axis on the left-hand side. The gray lines show the relic density (normalized to the observed value $0.12$) that is obtained with this coupling in the various other approximation, following the same convention for the line style and labels as in Fig.~\ref{fig:sigmav_eff}. The corresponding values are shown on the axis on the right-hand side.
}
  \label{fig:1Dslice}
\end{figure*}

The duration is further enhanced for a small relative mass splitting $\Delta m/m_\chi$, which implies that the equilibrium abundances of the mediator and $\chi$ are comparable until $x\sim m_\chi/\Delta m$ even if the conversion processes were fully efficient, \ie~in the usual coannihilation scenario. Eventually, the mediators decay via $\tilde q\to b\chi$, thereby transferring their remaining abundance to the population of $\chi$ particles.
For the chosen value of the coupling $\lambda_\chi$ for the two benchmark points shown in Fig.\,\ref{fig:benchmark}, the amount of conversions is sufficient to reduce the $\chi$ abundance to a final value that matches the observed relic density, $\Omega h^2=0.12$~\cite{Planck:2018vyg}.

\medskip
In Fig.\,\ref{fig:1Dslice}, we show the coupling $\lambda_\chi$ that is required to achieve $\Omega h^2=0.12$ as a function of the mass splitting, $\Delta m$, for fixed mass $m_\chi=1$\,TeV (left panel) and as a function of the dark matter mass, $m_\chi$, for fixed $\Delta m=5$\,GeV (right panel). The drastic change in the coupling at $\Delta m\simeq 35\,$GeV and $m_\chi\simeq 2850\,$GeV, respectively, is due to the transition between the conversion-driven freeze-out (to the left) and coannihilation regime (to the right), see below for details.

The gray lines in Fig.\,\ref{fig:1Dslice} show the impact on the relic density for various levels of approximation, relative to our fiducial choice with Sommerfeld enhancement and bound states up to $n=15$. The relic density differs up to a factor of order $10$ relative to the perturbative leading-order approximation, and for small $\Delta m/m_\chi$. Relative to the case when including Sommerfeld enhancement, we find differences of up to a factor of order five. The gray line labeled `BS, $n=1$' corresponds to the case when including the ground state. 
As apparent from the relatively small deviation of this curve from one, excited states with $n\leq 15$ only yield a comparably small correction for most of the shown parameter space. 

The kink at the transition between the conversion-driven freeze-out and coannihilation regime that can be seen in most curves arises due to the sudden increase of the coupling at this point that causes $\chi\chi$ and $\chi \tilde q$ annihilation processes to become relevant. Accordingly, in the latter regime, not only does the relative importance of non-perturbative effects on the effective mediator annihilation change but also the importance of the effective mediator annihilation with respect to $\chi\chi$ and $\chi \tilde q$ annihilation changes. This causes the quicker decrease of $\Omega h^2/0.12$ in the coannihilation regime most prominently seen in the left panel of Fig.\,\ref{fig:1Dslice}. Here, we can also observe that all curves approach unity toward large mass splittings as both the Boltzmann suppression of the mediator abundance during freeze-out and the larger coupling, $\lambda_\chi$, diminishes the relative importance of the mediator annihilation.

In the parameter slice with $m_\chi=1$\,TeV, chosen in the left panel, freeze-out mainly occurs while the system is still close to ionization equilibrium. This can also be seen from the gray dotted curve showing the result assuming ionization equilibrium (for all $n$). It only deviates significantly for low $\Delta m$ where freeze-out extends to large $x$. 
For even smaller relative mass splittings, $\Delta m/m_\chi$, considered in the right panel, this effect is even more pronounced as freeze-out extends to larger $x$ (even in the coannihilation region). Here, the result for ionization equilibrium differs by orders of magnitude from the one of our fiducial choice reaching $\Omega h^2/0.12\lesssim10^{-4}$ in the considered range of $m_\chi$ (outside the displayed range in Fig.\,\ref{fig:1Dslice}).

\subsection{Boundary between coannihilation and conversion-driven regime}\label{sec:boundary}

\begin{figure}
  \centering
  \includegraphics[width=0.45\textwidth]{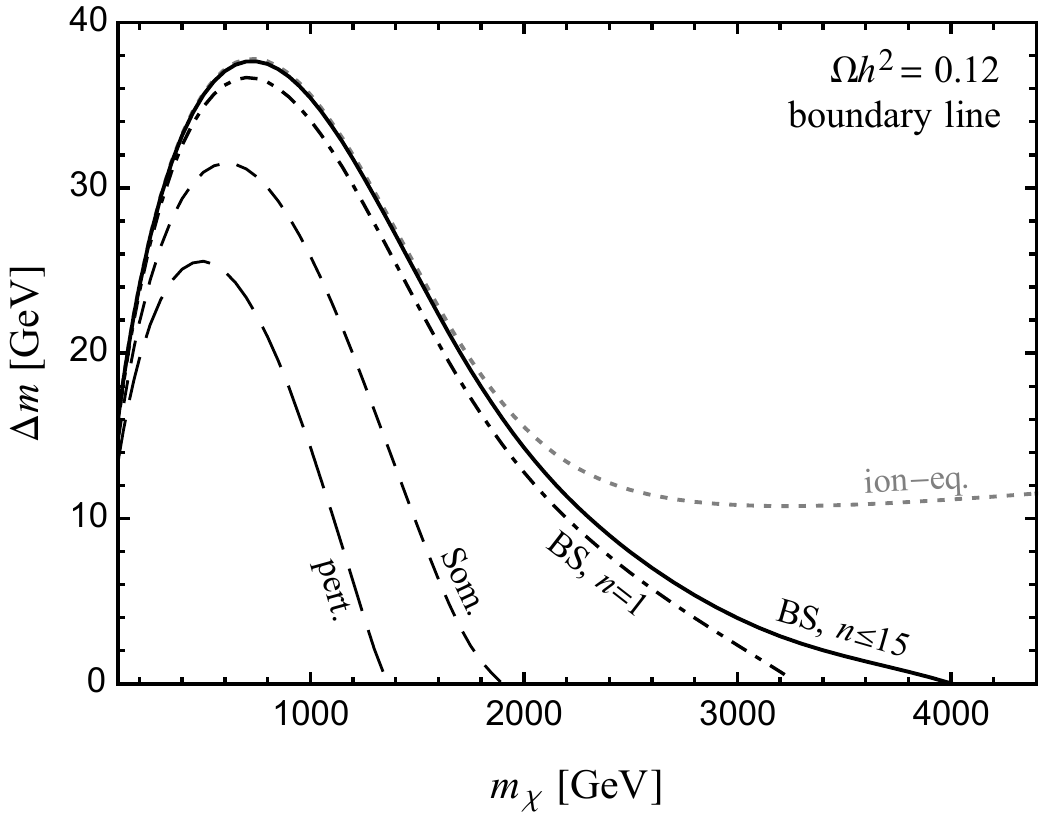}
  \caption{
 Boundary in the model parameter space $(m_\chi,\Delta m=m_{\tilde q}-m_\chi)$ between the regime of conversion-driven freeze-out (bottom left) and conventional coannihilation, when requiring the coupling $\lambda_\chi$ to be adjusted such that the relic abundance matches the observed dark matter density. The various lines correspond to the boundary obtained when successively including corrections to the perturbative $\tilde q\tilde q^\dag$ cross section (`pert.'), being Sommerfeld enhancement (`Som.'), the contribution from the $n=0$ bound state (`BS, $n=1$'), and excited states up to $n=15$ within the default approximation discussed in Sec.\,\ref{sec:sigeff} (`BS, $n\leq15$'). The gray dotted line shows the result that would be obtained when assuming ionization equilibrium for all (excited) states.
}
  \label{fig:boundary}
\end{figure}

In this section, we determine the part of parameter space of the model for which conversion-driven freeze-out is relevant.
For small mass splitting $\Delta m\equiv m_{\tilde q}-m_\chi$ and mass $m_\chi$, the $\tilde q\tilde q^\dag$ annihilation process becomes very efficient,
and would deplete the relic abundance below the observed dark matter density, if $\chi$ and $\tilde q$ were in chemical equilibrium during freeze-out.
Within the region of parameter space where this happens, the correct dark matter abundance can only be explained if the assumption of chemical equilibrium does \emph{not} hold.
The dynamics are described by conversion-driven freeze-out in this regime, and one obtains a viable relic density for couplings $\lambda_\chi\ll 1$. On the other hand, for points in parameter space where the $\tilde q\tilde q^\dag$ annihilation cross section is small enough, the standard scenario of coannihilation yields the observed dark matter abundance, with $\lambda_\chi\sim{\cal O}(1)$. The division between these regimes can effectively be obtained with high precision by solving the Boltzmann equation using the
conventional coannihilation approximation in the limit $\lambda_\chi\ll 1$. The relic density obtained in this limit matches the observed dark matter abundance along a line in the two-dimensional parameter space $(m_\chi,\Delta m)$, which we refer to as the \emph{boundary line}. 

Altogether, the correct relic density can be reproduced for any point within the two-dimensional parameter space for a suitable value of $\lambda_\chi$, via conversion-driven freeze-out below the boundary line, and via conventional coannihilation above the boundary line. Note that the effective $\tilde q\tilde q^\dag$ cross section, eq.~\eqref{eq:effgeneral}, including bound-state effects is relevant both in the coannihilation as well as the conversion-driven regimes, and therefore also for determining the boundary between them. 

In Fig.\,\ref{fig:boundary}, we show the boundary line in the $(m_\chi,\Delta m)$ plane obtained for various approximations, which successively include a number of effects. When using the perturbative tree-level $\tilde q\tilde q^\dag$ annihilation cross section only, one obtains the line labeled `pert.'\@. This is the result one would obtain when using standard tools for the relic density computation~\cite{Belanger:2018ccd,Ambrogi:2018jqj,Bringmann:2018lay} without further modification. The line labeled `Som.' is obtained when including Sommerfeld enhancement of $\tilde q\tilde q^\dag$ annihilation, and this approximation has been used in previous works in the context of conversion-driven freeze-out with colored mediators~\cite{Garny:2017rxs,Garny:2018icg}.\footnote{Note that the boundary in~\cite{Garny:2017rxs} slightly exceeds the one found here including Sommerfeld enhancement only. This is due to a slightly different choice of the running $\alpha_s$. Here, for definiteness and for a better comparison of the perturbative result to previous literature, we use the same $\alpha_s$ parametrization as in~\cite{Belanger:2018ccd}.}

The regime of conversion-driven freeze-out extends significantly when including the bound-state effects considered in this work. The line labeled `BS, $n=1$' in Fig.\,\ref{fig:boundary} corresponds to including the contribution from the ground state only. Finally, adding excited states up to $n=15$ within the default approximation discussed in Sec.\,\ref{sec:sigeff} yields the thick solid line. We observe that the conversion-driven freeze-out region reaches to significantly higher values of $m_\chi$ and also $\Delta m$ due to the impact of bound states. 

Let us briefly comment on the role of excited states. For $m_\chi/\Delta m\lesssim {\cal O}(10^2)$, freeze-out dominantly takes place in the regime of ionization equilibrium. In that case, excited states lead to a correction of the effective cross section of order 20\%, due to the additional available decay channels, see eq.~\eqref{eq:sigeffioneq2}. For $m_\chi/\Delta m\gtrsim {\cal O}(10^2)$, the freeze-out extends to lower temperatures. In this regime, a combination of two effects leads to a significant enhancement of the impact of excited states.
First, since ionization equilibrium breaks down for the ground state, its contribution to the effective cross section drops. 
Secondly, the bound-state formation rate for excited states exceeds the one of the ground state by many orders of magnitude at low temperatures.
Hence, excitations remain in ionization equilibrium toward smaller temperatures and dominate the effective cross section.\footnote{While in the considered scenario, very large values of $m_\chi/\Delta m$ only occur toward the `tail' of the boundary line, very large values of $x$ can naturally become relevant in the superWIMP scenario where dark matter is thermally decoupled and only produced through the late decay of the mediator particle. Indeed, already the effect of $n=1$ bound states is sizeable~\cite{Decant:2021mhj} in this scenario motivating further studies in the future.}

Potentially, the region of conversion-driven freeze-out could even become larger when including transitions between the bound states, which is beyond the scope of this work. To provide a maximal upper bound we show the result that would be obtained when assuming ionization equilibrium to hold during the entire freeze-out and including all $n$ using eq.~\eqref{eq:sigeffioneq2}, indicated by the gray dotted line. The full result when including transitions is expected to lie significantly below this line, and above the solid line, \cf~the respective results for $\avb{\sigma_{\tilde q\tilde q^\dagger} v}_\text{eff}^\text{BS}$ in the left panel of Fig.~\ref{fig:sigeff}. For the regime where the gray and thick solid lines differ from each other, ionization equilibrium breaks down during the freeze-out.
The boundary therefore becomes insensitive to uncertainties from transitions among bound states where both lines converge, \ie~for $m_\chi\lesssim 2\,$TeV.

\subsection{Coannihilation regime}\label{sec:coannihilation}

\begin{figure}
  \centering
  \includegraphics[width=0.45\textwidth]{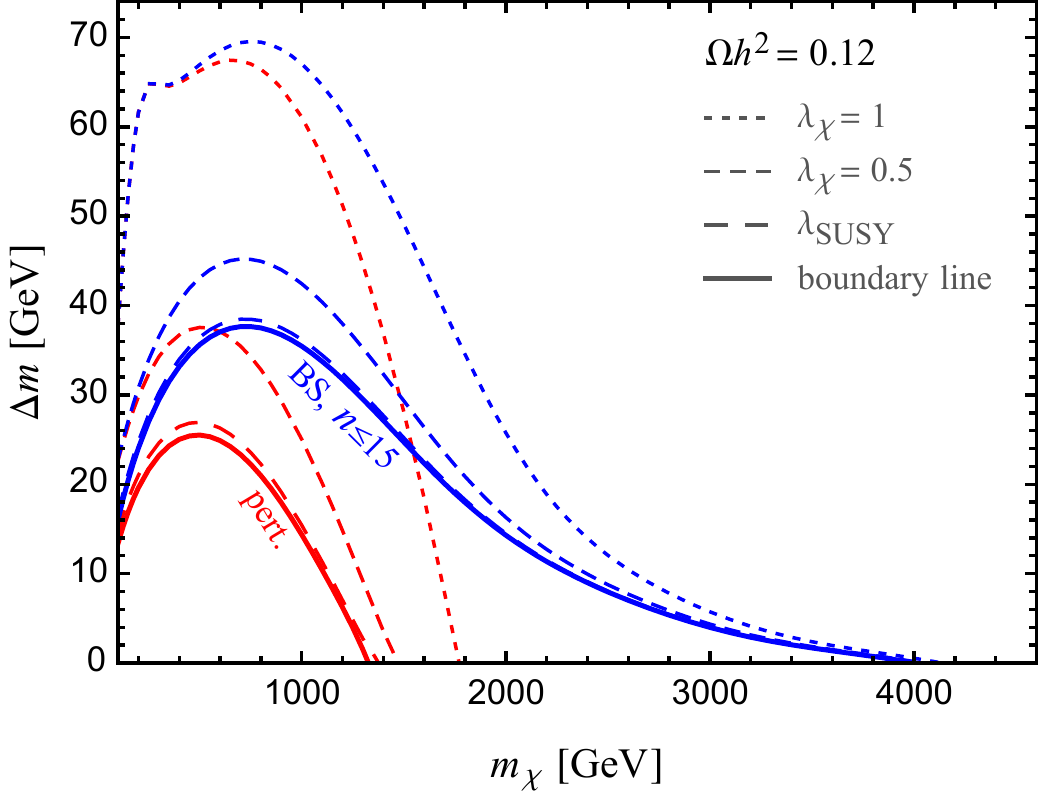}
  \caption{Contours for which dark matter coannihilation yields a relic abundance that matches the observed value, for three fixed values of the coupling $\lambda_\chi$.
  The red lines show the perturbative leading-order result, and the blue lines show the result when including bound states up to $n=15$ as well as Sommerfeld enhancement.
  For the relevance of individual corrections, we refer to Fig.\,\ref{fig:boundary}. The boundary to the conversion-driven regime is also shown.
  }
  \label{fig:lambdacont}
\end{figure}

While the main focus of this work is on the impact of bound states on conversion-driven freeze-out,
we also assess the relevance in the coannihilation regime. As is already apparent from Fig.\,\ref{fig:boundary},
bound states and Sommerfeld enhancement have a significant impact on the boundary, and therefore on
coannihilations as well. In Fig.\,\ref{fig:lambdacont},
we show the contours in the $(m_\chi,\Delta m)$ plane for which freeze-out in the coannihilation regime yields
the correct dark matter relic abundance for three values of the coupling, $\lambda_\chi=0.169, 0.5, 1$, respectively.
The former choice is motivated by supersymmetry, for which the $\chi$ particle can be viewed as the bino and the mediator as the right-handed sbottom quark within the MSSM\@.
In this case, the coupling is fixed by the bottom hypercharge. We note that for large $\lambda_\chi\gtrsim{\cal O}(1)$, additional annihilation diagrams for $\tilde q\tilde q^\dag\to b\bar b$
as well as $\tilde q\tilde q\to b b$ contribute, which are modified by bound-state formation. In this work, we are mainly interested in the case of small $\lambda_\chi$, and
therefore do not take these contributions into account, since their cross section scales as $\lambda_\chi^4$ and is subleading compared to the QCD contributions
to $\tilde q\tilde q^\dag$ annihilation.

The red lines in Fig.\,\ref{fig:lambdacont} correspond to the case with perturbative leading-order annihilation, and the blue lines correspond to our fiducial  
approximation that includes Sommerfeld enhancement and bound states up to $n=15$. 
It is apparent that the blue contours allow for significantly larger masses $m_\chi$ for a given $\lambda_\chi$.
For example, for $\lambda_\chi=0.5$ and $\Delta m=20$\,GeV, the mass for which the relic density matches the observed value shifts from $m_\chi\simeq 1.2$\,TeV to $2$\,TeV when including the aforementioned corrections. For the MSSM value, $\lambda_\chi=0.169$, the contour almost coincides with the boundary, and the mass shifts from $m_\chi\simeq 0.9$\,TeV to $1.8$\,TeV (for $\Delta m=20$\,GeV). In addition, for a very small mass splitting, including bound states allows for mediator masses in the multi-TeV regime, around $m_\chi=3$\,TeV for $\Delta m=5$\,GeV. This shift can be expected to be of major relevance for experimental searches for colored $t$-channel mediators within the coannihilation regime. It re-opens part of the parameter space that is constraint by conventional dark matter searches.

\subsection{Conversion-driven regime and collider limits}\label{sec:conversion}

\begin{figure}
  \centering
  \includegraphics[width=0.45\textwidth]{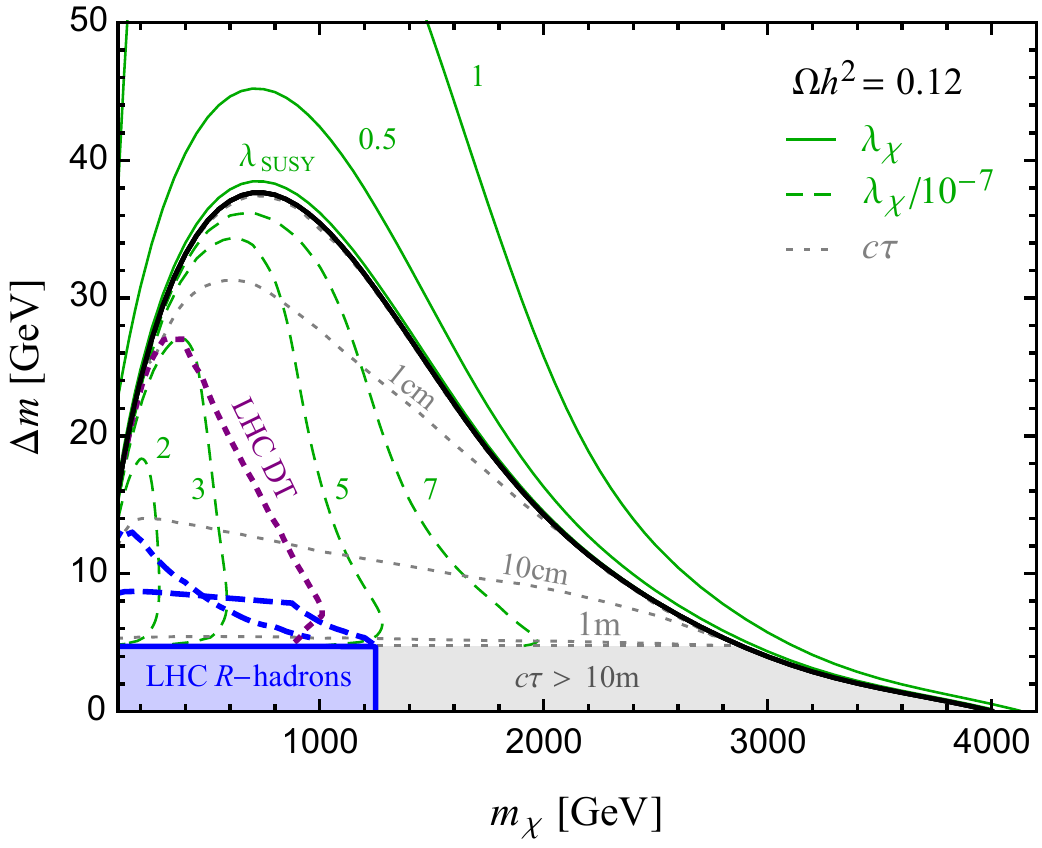}
  \caption{
Cosmologically allowed parameter space ($\Omega h^2 =0.12$) for conversion-driven freeze-out when taking bound states with $n\leq 15$ as well as Sommerfeld enhancement into account. Green dashed lines show contours of the coupling $\lambda_\chi$ in units of $10^{-7}$,
and gray lines show the contours of the mediator decay length. In addition, LHC bounds from $R$-hadron searches as well as disappearing track searches are shown, as well as the contours within the
coannihilation regime (see Fig.\,\ref{fig:lambdacont}).
}
  \label{fig:ParamSpaceContours}
\end{figure}

In Fig.\,\ref{fig:ParamSpaceContours} we show the viable parameter space within the regime of conversion-driven freeze-out.
The value of the coupling that is required to obtain the measured dark matter abundance is of order $10^{-6}\!-\!10^{-7}$ in that case.
We show several contours for $\lambda_\chi/10^{-7}=2,3,5,7$. The smallness of the coupling implies that this production mechanism
is compatible with null results from direct and indirect dark matter detection experiments, while still providing an explanation of the
abundance of dark matter that is insensitive to the initial conditions.

The decay length $c\tau$ of the mediator, where $\tau$ is its lifetime, is shown by the gray contour lines in Fig.\,\ref{fig:ParamSpaceContours}.
It is of the order of a few centimeters to $1\,\text{m}$ within most of the parameter space, going down to $1\,\text{mm}$ close to the boundary.
For the freeze-out computation, we limit ourselves to the parameter space where $\Delta m>m_b$, such that
the two-body decay $\tilde q\to \chi b$ is kinematically allowed. For even smaller mass splitting, conversions proceed via scatterings, and the mediator
would be stable on detector timescales. 

The primary signal of conversion-driven dark matter production with a colored mediator are searches for heavy, (meta-)stable colored particles at the LHC.
For $\Delta m<m_b$, the colored mediator becomes detector stable as its decay is four-body suppressed. We can directly apply the limit from the 13\,TeV ATLAS search~\cite{ATLAS:2019gqq} derived for an $R$-hadron containing a $b$-squark. It excludes masses below 1250\,GeV. The resulting limit is shown in Fig.\,\ref{fig:ParamSpaceContours} as a solid blue curve (and blue shaded exclusion region). For larger $\Delta m$ the decay length is in the range $1\,\text{mm}\!\sim\!1\,\text{m}$ such that a sizeable fraction of decays take place inside the inner detector. To estimate the reach of the same search for this case, we employ the reported cross section upper limits for the muon-system-agnostic analysis for a $b$-squark $R$-hadron. We rescale them by the relative suppression of the cross section upper limits toward small lifetimes reported in the similar ATLAS analysis~\cite{ATLAS:2018lob} where the case of a gluino $R$-hadron has been considered. Note that this introduces a certain level of approximation. A recasting of the search is, however, beyond the scope of this work. We use the cross-section predictions from~\cite{Beenakker:2016lwe}. The resulting limit is displayed as the blue, dashed curve in Fig.\,\ref{fig:ParamSpaceContours}. Furthermore, we display the limit from the recasting of the CMS 13\,TeV $R$-hadron search~\cite{CMS:2016ybj} performed in~\cite{Garny:2017rxs} as the blue, dot-dashed curve.  

Being only sensitive to the fraction of $R$-hadrons traversing a significant part of the detector, the sensitivity of these searches is exponentially suppressed for small lifetimes. Dedicated analyses exploiting the displaced nature of the decay are, hence, expected to greatly improve the sensitivity to this scenario. While several such analyses have been performed by the collaborations, their target model differs considerably from the one considered here, significantly reducing their reach or raising questions about their applicability as pointed out in~\cite{Brooijmans:2020yij} (contribution 7). For instance, the sensitivity of the displaced jets search~\cite{ATLAS:2017tny} considerably suffers from the imposed cut on the invariant mass of the displaced tracks. While the respective choice was optimized for the scenario considered in the search, it reduces the signal of the one considered here by around two orders of magnitude~\cite{Brooijmans:2020yij}. This is due to its relatively small mass splittings $\Delta m$ of order tens of GeV in our scenario, resulting in softer tracks. The search has been targeted to mass splittings of the order of hundreds of GeV. 

Another example of a potentially sensitive search is the one for disappearing tracks. The existing searches are targeted to charginos whose long lifetime arises due to a tiny mass splitting, ${\cal O} (100\,\text{MeV})$, to the dark matter particle. Accordingly, in the decay, an ultra-soft pion is emitted facilitating the use of a disappearance condition. In our scenario, the emitted $b$-jet is considerably harder than in the targeted model. 
However, the search is estimated to still provide sensitivity to the model considered here, as shown in the approximate recasting of~\cite{ATLAS:2017oal} performed in~\cite{Brooijmans:2020yij}. In this recasting, the probability of the $R$-hadron to cause a charged track was also taken into account. We overlay the respective limit as the purple dotted curve in Fig.\,\ref{fig:ParamSpaceContours}.

We conclude that, after including the impact of bound states, a wide part of the parameter space for conversion-driven freeze-out is still viable, and provides a clear target for long-lived particle searches at future LHC runs.

\section{Conclusion}\label{sec:conclusion}

In this work, we revisited the computation of the relic density in the presence of bound-state effects during dark matter freeze-out. With respect to previous work, we improved the calculations in various aspects and demonstrated the respective phenomenological implications on the cosmologically viable parameter space in the coannihilation and conversion-driven freeze-out scenario.

In the first part of this work, we reformulated the Boltzmann equations including arbitrary excitations of bound states and derived a general framework for incorporating their effects in terms of an effective annihilation cross section. While a full treatment of these effects requires the knowledge of all involved bound-state formation, decay, and transition rates, we introduced meaningful limiting cases when assuming fully efficient or non-efficient transitions. We provided simple analytical expressions for the effective cross section in these limits, as well as a general result. Furthermore, we showed that for an arbitrary set of bound states in ionization equilibrium, the effective cross section is independent of bound-state formation and transition rates, and only depends on a weighted sum of bound-state decay rates.

For the case of a colored coannihilator, we computed the radiative bound-state formation rates for arbitrary excitations with quantum numbers $n,\ell$, and estimate the lowest order transition rates. 
Furthermore, we investigated the impact of NLO corrections to bound-state decays. We further discuss the relevance of NLO effects on bound-state formation and decay in App.\,\ref{sec:NLOcorr}.

We then solved the coupled Boltzmann equation for the mediator and the dark matter particle in a $t$-channel model and assessed the impact of bound states for coannihilations as well as conversion-driven freeze-out. On the one hand, in ionization equilibrium, the effective mediator annihilation cross section is insensitive to the bound-state formation but directly proportional to the bound-state decay rates. Including excited states increases the effective cross section by about 20\% in that case. 
On the other hand, after the breakdown of ionization equilibrium of the ground state, higher excitations become increasingly important. At the same time, a large bound-state formation rate 
extends the duration of ionization equilibrium down to smaller temperatures. Nevertheless, we found that freeze-out significantly extends beyond the period of ionization equilibrium for small relative mass splittings between the mediator and dark matter, phenomenologically most relevant in the region of high masses, $m_\chi\gtrsim 2$\,TeV. In this region of parameter space, our fiducial approximation that neglects bound-state transitions is expected to underestimate the effects of excited bound states, motivating further studies. In addition, we demonstrated that NLO corrections to the bound-state formation rate itself play only a moderate role in the setup considered here.

Evaluating the cosmologically viable parameter space, we found that the region for which conversion-driven freeze-out is relevant extends significantly when including bound-state effects, ranging up to the multi-TeV region. In addition, our findings imply that significantly higher dark matter masses are viable also within the coannihilation region. This has immediate consequences for dark matter searches. For instance, considering a mass splitting of 20 GeV and a coupling of $\sim 0.169$,  as predicted in the MSSM, the dark matter mass that matches the relic density is shifted from around 900 GeV to 1.8\,TeV by the inclusion of the discussed effects. On the other hand, when keeping the masses fixed at $m_\chi=900$ GeV and $\Delta m = 20$ GeV, the coupling would change from $0.169$  to around $5\times 10^{-7}$ as it lies in the conversion-driven freeze-out regime.

Dark matter produced via conversion-driven freeze-out is compatible with (in)direct detection limits due to a very weak coupling but yields signatures of long-lived particles at the LHC\@. We discussed the applicability of existing searches for $R$-hadrons, disappearing tracks and displaced jets, which exclude masses below about $0.6-1.2$\,TeV. Because of the increase of the viable parameter space for conversion-driven freeze-out, extending into the multi-TeV region, the scenario provides great prospects for long-lived particle searches at future LHC runs.

The computations considered here can be improved in future work in several ways, regarding the description of transitions among bound states, the decay of excited states with angular momentum, as well as the inclusion of thermal corrections.

\section*{Acknowledgments}

We thank Martin Beneke, Stefan Lederer, Kai Urban and Stefan Vogl
for discussions as well as Tobias Binder for pointing us to a correction of the zero-temperature NLO result provided in~\cite{Binder:2021otw}.
This work was supported by the DFG Collaborative Research Institution Neutrinos and Dark
Matter in Astro- and Particle Physics (SFB 1258) and the Collaborative Research Center TRR 257. Furthermore, JH acknowledges support by the F.R.S.-FNRS via the Charg\'e de recherches fellowship.

{\bf Note added:} While this work was being completed, \cite{Binder:2021vfo} appeared that discusses the inclusion
of a set of multiple bound states and transitions among them in terms of an effective cross section in analogy to the material presented in Sec.\,\ref{sec:bound} in this work.

\begin{appendix}

\section{Bound-state formation cross section}\label{sec:bsfapp}

In this appendix we sketch the derivation of the recombination cross section, eq.~\eqref{eq:sigmaBSFnl}.
We use hydrogen-like wave functions for the scattering and bound states, with normalization
\bea
  \int \diff^3r \psi^*_{\vec p_\text{rel}}(\vec r)\psi_{\vec p'_\text{rel}}(\vec r)=(2\pi)^3\delta^{(3)}(\vec p_\text{rel}-\vec p'_\text{rel})\,,\nn\\
  \int \diff^3r \psi^*_{n\ell m}(\vec r)\psi_{n'\ell' m'}(\vec r)=\delta_{nn'}\delta_{\ell\ell'}\delta_{mm'}\,.
\eea
The scattering state has an energy eigenvalue $\vec p_\text{rel}^2/(2\mu)$, where $\mu=m_{\tilde q}/2$ is the reduced mass, and satisfies the Schr{\"o}dinger equation
with potential $V=-\alpha^\text{eff}_s/r$. For the bound state, we assume $V=-\alpha^\text{eff}_b/r$, with a different effective coupling, and eigenvalue given by
$-E_{{\cal B}_{n\ell}}=-(\alpha^\text{eff}_b)^2\mu/(2n^2)$.
We omit labels for the $SU(N_c)$ representation, with it being understood that the scattering state is evaluated for the effective strong coupling of the octet (singlet) for the gluonic (electromagnetic) recombination process, while the bound state is always a singlet. The derivation is general and the representation enters only via the effective coupling strengths.

The scattering-state wave function is given by (see e.g.~\cite{Yao:2020xzw})
\be
  \psi_{\vec p_\text{rel}}(\vec r) = 4\pi\sum_{\ell,m} i^\ell \E^{i\delta_\ell}\frac{F_\ell(\rho)}{\rho}Y_{\ell m}(\hat r)Y_{\ell m}^*(\hat p_\text{rel})\,,
\ee
with $\zeta_s=\alpha^\text{eff}_s/v_\text{rel}$ and
\bea
  \rho&=&p_\text{rel} r\,,\nn\\
  \delta_\ell &=& \text{arg}\!\left(\Gamma(1+\ell-i\zeta_s)\right)\,,\nn\\
  F_\ell(\rho) &=& \frac{2^\ell \E^{\pi\zeta_s/2}\left|\Gamma(1+\ell-i\zeta_s)\right|}{(2\ell+1)!}\,\rho^{\ell+1}\E^{i\rho}\,, \nn\\
  && {} \times {}_1F_1(\ell+1-i\zeta_s,2\ell+2;-2i\rho)\,.
\eea
The bound-state wave function is given by (see e.g.~\cite{Harz:2018csl})
\be
  \psi_{n\ell m}(\vec r) = F_{n\ell}(r) Y_{\ell m}(\hat r)\,,
\ee
where $\kappa=\mu\alpha_b^\text{eff}=p_\text{rel}\zeta_b$, and the radial part is
\bea
   F_{n\ell}(r) &=& \kappa^{3/2}\sqrt{\frac{4(n-\ell-1)!}{n^4(n+\ell)!}}\left(\frac{2\kappa r}{n}\right)^\ell \nn\\
   && {} \times L^{(2\ell+1)}_{n-\ell-1}\left(\frac{2\kappa r}{n}\right)\E^{-\kappa r/n}\,.
\eea
We are interested in
\bea
  |\langle\psi_{n\ell}|\vec r|\psi_{\vec p_\text{rel}}\rangle|^2 &=& \frac{1}{2\ell+1}\sum_m\int \diff^3r\,\diff^3r'\, \vec r\cdot\vec r'\nn\\
  && {} \times \psi_{n\ell m}(\vec r)\psi^*_{n\ell m}(\vec r')\psi_{\vec p_\text{rel}}^*(\vec r)\psi_{\vec p_\text{rel}}(\vec r')\,.\nn\\
\eea
To separate radial and angular parts we use
\be
  \vec r\cdot\vec r'=rr'\frac{4\pi}{3}\sum_{\tilde m=-1}^1Y_{1\tilde m}(\hat r)Y_{1\tilde m}^*(\hat r')\,.
\ee
The angular integral, for given partial wave contribution $\ell'm'$ in $\psi_{\vec p_\text{rel}}^*(\vec r)$, and $\ell''m''$ in $\psi_{\vec p_\text{rel}}(\vec r')$, respectively,
can be computed using standard relations for Wigner $3j$-symbols,
\bea
 I_A &\equiv& \frac{4\pi}{3}\sum_{m=-\ell}^\ell\sum_{\tilde m=-1}^1\int d\Omega_rd\Omega_{r'}Y_{\ell m}(\hat r)Y_{\ell m}^*(\hat r') \nn\\
 && {} \times Y_{\ell'm'}^*(\hat r)Y_{\ell''m''}(\hat r')Y_{1\tilde m}(\hat r)Y_{1\tilde m}^*(\hat r')\nn\\
 &=& \delta_{\ell'\ell''}\delta_{m'm''}\frac{\ell'\delta_{\ell',\ell+1}+\ell\delta_{\ell,\ell'+1}}{2\ell'+1}\,.
\eea
This gives
\be
  |\langle\psi_{n\ell}|\vec r|\psi_{\vec p_\text{rel}}\rangle|^2=
  \frac{4\pi}{p_\text{rel}^5}\sum_{\ell'}\frac{\ell'\delta_{\ell',\ell+1}+\ell\delta_{\ell,\ell'+1}}{2\ell+1}|I_R|^2\,,
\ee
with the radial overlap integral
\be
  I_R \equiv \int_0^\infty d\rho \rho^2 f_{n\ell}(\rho)F_{\ell'}^*(\rho)\,,
\ee
where $f_{n\ell}(\rho)=F_{n\ell}(r)/p_\text{rel}^{3/2}$ is the dimensionless radial wave function of the bound state.

To compute the radial integral we use an integral representation of the hypergeometric function that appears in the scattering wave function,
\bea
  \lefteqn{ {}_1F_1(\ell'+1-i\zeta_s,2\ell'+2;-2i\rho)= }\nn\\
  && \frac{(2\ell'+1)!}{|\Gamma(1+\ell'-i\zeta_s)|^2}\int_0^1 ds s^{\ell'-i\zeta_s}(1-s)^{\ell'+i\zeta_s}\E^{-2i\rho s}\,.\nn\\
\eea
Note that by substituting $s\to 1-s$ one finds that $F_{\ell'}(\rho)$ is real, such that we can drop the complex conjugate in $I_R$. In addition, we use the generating function of the Laguerre polynomials for the bound-state wave function, to write
\be
  L_n^{(\alpha)}(x) = \frac{1}{n!}\left(\frac{d}{dt}\right)^n\frac{\E^{-x\frac{t}{1-t}}}{(1-t)^{1+\alpha}}\Big|_{t=0}\,.
\ee
The $\rho$ integration can be performed using the definition of the $\Gamma$ function, and we obtain
\bea
  I_R &=& \frac{2\zeta_b^{3/2}(\ell+\ell'+3)!}{n^2\sqrt{(n-\ell-1)!(n+\ell)!}}\left(\frac{2\zeta_b}{n}\right)^\ell\frac{2^{\ell'}\E^{\pi\zeta_s/2}}{|\Gamma(1+\ell'-i\zeta_s)|}\nn\\
  && \left(\frac{d}{dt}\right)^{n-\ell-1}\frac{1}{(1-t)^{2\ell+2}}\nn\\
  && \int_0^1 ds \frac{s^{\ell'-i\zeta_s}(1-s)^{\ell'+i\zeta_s}}{\left(\frac{\zeta_b}{n}\frac{1+t}{1-t}+i(2s-1)\right)^{\ell+\ell'+4}}\Big|_{t=0}\,.
\eea
We find, setting $a\equiv i\zeta_s,b\equiv i\frac{\zeta_b}{n}\frac{1+t}{1-t}$,
\bea
  \lefteqn{ \int_0^1 ds \frac{s^{\ell'-a}(1-s)^{\ell'+a}}{s-\frac12-b/2}= }\nn\\
  && \frac{\pi}{\sin(a\pi)}\left[\left(\frac{b-1}{b+1}\right)^a\frac{(1-b^2)^{\ell'}}{2^{2\ell'}}+\sum_{r=0}^{2\ell'}a^{2\ell'-r}b^rc_r\right]\,,\nn\\
\eea
with some rational coefficients $c_r$, that will be unimportant in the following.
We can generate the required integral by differentiating $\ell+\ell'+3$ with respect to $b$. Because of the selection rule, $\ell+\ell'+3\geq 2\ell'+2>2\ell'$, such that
the sum over $r$ in the square bracket drops out, as announced. Using
\be
  \left(\frac{b-1}{b+1}\right)^a=\exp\left(-2\zeta_s \text{arccot} \frac{\zeta_b}{n}\frac{1+t}{1-t} \right)\,,
\ee
setting $z\equiv \frac{\zeta_b}{n}\frac{1+t}{1-t}=-ib$ and using $\sin(a\pi)=i\sinh(\zeta_s\pi)$ yields
\bea
  \lefteqn{ \int_0^1 ds \frac{s^{\ell'-i\zeta_s}(1-s)^{\ell'+i\zeta_s}}{\left(z+i(2s-1)\right)^{\ell+\ell'+4}}= }\nn\\
  && \frac{\pi}{\sinh(\pi\zeta_s)}\frac{(-1)^{\ell+\ell'}}{2^{2\ell'+1}(\ell+\ell'+3)!}\nn\\
  && {} \times \left(\frac{d}{dz}\right)^{\ell+\ell'+3}(1+z^2)^{\ell'}\E^{-2\zeta_s \text{arccot}(z)}\,,
\eea
which allows us to evaluate the radial integral.
Using
\bea
   |\Gamma(1+\ell'-i\zeta_s)|^2 &=& \frac{\pi}{\zeta_s\sinh(\pi\zeta_s)} \times \zeta_s^2\times(1+\zeta_s^2) \nn\\
  && {} \times(2+\zeta_s^2)\times\cdots\times({\ell'}^2+\zeta_s^2)\,,\nn\\
\eea
finally gives the result, eq.~\eqref{eq:IR}, for the radial integral.

\section{NLO corrections}\label{sec:NLOcorr}

Here, we discuss the impact of NLO corrections to bound-state formation effects.
In general, there are various sources of potential higher-order corrections for the complete effective cross section, eq.~\eqref{eq:effgeneral}, including
\begin{enumerate}
\item\label{deccorr} the bound-state decay rate (relevant in ionization equilibrium), including (i) virtual corrections to the ${\cal B}_{n0}\to gg$ decay, (ii) real corrections, that is, three-body decays into $ggg$ and $g\bar q q$, (iii) scattering processes such as $q{\cal B}_{n0}\to qg$, (iv) decays of $\ell\not=0$ bound states,
\item\label{bsfcorr} the bound-state formation rate (relevant out of ionization equilibrium), including (i) transition operators beyond the color-electric dipole term, (ii) virtual and real corrections to the $\tilde q\tilde q^\dag\to {\cal B}_{n\ell} g$ transition, involving $3\to 2$ and $2\to 3$ processes (collisional bound-state formation), and
\item the transition rates between bound states (also relevant out of ionization equilibrium).
\end{enumerate}
A complete treatment of all NLO corrections in $\alpha_s$ would be interesting but is not available at the moment.
In Sec.\,\ref{sec:decay}, we briefly discussed the impact of NLO corrections to bound-state decay [related to point \ref{deccorr}(i/ii)], and discussed also point \ref{deccorr}(iii). 
In this appendix, we investigate the quantitative impact of NLO corrections to the decay [\cf~\ref{deccorr}(i/ii)] as well as
the NLO corrections considered in~\cite{Binder:2021otw} (see also~\cite{Binder:2020efn}), that are related to point \ref{bsfcorr}(ii).

\subsection{NLO corrections to bound-state formation}\label{sec:nlobsf}

\begin{figure*}
  \centering
  \includegraphics[width=0.42\textwidth, trim= {0cm -0.08cm 0cm 0cm}, clip]{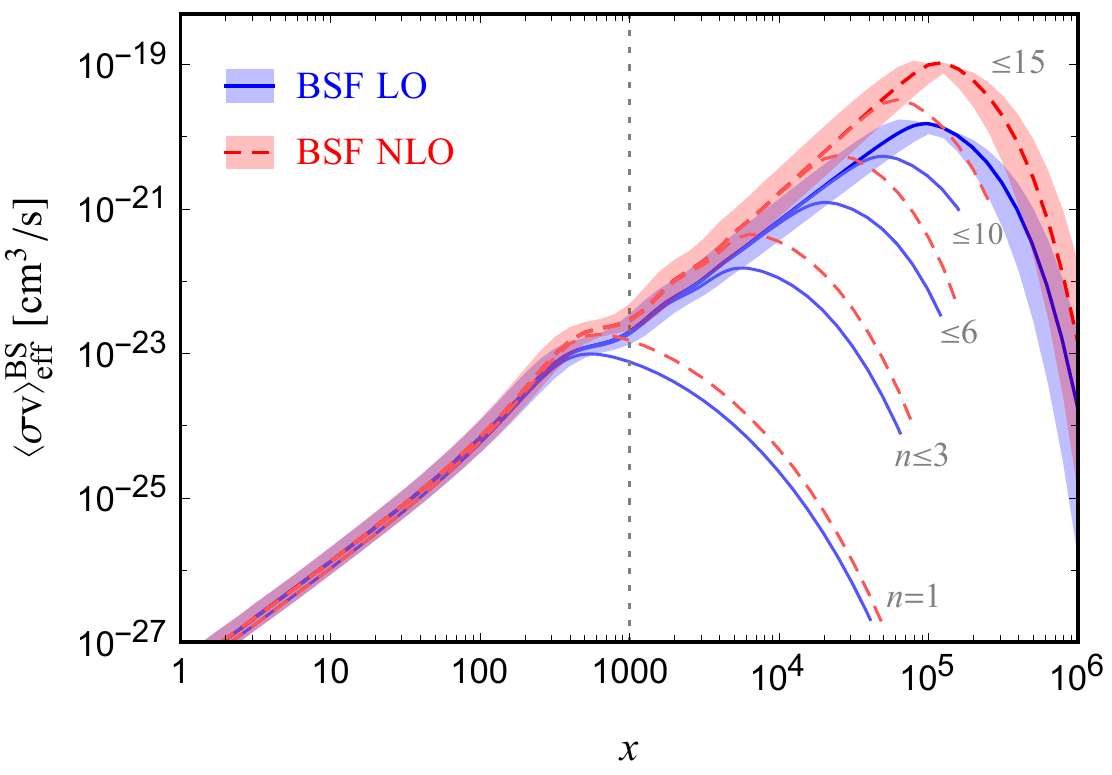}
    \hspace*{5mm}
  \includegraphics[width=0.44\textwidth, trim= {0cm 0.1cm 0cm 0cm}, clip]{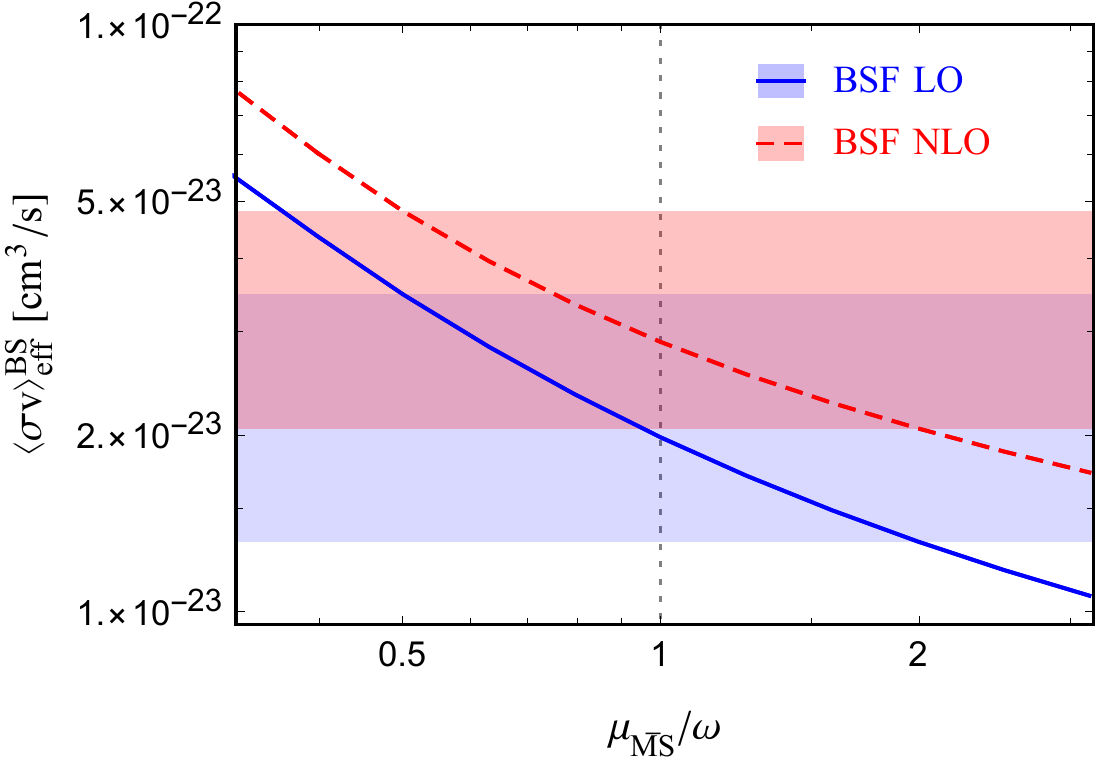}
  \caption{Impact of the class of NLO corrections presented in~\cite{Binder:2021otw} on the contribution to the effective cross section, eq.~\eqref{eq:effnfirst}, from bound states. The corrections capture collisional bound-state formation processes. The left panel shows the dependence on the temperature parameter $x$. The bands show the uncertainty from the scale choice of the strong coupling (increased or decreased by a factor two relative to the fiducial choice discussed in the main text) when taking excited states up to $n=15$ into account. The right panel shows the dependence on the renormalization scale for $x=10^3$. We consider the benchmark point $m_\chi=1$\,TeV, $\Delta m =20\,$GeV.
}
  \label{fig:scaledep}
\end{figure*}

The impact of including the vacuum and finite temperature correction as given in~\cite{Binder:2021otw} on the effective cross section, eq.~\eqref{eq:effgeneral}, in the no-transition limit, eq.~\eqref{eq:effnfirst}, is shown in Fig.\,\ref{fig:scaledep}. In~\cite{Binder:2020efn,Binder:2021otw} it was pointed out that the correction to the bound-state formation cross section
becomes very large for small enough $x$, corresponding to $T\gtrsim E_{{\cal B}_{n\ell}}$. Nevertheless, for these temperatures, ionization equilibrium holds to a large extent. In ionization equilibrium, the effective cross section becomes insensitive to the bound-state formation cross section. Therefore, the effect of the NLO corrections considered in~\cite{Binder:2021otw} on the effective cross section is almost negligible for small $x$ (left part of the left panel in Fig.\,\ref{fig:scaledep}). For large $x$, on the other hand, the temperature is so small that the finite-temperature contribution of the NLO corrections gives a negligible contribution. In this region, the zero-temperature correction dominates. This is the reason why the difference between LO and the NLO correction considered in~\cite{Binder:2021otw} is moderate in the right part of the left panel in Fig.\,\ref{fig:scaledep}. However, it becomes more relevant for excited states, due to the larger effective strong coupling, given our scale choice eq.~\eqref{eq:alphaeff}.

In order to further assess the impact of NLO corrections, we show the dependence of the effective cross section when changing all scales at which the strong coupling is evaluated by a factor of two or a half, respectively, by the colored bands in Fig.\,\ref{fig:scaledep}. Within the perturbative uncertainty, both results are consistent with each other. We observe that including the NLO corrections considered in~\cite{Binder:2021otw} leads only to a small reduction of the scale uncertainty (right panel of Fig.\,\ref{fig:scaledep}). This indicates that further sources of higher-order corrections, including those listed above, would have to be taken into account for a complete NLO analysis.

\begin{figure}
  \centering
  \includegraphics[width=0.42\textwidth, trim= {0cm -0.08cm 0cm 0cm}, clip]{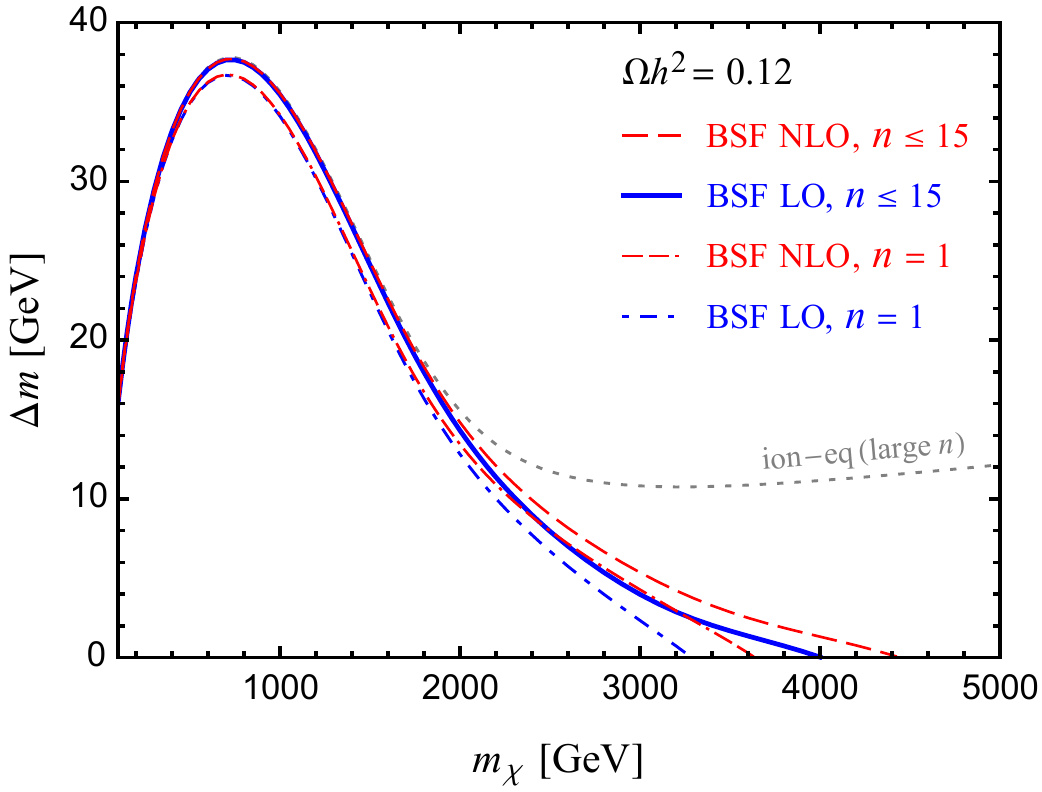}
  \caption{Impact of the class of NLO corrections presented in~\cite{Binder:2021otw} on the boundary between the coannihilation and conversion-driven regime (red: with BSF NLO correction, blue: without). The impact on the boundary is smaller than the difference that arises when including excited states ($n\leq 15$) as opposed to the ground state only ($n=1$), which is shown for comparison for both cases, respectively. 
}
  \label{fig:boundaryNLO}
\end{figure}

The effect of the NLO corrections on the boundary between the coannihilation and conversion-driven regime is shown in Fig.\,\ref{fig:boundaryNLO}, and compared to the impact of taking excited states into account. We find that the latter is significantly more important. 

\subsection{NLO corrections to bound-state decay}\label{sec:nlodecay}

\begin{figure*}
  \centering
  \includegraphics[width=0.436\textwidth, trim= {0cm 0.05cm 0cm 0cm}, clip]{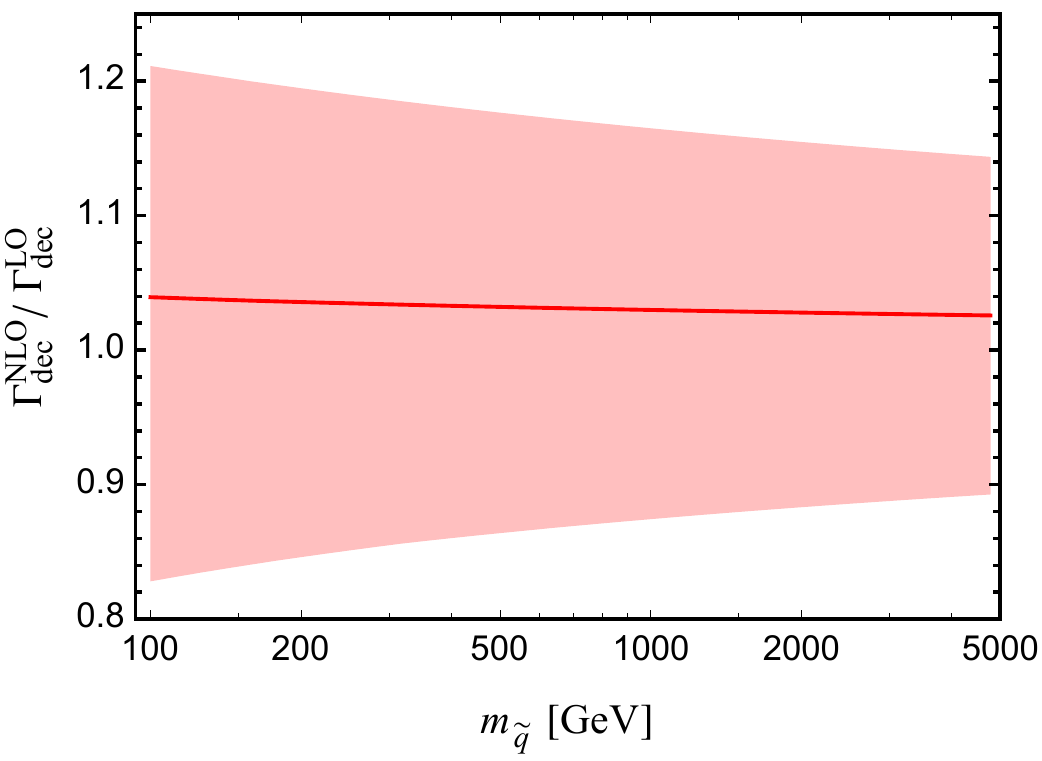}
    \hspace*{5mm}
  \includegraphics[width=0.44\textwidth, trim= {0cm 0.05cm 0cm 0cm}, clip]{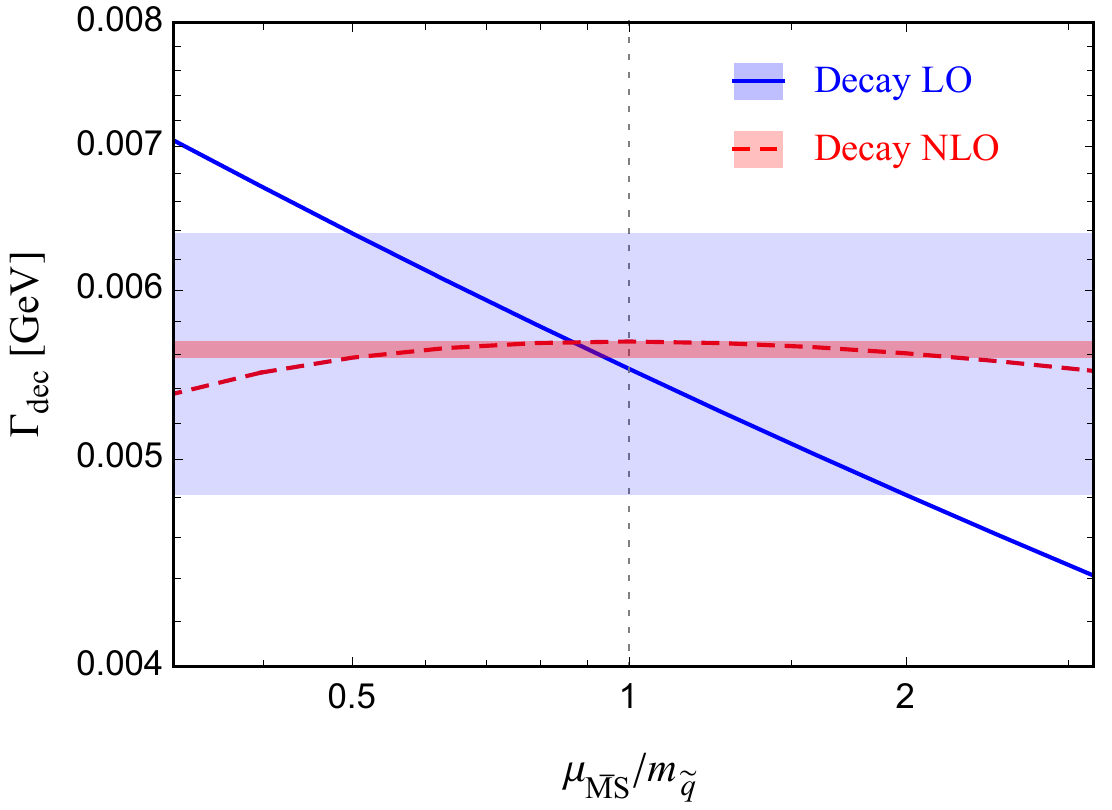}
  \caption{
 Impact of the NLO correction, eq.~\eqref{eq:BdecNLO}, to the bound-state decay rate. The left panel shows the NLO correction as a function of $m_{\tilde q}$. The central line corresponds to $\mu_{\overline{\mathrm{MS}}}=m_{\tilde q}$ while the lower and upper boundaries of the red shaded band corresponds to the choices $1/2\,m_{\tilde q}$ and $2m_{\tilde q}$, respectively. In the right panel, we display the scale dependence of the LO and NLO decay rates for the benchmark point with $m_{\tilde q}=1020$\,GeV.
}
  \label{fig:decNLO}
\end{figure*}

Real and virtual correction to the decay ${\cal B}_{10}\to gg$ have been computed in~\cite{Martin:2009dj}.
The relative correction at NLO in the limit of massless quarks is given by eq.~\eqref{eq:BdecNLO} in the main text.
Note that collinear singularities in the real correction cancel when including the virtual piece~\cite{Martin:2009dj}, analogously to heavy quarkonium decay~\cite{Barbieri:1979be,Hagiwara:1980nv,Petrelli:1997ge}.

In Fig.\,\ref{fig:decNLO} (left panel) we show the ratio $\Gamma_\mathrm{dec}^\mathrm{NLO}/\Gamma_\mathrm{dec}^\mathrm{LO}$ versus $m_{\tilde q}$ for different choices of the $\overline{\mathrm{MS}}$ renormalization scale. The central line corresponds to $\mu_{\overline{\mathrm{MS}}}=m_{\tilde q}$ while the lower and upper boundaries of the red shaded band correspond to the choices
$\mu_{\overline{\mathrm{MS}}}/m_{\tilde q}=1/2$ and 2, respectively. We adopted $\mu_{\overline{\mathrm{MS}}}=m_{\tilde q}$ in the main text, while $\mu_{\overline{\mathrm{MS}}}=2m_{\tilde q}$ is used \eg~in~\cite{Martin:2009dj}. We observe that the NLO correction is significantly smaller for $\mu_{\overline{\mathrm{MS}}}=m_{\tilde q}$, at the level of a few percent. This justifies using the LO decay rate in our main analysis for this scale choice. 

In Fig.\,\ref{fig:decNLO} (right panel) we show the dependence of the decay rate on $\mu_{\overline{\mathrm{MS}}}$ at LO and NLO, respectively. As expected, the NLO result is significantly less sensitive to the scale choice. Note that for these figures we have set $n_f=5$ and neglected the contribution from the top quark since the use of the massless approximation is in general not well justified in that case. Using the expressions for the real corrections for massive quarks obtained in~\cite{Martin:2009dj} confirms that the top quark contribution would amount to a small change of the already small NLO correction. Note that in Fig.\,\ref{fig:decNLO} we only vary $\alpha_s^\text{ann}$ while keeping $\alpha_b^\text{eff}$ fixed.

\begin{figure}
  \centering
  \includegraphics[width=0.42\textwidth, trim= {0cm -0.08cm 0cm 0cm}, clip]{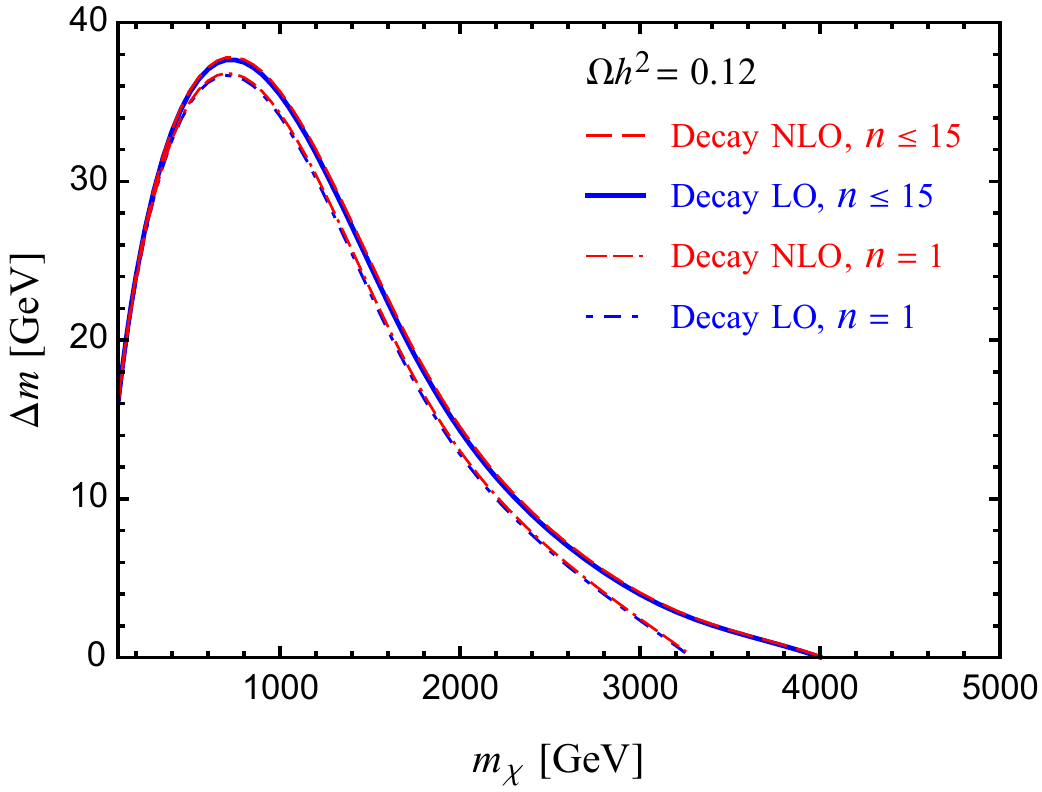}
  \caption{Impact of the NLO correction to the bound-state decay rate on the boundary line between the coannihilation and conversion-driven regime.
}
  \label{fig:boundaryDecNLO}
\end{figure}

In Fig.\,\ref{fig:boundaryDecNLO} we show the impact on the boundary line between conversion-driven freeze-out and coannihilation when taking into account NLO corrections to the decay. As expected, their impact is very small both for the ground state only and when taking into account excitations. Note that to obtain the NLO line when taking excited states into account we have assumed that $\Gamma_\mathrm{dec}^\mathrm{NLO}/\Gamma_\mathrm{dec}^\mathrm{LO}$ is identical for all states with arbitrary $n$ and $\ell=0$.

\end{appendix}
\bibliography{bibliography}{}

\begin{thebibliography}{56}
\expandafter\ifx\csname natexlab\endcsname\relax\def\natexlab#1{#1}\fi
\expandafter\ifx\csname bibnamefont\endcsname\relax
  \def\bibnamefont#1{#1}\fi
\expandafter\ifx\csname bibfnamefont\endcsname\relax
  \def\bibfnamefont#1{#1}\fi
\expandafter\ifx\csname citenamefont\endcsname\relax
  \def\citenamefont#1{#1}\fi
\expandafter\ifx\csname url\endcsname\relax
  \def\url#1{\texttt{#1}}\fi
\expandafter\ifx\csname urlprefix\endcsname\relax\def\urlprefix{URL }\fi
\providecommand{\bibinfo}[2]{#2}
\providecommand{\eprint}[2][]{\url{#2}}

\bibitem[{\citenamefont{Kahlhoefer}(2017)}]{Kahlhoefer:2017dnp}
\bibinfo{author}{\bibfnamefont{F.}~\bibnamefont{Kahlhoefer}},
  \bibinfo{journal}{Int. J. Mod. Phys. A} \textbf{\bibinfo{volume}{32}},
  \bibinfo{pages}{1730006} (\bibinfo{year}{2017}), \eprint{1702.02430}.

\bibitem[{\citenamefont{Marrod\'an~Undagoitia and
  Rauch}(2016)}]{Undagoitia:2015gya}
\bibinfo{author}{\bibfnamefont{T.}~\bibnamefont{Marrod\'an~Undagoitia}}
  \bibnamefont{and} \bibinfo{author}{\bibfnamefont{L.}~\bibnamefont{Rauch}},
  \bibinfo{journal}{J. Phys. G} \textbf{\bibinfo{volume}{43}},
  \bibinfo{pages}{013001} (\bibinfo{year}{2016}), \eprint{1509.08767}.

\bibitem[{\citenamefont{Gaskins}(2016)}]{Gaskins:2016cha}
\bibinfo{author}{\bibfnamefont{J.~M.} \bibnamefont{Gaskins}},
  \bibinfo{journal}{Contemp. Phys.} \textbf{\bibinfo{volume}{57}},
  \bibinfo{pages}{496} (\bibinfo{year}{2016}), \eprint{1604.00014}.

\bibitem[{\citenamefont{Griest and Seckel}(1991)}]{Griest:1990kh}
\bibinfo{author}{\bibfnamefont{K.}~\bibnamefont{Griest}} \bibnamefont{and}
  \bibinfo{author}{\bibfnamefont{D.}~\bibnamefont{Seckel}},
  \bibinfo{journal}{Phys. Rev.} \textbf{\bibinfo{volume}{D43}},
  \bibinfo{pages}{3191} (\bibinfo{year}{1991}).

\bibitem[{\citenamefont{Edsjo and Gondolo}(1997)}]{Edsjo:1997bg}
\bibinfo{author}{\bibfnamefont{J.}~\bibnamefont{Edsjo}} \bibnamefont{and}
  \bibinfo{author}{\bibfnamefont{P.}~\bibnamefont{Gondolo}},
  \bibinfo{journal}{Phys. Rev.} \textbf{\bibinfo{volume}{D56}},
  \bibinfo{pages}{1879} (\bibinfo{year}{1997}), \eprint{hep-ph/9704361}.

\bibitem[{\citenamefont{Ellis et~al.}(2000)\citenamefont{Ellis, Falk, Olive,
  and Srednicki}}]{Ellis:1999mm}
\bibinfo{author}{\bibfnamefont{J.~R.} \bibnamefont{Ellis}},
  \bibinfo{author}{\bibfnamefont{T.}~\bibnamefont{Falk}},
  \bibinfo{author}{\bibfnamefont{K.~A.} \bibnamefont{Olive}}, \bibnamefont{and}
  \bibinfo{author}{\bibfnamefont{M.}~\bibnamefont{Srednicki}},
  \bibinfo{journal}{Astropart. Phys.} \textbf{\bibinfo{volume}{13}},
  \bibinfo{pages}{181} (\bibinfo{year}{2000}), \bibinfo{note}{[Erratum:
  Astropart.Phys. 15, 413--414 (2001)]}, \eprint{hep-ph/9905481}.

\bibitem[{\citenamefont{Boehm et~al.}(2000)\citenamefont{Boehm, Djouadi, and
  Drees}}]{Boehm:1999bj}
\bibinfo{author}{\bibfnamefont{C.}~\bibnamefont{Boehm}},
  \bibinfo{author}{\bibfnamefont{A.}~\bibnamefont{Djouadi}}, \bibnamefont{and}
  \bibinfo{author}{\bibfnamefont{M.}~\bibnamefont{Drees}},
  \bibinfo{journal}{Phys. Rev. D} \textbf{\bibinfo{volume}{62}},
  \bibinfo{pages}{035012} (\bibinfo{year}{2000}), \eprint{hep-ph/9911496}.

\bibitem[{\citenamefont{Ellis et~al.}(2003)\citenamefont{Ellis, Olive, and
  Santoso}}]{Ellis:2001nx}
\bibinfo{author}{\bibfnamefont{J.~R.} \bibnamefont{Ellis}},
  \bibinfo{author}{\bibfnamefont{K.~A.} \bibnamefont{Olive}}, \bibnamefont{and}
  \bibinfo{author}{\bibfnamefont{Y.}~\bibnamefont{Santoso}},
  \bibinfo{journal}{Astropart. Phys.} \textbf{\bibinfo{volume}{18}},
  \bibinfo{pages}{395} (\bibinfo{year}{2003}), \eprint{hep-ph/0112113}.

\bibitem[{\citenamefont{Garny et~al.}(2015)\citenamefont{Garny, Ibarra, and
  Vogl}}]{Garny:2015wea}
\bibinfo{author}{\bibfnamefont{M.}~\bibnamefont{Garny}},
  \bibinfo{author}{\bibfnamefont{A.}~\bibnamefont{Ibarra}}, \bibnamefont{and}
  \bibinfo{author}{\bibfnamefont{S.}~\bibnamefont{Vogl}},
  \bibinfo{journal}{Int. J. Mod. Phys.} \textbf{\bibinfo{volume}{D24}},
  \bibinfo{pages}{1530019} (\bibinfo{year}{2015}), \eprint{1503.01500}.

\bibitem[{\citenamefont{Ibarra et~al.}(2015)\citenamefont{Ibarra, Pierce, Shah,
  and Vogl}}]{Ibarra:2015nca}
\bibinfo{author}{\bibfnamefont{A.}~\bibnamefont{Ibarra}},
  \bibinfo{author}{\bibfnamefont{A.}~\bibnamefont{Pierce}},
  \bibinfo{author}{\bibfnamefont{N.~R.} \bibnamefont{Shah}}, \bibnamefont{and}
  \bibinfo{author}{\bibfnamefont{S.}~\bibnamefont{Vogl}},
  \bibinfo{journal}{Phys. Rev.} \textbf{\bibinfo{volume}{D91}},
  \bibinfo{pages}{095018} (\bibinfo{year}{2015}), \eprint{1501.03164}.

\bibitem[{\citenamefont{Delgado et~al.}(2017)\citenamefont{Delgado, Martin, and
  Raj}}]{Delgado:2016umt}
\bibinfo{author}{\bibfnamefont{A.}~\bibnamefont{Delgado}},
  \bibinfo{author}{\bibfnamefont{A.}~\bibnamefont{Martin}}, \bibnamefont{and}
  \bibinfo{author}{\bibfnamefont{N.}~\bibnamefont{Raj}},
  \bibinfo{journal}{Phys. Rev.} \textbf{\bibinfo{volume}{D95}},
  \bibinfo{pages}{035002} (\bibinfo{year}{2017}), \eprint{1608.05345}.

\bibitem[{\citenamefont{Garny et~al.}(2018)\citenamefont{Garny, Heisig,
  Hufnagel, and L{\"u}lf}}]{Garny:2018icg}
\bibinfo{author}{\bibfnamefont{M.}~\bibnamefont{Garny}},
  \bibinfo{author}{\bibfnamefont{J.}~\bibnamefont{Heisig}},
  \bibinfo{author}{\bibfnamefont{M.}~\bibnamefont{Hufnagel}}, \bibnamefont{and}
  \bibinfo{author}{\bibfnamefont{B.}~\bibnamefont{L{\"u}lf}},
  \bibinfo{journal}{Phys. Rev.} \textbf{\bibinfo{volume}{D97}},
  \bibinfo{pages}{075002} (\bibinfo{year}{2018}), \eprint{1802.00814}.

\bibitem[{\citenamefont{Arina et~al.}(2020)\citenamefont{Arina, Fuks, and
  Mantani}}]{Arina:2020udz}
\bibinfo{author}{\bibfnamefont{C.}~\bibnamefont{Arina}},
  \bibinfo{author}{\bibfnamefont{B.}~\bibnamefont{Fuks}}, \bibnamefont{and}
  \bibinfo{author}{\bibfnamefont{L.}~\bibnamefont{Mantani}},
  \bibinfo{journal}{Eur. Phys. J. C} \textbf{\bibinfo{volume}{80}},
  \bibinfo{pages}{409} (\bibinfo{year}{2020}), \eprint{2001.05024}.

\bibitem[{\citenamefont{Arina et~al.}(2021)\citenamefont{Arina, Fuks, Mantani,
  Mies, Panizzi, and Salko}}]{Arina:2020tuw}
\bibinfo{author}{\bibfnamefont{C.}~\bibnamefont{Arina}},
  \bibinfo{author}{\bibfnamefont{B.}~\bibnamefont{Fuks}},
  \bibinfo{author}{\bibfnamefont{L.}~\bibnamefont{Mantani}},
  \bibinfo{author}{\bibfnamefont{H.}~\bibnamefont{Mies}},
  \bibinfo{author}{\bibfnamefont{L.}~\bibnamefont{Panizzi}}, \bibnamefont{and}
  \bibinfo{author}{\bibfnamefont{J.}~\bibnamefont{Salko}},
  \bibinfo{journal}{Phys. Lett. B} \textbf{\bibinfo{volume}{813}},
  \bibinfo{pages}{136038} (\bibinfo{year}{2021}), \eprint{2010.07559}.

\bibitem[{\citenamefont{Garny et~al.}(2017)\citenamefont{Garny, Heisig,
  L{\"u}lf, and Vogl}}]{Garny:2017rxs}
\bibinfo{author}{\bibfnamefont{M.}~\bibnamefont{Garny}},
  \bibinfo{author}{\bibfnamefont{J.}~\bibnamefont{Heisig}},
  \bibinfo{author}{\bibfnamefont{B.}~\bibnamefont{L{\"u}lf}}, \bibnamefont{and}
  \bibinfo{author}{\bibfnamefont{S.}~\bibnamefont{Vogl}},
  \bibinfo{journal}{Phys. Rev.} \textbf{\bibinfo{volume}{D96}},
  \bibinfo{pages}{103521} (\bibinfo{year}{2017}), \eprint{1705.09292}.

\bibitem[{\citenamefont{D'Agnolo et~al.}(2017)\citenamefont{D'Agnolo,
  Pappadopulo, and Ruderman}}]{DAgnolo:2017dbv}
\bibinfo{author}{\bibfnamefont{R.~T.} \bibnamefont{D'Agnolo}},
  \bibinfo{author}{\bibfnamefont{D.}~\bibnamefont{Pappadopulo}},
  \bibnamefont{and} \bibinfo{author}{\bibfnamefont{J.~T.}
  \bibnamefont{Ruderman}}, \bibinfo{journal}{Phys. Rev. Lett.}
  \textbf{\bibinfo{volume}{119}}, \bibinfo{pages}{061102}
  (\bibinfo{year}{2017}), \eprint{1705.08450}.

\bibitem[{\citenamefont{Junius et~al.}(2019)\citenamefont{Junius,
  Lopez-Honorez, and Mariotti}}]{Junius:2019dci}
\bibinfo{author}{\bibfnamefont{S.}~\bibnamefont{Junius}},
  \bibinfo{author}{\bibfnamefont{L.}~\bibnamefont{Lopez-Honorez}},
  \bibnamefont{and} \bibinfo{author}{\bibfnamefont{A.}~\bibnamefont{Mariotti}},
  \bibinfo{journal}{JHEP} \textbf{\bibinfo{volume}{07}}, \bibinfo{pages}{136}
  (\bibinfo{year}{2019}), \eprint{1904.07513}.

\bibitem[{\citenamefont{Br\"ummer}(2020)}]{Brummer:2019inq}
\bibinfo{author}{\bibfnamefont{F.}~\bibnamefont{Br\"ummer}},
  \bibinfo{journal}{JHEP} \textbf{\bibinfo{volume}{01}}, \bibinfo{pages}{113}
  (\bibinfo{year}{2020}), \eprint{1910.01549}.

\bibitem[{\citenamefont{Maity and Ray}(2020)}]{Maity:2019hre}
\bibinfo{author}{\bibfnamefont{T.~N.} \bibnamefont{Maity}} \bibnamefont{and}
  \bibinfo{author}{\bibfnamefont{T.~S.} \bibnamefont{Ray}},
  \bibinfo{journal}{Phys. Rev. D} \textbf{\bibinfo{volume}{101}},
  \bibinfo{pages}{103013} (\bibinfo{year}{2020}), \eprint{1908.10343}.

\bibitem[{\citenamefont{Blekman et~al.}(2020)\citenamefont{Blekman, Desai,
  Filimonova, Sahasransu, and Westhoff}}]{Blekman:2020hwr}
\bibinfo{author}{\bibfnamefont{F.}~\bibnamefont{Blekman}},
  \bibinfo{author}{\bibfnamefont{N.}~\bibnamefont{Desai}},
  \bibinfo{author}{\bibfnamefont{A.}~\bibnamefont{Filimonova}},
  \bibinfo{author}{\bibfnamefont{A.~R.} \bibnamefont{Sahasransu}},
  \bibnamefont{and} \bibinfo{author}{\bibfnamefont{S.}~\bibnamefont{Westhoff}},
  \bibinfo{journal}{JHEP} \textbf{\bibinfo{volume}{11}}, \bibinfo{pages}{112}
  (\bibinfo{year}{2020}), \eprint{2007.03708}.

\bibitem[{\citenamefont{B\'elanger et~al.}(2022)}]{Belanger:2021smw}
\bibinfo{author}{\bibfnamefont{G.}~\bibnamefont{B\'elanger}}
  \bibnamefont{et~al.}, \bibinfo{journal}{JHEP} \textbf{\bibinfo{volume}{02}},
  \bibinfo{pages}{042} (\bibinfo{year}{2022}), \eprint{2111.08027}.

\bibitem[{\citenamefont{Herms and Ibarra}(2021)}]{Herms:2021fql}
\bibinfo{author}{\bibfnamefont{J.}~\bibnamefont{Herms}} \bibnamefont{and}
  \bibinfo{author}{\bibfnamefont{A.}~\bibnamefont{Ibarra}},
  \bibinfo{journal}{JCAP} \textbf{\bibinfo{volume}{10}}, \bibinfo{pages}{026}
  (\bibinfo{year}{2021}), \eprint{2103.10392}.

\bibitem[{\citenamefont{Petraki et~al.}(2015)\citenamefont{Petraki, Postma, and
  Wiechers}}]{Petraki:2015hla}
\bibinfo{author}{\bibfnamefont{K.}~\bibnamefont{Petraki}},
  \bibinfo{author}{\bibfnamefont{M.}~\bibnamefont{Postma}}, \bibnamefont{and}
  \bibinfo{author}{\bibfnamefont{M.}~\bibnamefont{Wiechers}},
  \bibinfo{journal}{JHEP} \textbf{\bibinfo{volume}{06}}, \bibinfo{pages}{128}
  (\bibinfo{year}{2015}), \eprint{1505.00109}.

\bibitem[{\citenamefont{Asadi et~al.}(2017)\citenamefont{Asadi, Baumgart,
  Fitzpatrick, Krupczak, and Slatyer}}]{Asadi:2016ybp}
\bibinfo{author}{\bibfnamefont{P.}~\bibnamefont{Asadi}},
  \bibinfo{author}{\bibfnamefont{M.}~\bibnamefont{Baumgart}},
  \bibinfo{author}{\bibfnamefont{P.~J.} \bibnamefont{Fitzpatrick}},
  \bibinfo{author}{\bibfnamefont{E.}~\bibnamefont{Krupczak}}, \bibnamefont{and}
  \bibinfo{author}{\bibfnamefont{T.~R.} \bibnamefont{Slatyer}},
  \bibinfo{journal}{JCAP} \textbf{\bibinfo{volume}{02}}, \bibinfo{pages}{005}
  (\bibinfo{year}{2017}), \eprint{1610.07617}.

\bibitem[{\citenamefont{Mitridate et~al.}(2017)\citenamefont{Mitridate, Redi,
  Smirnov, and Strumia}}]{Mitridate:2017izz}
\bibinfo{author}{\bibfnamefont{A.}~\bibnamefont{Mitridate}},
  \bibinfo{author}{\bibfnamefont{M.}~\bibnamefont{Redi}},
  \bibinfo{author}{\bibfnamefont{J.}~\bibnamefont{Smirnov}}, \bibnamefont{and}
  \bibinfo{author}{\bibfnamefont{A.}~\bibnamefont{Strumia}},
  \bibinfo{journal}{JCAP} \textbf{\bibinfo{volume}{1705}}, \bibinfo{pages}{006}
  (\bibinfo{year}{2017}), \eprint{1702.01141}.

\bibitem[{\citenamefont{Harz and Petraki}(2018)}]{Harz:2018csl}
\bibinfo{author}{\bibfnamefont{J.}~\bibnamefont{Harz}} \bibnamefont{and}
  \bibinfo{author}{\bibfnamefont{K.}~\bibnamefont{Petraki}},
  \bibinfo{journal}{JHEP} \textbf{\bibinfo{volume}{07}}, \bibinfo{pages}{096}
  (\bibinfo{year}{2018}), \eprint{1805.01200}.

\bibitem[{\citenamefont{Binder et~al.}(2020)\citenamefont{Binder, Blobel, Harz,
  and Mukaida}}]{Binder:2020efn}
\bibinfo{author}{\bibfnamefont{T.}~\bibnamefont{Binder}},
  \bibinfo{author}{\bibfnamefont{B.}~\bibnamefont{Blobel}},
  \bibinfo{author}{\bibfnamefont{J.}~\bibnamefont{Harz}}, \bibnamefont{and}
  \bibinfo{author}{\bibfnamefont{K.}~\bibnamefont{Mukaida}},
  \bibinfo{journal}{JHEP} \textbf{\bibinfo{volume}{09}}, \bibinfo{pages}{086}
  (\bibinfo{year}{2020}), \eprint{2002.07145}.

\bibitem[{\citenamefont{Brambilla et~al.}(2011)\citenamefont{Brambilla,
  Escobedo, Ghiglieri, and Vairo}}]{Brambilla:2011sg}
\bibinfo{author}{\bibfnamefont{N.}~\bibnamefont{Brambilla}},
  \bibinfo{author}{\bibfnamefont{M.~A.} \bibnamefont{Escobedo}},
  \bibinfo{author}{\bibfnamefont{J.}~\bibnamefont{Ghiglieri}},
  \bibnamefont{and} \bibinfo{author}{\bibfnamefont{A.}~\bibnamefont{Vairo}},
  \bibinfo{journal}{JHEP} \textbf{\bibinfo{volume}{12}}, \bibinfo{pages}{116}
  (\bibinfo{year}{2011}), \eprint{1109.5826}.

\bibitem[{\citenamefont{Yao and M\"uller}(2019)}]{Yao:2018sgn}
\bibinfo{author}{\bibfnamefont{X.}~\bibnamefont{Yao}} \bibnamefont{and}
  \bibinfo{author}{\bibfnamefont{B.}~\bibnamefont{M\"uller}},
  \bibinfo{journal}{Phys. Rev. D} \textbf{\bibinfo{volume}{100}},
  \bibinfo{pages}{014008} (\bibinfo{year}{2019}), \eprint{1811.09644}.

\bibitem[{\citenamefont{Binder et~al.}(2022)\citenamefont{Binder, Mukaida,
  Scheihing-Hitschfeld, and Yao}}]{Binder:2021otw}
\bibinfo{author}{\bibfnamefont{T.}~\bibnamefont{Binder}},
  \bibinfo{author}{\bibfnamefont{K.}~\bibnamefont{Mukaida}},
  \bibinfo{author}{\bibfnamefont{B.}~\bibnamefont{Scheihing-Hitschfeld}},
  \bibnamefont{and} \bibinfo{author}{\bibfnamefont{X.}~\bibnamefont{Yao}},
  \bibinfo{journal}{JHEP} \textbf{\bibinfo{volume}{01}}, \bibinfo{pages}{137}
  (\bibinfo{year}{2022}), \eprint{2107.03945}.

\bibitem[{\citenamefont{Liew and Luo}(2017)}]{Liew:2016hqo}
\bibinfo{author}{\bibfnamefont{S.~P.} \bibnamefont{Liew}} \bibnamefont{and}
  \bibinfo{author}{\bibfnamefont{F.}~\bibnamefont{Luo}},
  \bibinfo{journal}{JHEP} \textbf{\bibinfo{volume}{02}}, \bibinfo{pages}{091}
  (\bibinfo{year}{2017}), \eprint{1611.08133}.

\bibitem[{\citenamefont{Biondini and Laine}(2018)}]{Biondini:2018pwp}
\bibinfo{author}{\bibfnamefont{S.}~\bibnamefont{Biondini}} \bibnamefont{and}
  \bibinfo{author}{\bibfnamefont{M.}~\bibnamefont{Laine}},
  \bibinfo{journal}{JHEP} \textbf{\bibinfo{volume}{04}}, \bibinfo{pages}{072}
  (\bibinfo{year}{2018}), \eprint{1801.05821}.

\bibitem[{\citenamefont{Biondini and Vogl}(2019)}]{Biondini:2018ovz}
\bibinfo{author}{\bibfnamefont{S.}~\bibnamefont{Biondini}} \bibnamefont{and}
  \bibinfo{author}{\bibfnamefont{S.}~\bibnamefont{Vogl}},
  \bibinfo{journal}{JHEP} \textbf{\bibinfo{volume}{02}}, \bibinfo{pages}{016}
  (\bibinfo{year}{2019}), \eprint{1811.02581}.

\bibitem[{\citenamefont{Ellis et~al.}(2015)\citenamefont{Ellis, Luo, and
  Olive}}]{Ellis:2015vaa}
\bibinfo{author}{\bibfnamefont{J.}~\bibnamefont{Ellis}},
  \bibinfo{author}{\bibfnamefont{F.}~\bibnamefont{Luo}}, \bibnamefont{and}
  \bibinfo{author}{\bibfnamefont{K.~A.} \bibnamefont{Olive}},
  \bibinfo{journal}{JHEP} \textbf{\bibinfo{volume}{09}}, \bibinfo{pages}{127}
  (\bibinfo{year}{2015}), \eprint{1503.07142}.

\bibitem[{\citenamefont{Binder}(2019)}]{binderPhD}
\bibinfo{author}{\bibfnamefont{T.}~\bibnamefont{Binder}}, Ph.D. thesis,
  \bibinfo{school}{University of Gottingen} (\bibinfo{year}{2019}),
  \urlprefix\url{http://hdl.handle.net/11858/00-1735-0000-002E-E5E9-2}.

\bibitem[{\citenamefont{Bethe and Salpeter}(1957)}]{Bethe:1957ncq}
\bibinfo{author}{\bibfnamefont{H.~A.} \bibnamefont{Bethe}} \bibnamefont{and}
  \bibinfo{author}{\bibfnamefont{E.~E.} \bibnamefont{Salpeter}},
  \emph{\bibinfo{title}{{Quantum Mechanics of One- and Two-Electron Atoms}}}
  (\bibinfo{year}{1957}).

\bibitem[{\citenamefont{Alwall et~al.}(2014)\citenamefont{Alwall, Frederix,
  Frixione, Hirschi, Maltoni, Mattelaer, Shao, Stelzer, Torrielli, and
  Zaro}}]{Alwall:2014hca}
\bibinfo{author}{\bibfnamefont{J.}~\bibnamefont{Alwall}},
  \bibinfo{author}{\bibfnamefont{R.}~\bibnamefont{Frederix}},
  \bibinfo{author}{\bibfnamefont{S.}~\bibnamefont{Frixione}},
  \bibinfo{author}{\bibfnamefont{V.}~\bibnamefont{Hirschi}},
  \bibinfo{author}{\bibfnamefont{F.}~\bibnamefont{Maltoni}},
  \bibinfo{author}{\bibfnamefont{O.}~\bibnamefont{Mattelaer}},
  \bibinfo{author}{\bibfnamefont{H.~S.} \bibnamefont{Shao}},
  \bibinfo{author}{\bibfnamefont{T.}~\bibnamefont{Stelzer}},
  \bibinfo{author}{\bibfnamefont{P.}~\bibnamefont{Torrielli}},
  \bibnamefont{and} \bibinfo{author}{\bibfnamefont{M.}~\bibnamefont{Zaro}},
  \bibinfo{journal}{JHEP} \textbf{\bibinfo{volume}{07}}, \bibinfo{pages}{079}
  (\bibinfo{year}{2014}), \eprint{1405.0301}.

\bibitem[{\citenamefont{Barbieri et~al.}(1979)\citenamefont{Barbieri, d'Emilio,
  Curci, and Remiddi}}]{Barbieri:1979be}
\bibinfo{author}{\bibfnamefont{R.}~\bibnamefont{Barbieri}},
  \bibinfo{author}{\bibfnamefont{E.}~\bibnamefont{d'Emilio}},
  \bibinfo{author}{\bibfnamefont{G.}~\bibnamefont{Curci}}, \bibnamefont{and}
  \bibinfo{author}{\bibfnamefont{E.}~\bibnamefont{Remiddi}},
  \bibinfo{journal}{Nucl. Phys. B} \textbf{\bibinfo{volume}{154}},
  \bibinfo{pages}{535} (\bibinfo{year}{1979}).

\bibitem[{\citenamefont{Hagiwara et~al.}(1981)\citenamefont{Hagiwara, Kim, and
  Yoshino}}]{Hagiwara:1980nv}
\bibinfo{author}{\bibfnamefont{K.}~\bibnamefont{Hagiwara}},
  \bibinfo{author}{\bibfnamefont{C.~B.} \bibnamefont{Kim}}, \bibnamefont{and}
  \bibinfo{author}{\bibfnamefont{T.}~\bibnamefont{Yoshino}},
  \bibinfo{journal}{Nucl. Phys. B} \textbf{\bibinfo{volume}{177}},
  \bibinfo{pages}{461} (\bibinfo{year}{1981}).

\bibitem[{\citenamefont{Petrelli et~al.}(1998)\citenamefont{Petrelli, Cacciari,
  Greco, Maltoni, and Mangano}}]{Petrelli:1997ge}
\bibinfo{author}{\bibfnamefont{A.}~\bibnamefont{Petrelli}},
  \bibinfo{author}{\bibfnamefont{M.}~\bibnamefont{Cacciari}},
  \bibinfo{author}{\bibfnamefont{M.}~\bibnamefont{Greco}},
  \bibinfo{author}{\bibfnamefont{F.}~\bibnamefont{Maltoni}}, \bibnamefont{and}
  \bibinfo{author}{\bibfnamefont{M.~L.} \bibnamefont{Mangano}},
  \bibinfo{journal}{Nucl. Phys. B} \textbf{\bibinfo{volume}{514}},
  \bibinfo{pages}{245} (\bibinfo{year}{1998}), \eprint{hep-ph/9707223}.

\bibitem[{\citenamefont{Martin and Younkin}(2009)}]{Martin:2009dj}
\bibinfo{author}{\bibfnamefont{S.~P.} \bibnamefont{Martin}} \bibnamefont{and}
  \bibinfo{author}{\bibfnamefont{J.~E.} \bibnamefont{Younkin}},
  \bibinfo{journal}{Phys. Rev. D} \textbf{\bibinfo{volume}{80}},
  \bibinfo{pages}{035026} (\bibinfo{year}{2009}), \eprint{0901.4318}.

\bibitem[{\citenamefont{Le~Bellac}(1996)}]{le_bellac_1996}
\bibinfo{author}{\bibfnamefont{M.}~\bibnamefont{Le~Bellac}},
  \emph{\bibinfo{title}{Thermal Field Theory}}, Cambridge Monographs on
  Mathematical Physics (\bibinfo{publisher}{Cambridge University Press},
  \bibinfo{year}{1996}).

\bibitem[{\citenamefont{Aghanim et~al.}(2020)}]{Planck:2018vyg}
\bibinfo{author}{\bibfnamefont{N.}~\bibnamefont{Aghanim}} \bibnamefont{et~al.}
  (\bibinfo{collaboration}{Planck}), \bibinfo{journal}{Astron. Astrophys.}
  \textbf{\bibinfo{volume}{641}}, \bibinfo{pages}{A6} (\bibinfo{year}{2020}),
  \bibinfo{note}{[Erratum: Astron.Astrophys. 652, C4 (2021)]},
  \eprint{1807.06209}.

\bibitem[{\citenamefont{B{\'e}langer et~al.}(2018)\citenamefont{B{\'e}langer,
  Boudjema, Goudelis, Pukhov, and Zaldivar}}]{Belanger:2018ccd}
\bibinfo{author}{\bibfnamefont{G.}~\bibnamefont{B{\'e}langer}},
  \bibinfo{author}{\bibfnamefont{F.}~\bibnamefont{Boudjema}},
  \bibinfo{author}{\bibfnamefont{A.}~\bibnamefont{Goudelis}},
  \bibinfo{author}{\bibfnamefont{A.}~\bibnamefont{Pukhov}}, \bibnamefont{and}
  \bibinfo{author}{\bibfnamefont{B.}~\bibnamefont{Zaldivar}},
  \bibinfo{journal}{Comput. Phys. Commun.} \textbf{\bibinfo{volume}{231}},
  \bibinfo{pages}{173} (\bibinfo{year}{2018}), \eprint{1801.03509}.

\bibitem[{\citenamefont{Ambrogi et~al.}(2019)\citenamefont{Ambrogi, Arina,
  Backovic, Heisig, Maltoni, Mantani, Mattelaer, and
  Mohlabeng}}]{Ambrogi:2018jqj}
\bibinfo{author}{\bibfnamefont{F.}~\bibnamefont{Ambrogi}},
  \bibinfo{author}{\bibfnamefont{C.}~\bibnamefont{Arina}},
  \bibinfo{author}{\bibfnamefont{M.}~\bibnamefont{Backovic}},
  \bibinfo{author}{\bibfnamefont{J.}~\bibnamefont{Heisig}},
  \bibinfo{author}{\bibfnamefont{F.}~\bibnamefont{Maltoni}},
  \bibinfo{author}{\bibfnamefont{L.}~\bibnamefont{Mantani}},
  \bibinfo{author}{\bibfnamefont{O.}~\bibnamefont{Mattelaer}},
  \bibnamefont{and}
  \bibinfo{author}{\bibfnamefont{G.}~\bibnamefont{Mohlabeng}},
  \bibinfo{journal}{Phys. Dark Univ.} \textbf{\bibinfo{volume}{24}},
  \bibinfo{pages}{100249} (\bibinfo{year}{2019}), \eprint{1804.00044}.

\bibitem[{\citenamefont{Bringmann et~al.}(2018)\citenamefont{Bringmann,
  Edsj\"o, Gondolo, Ullio, and Bergstr\"om}}]{Bringmann:2018lay}
\bibinfo{author}{\bibfnamefont{T.}~\bibnamefont{Bringmann}},
  \bibinfo{author}{\bibfnamefont{J.}~\bibnamefont{Edsj\"o}},
  \bibinfo{author}{\bibfnamefont{P.}~\bibnamefont{Gondolo}},
  \bibinfo{author}{\bibfnamefont{P.}~\bibnamefont{Ullio}}, \bibnamefont{and}
  \bibinfo{author}{\bibfnamefont{L.}~\bibnamefont{Bergstr\"om}},
  \bibinfo{journal}{JCAP} \textbf{\bibinfo{volume}{07}}, \bibinfo{pages}{033}
  (\bibinfo{year}{2018}), \eprint{1802.03399}.

\bibitem[{\citenamefont{Decant et~al.}(2022)\citenamefont{Decant, Heisig,
  Hooper, and Lopez-Honorez}}]{Decant:2021mhj}
\bibinfo{author}{\bibfnamefont{Q.}~\bibnamefont{Decant}},
  \bibinfo{author}{\bibfnamefont{J.}~\bibnamefont{Heisig}},
  \bibinfo{author}{\bibfnamefont{D.~C.} \bibnamefont{Hooper}},
  \bibnamefont{and}
  \bibinfo{author}{\bibfnamefont{L.}~\bibnamefont{Lopez-Honorez}},
  \bibinfo{journal}{JCAP} \textbf{\bibinfo{volume}{03}}, \bibinfo{pages}{041}
  (\bibinfo{year}{2022}), \eprint{2111.09321}.

\bibitem[{\citenamefont{Aaboud et~al.}(2019{\natexlab{a}})}]{ATLAS:2019gqq}
\bibinfo{author}{\bibfnamefont{M.}~\bibnamefont{Aaboud}} \bibnamefont{et~al.}
  (\bibinfo{collaboration}{ATLAS}), \bibinfo{journal}{Phys. Rev. D}
  \textbf{\bibinfo{volume}{99}}, \bibinfo{pages}{092007}
  (\bibinfo{year}{2019}{\natexlab{a}}), \eprint{1902.01636}.

\bibitem[{\citenamefont{Aaboud et~al.}(2019{\natexlab{b}})}]{ATLAS:2018lob}
\bibinfo{author}{\bibfnamefont{M.}~\bibnamefont{Aaboud}} \bibnamefont{et~al.}
  (\bibinfo{collaboration}{ATLAS}), \bibinfo{journal}{Phys. Lett. B}
  \textbf{\bibinfo{volume}{788}}, \bibinfo{pages}{96}
  (\bibinfo{year}{2019}{\natexlab{b}}), \eprint{1808.04095}.

\bibitem[{\citenamefont{Beenakker et~al.}(2016)\citenamefont{Beenakker,
  Borschensky, Kr\"amer, Kulesza, and Laenen}}]{Beenakker:2016lwe}
\bibinfo{author}{\bibfnamefont{W.}~\bibnamefont{Beenakker}},
  \bibinfo{author}{\bibfnamefont{C.}~\bibnamefont{Borschensky}},
  \bibinfo{author}{\bibfnamefont{M.}~\bibnamefont{Kr\"amer}},
  \bibinfo{author}{\bibfnamefont{A.}~\bibnamefont{Kulesza}}, \bibnamefont{and}
  \bibinfo{author}{\bibfnamefont{E.}~\bibnamefont{Laenen}},
  \bibinfo{journal}{JHEP} \textbf{\bibinfo{volume}{12}}, \bibinfo{pages}{133}
  (\bibinfo{year}{2016}), \eprint{1607.07741}.

\bibitem[{\citenamefont{{CMS Collaboration}}(2016)}]{CMS:2016ybj}
\bibinfo{author}{\bibnamefont{{CMS Collaboration}}},
  \bibinfo{journal}{CMS-PAS-EXO-16-036}  (\bibinfo{year}{2016}).

\bibitem[{\citenamefont{Brooijmans et~al.}(2020)}]{Brooijmans:2020yij}
\bibinfo{author}{\bibfnamefont{G.}~\bibnamefont{Brooijmans}}
  \bibnamefont{et~al.}, in \emph{\bibinfo{booktitle}{{11th Les Houches Workshop
  on Physics at TeV Colliders}: {PhysTeV Les Houches}}} (\bibinfo{year}{2020}),
  \eprint{2002.12220}.

\bibitem[{\citenamefont{Aaboud et~al.}(2018{\natexlab{a}})}]{ATLAS:2017tny}
\bibinfo{author}{\bibfnamefont{M.}~\bibnamefont{Aaboud}} \bibnamefont{et~al.}
  (\bibinfo{collaboration}{ATLAS}), \bibinfo{journal}{Phys. Rev. D}
  \textbf{\bibinfo{volume}{97}}, \bibinfo{pages}{052012}
  (\bibinfo{year}{2018}{\natexlab{a}}), \eprint{1710.04901}.

\bibitem[{\citenamefont{Aaboud et~al.}(2018{\natexlab{b}})}]{ATLAS:2017oal}
\bibinfo{author}{\bibfnamefont{M.}~\bibnamefont{Aaboud}} \bibnamefont{et~al.}
  (\bibinfo{collaboration}{ATLAS}), \bibinfo{journal}{JHEP}
  \textbf{\bibinfo{volume}{06}}, \bibinfo{pages}{022}
  (\bibinfo{year}{2018}{\natexlab{b}}), \eprint{1712.02118}.

\bibitem[{\citenamefont{Binder et~al.}(2021)\citenamefont{Binder, Filimonova,
  Petraki, and White}}]{Binder:2021vfo}
\bibinfo{author}{\bibfnamefont{T.}~\bibnamefont{Binder}},
  \bibinfo{author}{\bibfnamefont{A.}~\bibnamefont{Filimonova}},
  \bibinfo{author}{\bibfnamefont{K.}~\bibnamefont{Petraki}}, \bibnamefont{and}
  \bibinfo{author}{\bibfnamefont{G.}~\bibnamefont{White}}
  (\bibinfo{year}{2021}), \eprint{2112.00042}.

\bibitem[{\citenamefont{Yao et~al.}(2021)\citenamefont{Yao, Ke, Xu, Bass, and
  M\"uller}}]{Yao:2020xzw}
\bibinfo{author}{\bibfnamefont{X.}~\bibnamefont{Yao}},
  \bibinfo{author}{\bibfnamefont{W.}~\bibnamefont{Ke}},
  \bibinfo{author}{\bibfnamefont{Y.}~\bibnamefont{Xu}},
  \bibinfo{author}{\bibfnamefont{S.~A.} \bibnamefont{Bass}}, \bibnamefont{and}
  \bibinfo{author}{\bibfnamefont{B.}~\bibnamefont{M\"uller}},
  \bibinfo{journal}{JHEP} \textbf{\bibinfo{volume}{01}}, \bibinfo{pages}{046}
  (\bibinfo{year}{2021}), \eprint{2004.06746}.

\end{thebibliography}

\end{document}